\documentclass[bibyear]{aa} 

\usepackage{natbib,twoopt}
\usepackage[breaklinks=true,colorlinks]{hyperref} 
\bibpunct{(}{)}{;}{a}{}{,}             
\makeatletter
  \newcommandtwoopt{\citeads}[3][][]{\href{http://adsabs.harvard.edu/abs/#3}%
    {\def\hyper@linkstart##1##2{}%
     \let\hyper@linkend\@empty\citealp[#1][#2]{#3}}}
  \newcommandtwoopt{\citealtads}[3][][]{\href{http://adsabs.harvard.edu/abs/#3}%
    {\def\hyper@linkstart##1##2{}%
     \let\hyper@linkend\@empty\citealt[#1][#2]{#3}}}
  \newcommandtwoopt{\citepads}[3][][]{\href{http://adsabs.harvard.edu/abs/#3}%
    {\def\hyper@linkstart##1##2{}%
     \let\hyper@linkend\@empty\citep[#1][#2]{#3}}}
  \newcommandtwoopt{\citetads}[3][][]{\href{http://adsabs.harvard.edu/abs/#3}%
    {\def\hyper@linkstart##1##2{}%
     \let\hyper@linkend\@empty\citet[#1][#2]{#3}}}
  \newcommandtwoopt{\citeyearads}[3][][]%
    {\href{http://adsabs.harvard.edu/abs/#3}
    {\def\hyper@linkstart##1##2{}%
     \let\hyper@linkend\@empty\citeyear[#1][#2]{#3}}}
\makeatother
\usepackage{graphicx}
\usepackage{txfonts}
\usepackage{lscape}
\hypersetup{
    citecolor = {blue},
    urlcolor = {blue},
    linkcolor = {cyan}}
\makeatletter
\renewcommand*\aa@pageof{, page \thepage{} of \pageref*{LastPage}}
\makeatother

\defcitealias{2021A&A...645A.133R}{Paper I}

\begin{document}

   \title{Diagnosing the interstellar medium of galaxies with far-infrared emission lines}

   \subtitle{II. {[\ion{C}{II}]}, {[\ion{O}{I}]}, {[\ion{O}{III}]}, {[\ion{N}{II}]} and {[\ion{N}{III}]} up to z=6}

   \author{Andr\'{e}s F. Ramos Padilla \thanks{\email{ramos@astro.rug.nl}}
          \inst{1,2}
          \and
        L. Wang\inst{1,2} 
        \and
        F. F. S. van der Tak
        \inst{1,2}
        \and 
        S. C. Trager\inst{1}
          }

   \institute{Kapteyn Astronomical Institute, University of Groningen, Landleven 12, 9747 AD, Groningen, the Netherlands \\\
         \and
             SRON Netherlands Institute for Space Research, Landleven 12, 9747 AD, Groningen, the Netherlands \\
             }

   \date{Received September 15, 1996; accepted March 16, 1997}

 
  \abstract
   {Gas cooling processes in the interstellar medium (ISM) are key to understanding how star-formation processes occur in galaxies. Far-infrared (FIR) fine-structure emission lines can be used as a tool to understand the gas conditions and trace the different phases of the ISM.}
   {We model the most important far-infrared (FIR) emission lines throughout cosmic time back to $z=6$ with cosmological hydrodynamical simulations. We study how different physical parameters, such as the interstellar radiation field (ISRF) and metallicity, impact the ISM phases traced by FIR line luminosities and connect those with the star-formation rate (SFR).}
   {We implement a physically motivated multi-phase model of the ISM by post-processing \textsc{EAGLE} cosmological simulation with \textsc{Cloudy} look-up tables. In this model, we assume four phases of the ISM: dense molecular gas, neutral atomic gas, diffuse ionised gas (DIG) and \textsc{Hii} regions.}
   {Our model shows good agreement with the observed luminosity--SFR relation up to $z=6$ in the FIR emission lines analysed and we also provide linear fits. Our predictions also agree with observations in terms of diagnostic diagrams involving various line ratios.}
   {We find that [\ion{C}{II}] is the best SFR tracer of the FIR lines even though it traces multiple ISM phases, while [\ion{O}{III}] and [\ion{N}{II}] can be used to understand the DIG-\textsc{Hii} balance in the ionised phase. In addition, line ratios like [\ion{C}{II}]/[\ion{O}{III}] and [\ion{N}{II}]/[\ion{O}{I}] are useful to trace parameters such as ISRF, metallicity and specific star-formation rate. These results help to interpret observations of FIR line emission from the local Universe to high-$z$ galaxies.}


   \maketitle
%

\section{Introduction}

After the Universe was largely ionised (the period known as ``cosmic dawn'' at $z\lesssim6$--$8$), the composition of the gas and its cooling gradually changed, affecting the star-formation processes in galaxies \citepads{2018PhR...780....1D}. Since then, star formation and black hole accretion processes co-evolved with cosmic time and shaped the evolution of galaxies \citepads{2014ARA&A..52..415M}. The study of the interstellar medium (ISM) in the local Universe allows us to comprehend the current star-formation processes, but the ISM gas cooling budget may not be the same at earlier cosmic epochs \citepads{2013ARA&A..51..105C}. Recent observational data have opened new paths for describing and understanding the ISM gas processes of local galaxies \citep[e.g.][]{2001ApJ...561..766M,2014A&A...568A..62D,2015A&A...578A..53C}. However, the complete picture of how the ISM evolves over cosmic time and how its conditions are connected with star formation in galaxies is not well understood.

Far-infrared (FIR) fine-structure emission lines (Table~\ref{tab:lines}) are dominant in the gas cooling of the ISM and can help us to understand the star-formation processes, from a theoretical \citep[e.g.][]{1985ApJ...291..722T,2012ApJS..203...13G,2022arXiv220205867W} and observational perspective \citep[e.g.][]{2010ApJ...714L.147F,2013ARA&A..51..105C, 2015A&A...578A..53C,2017ApJ...846...32D,2018ApJ...861...95H}. These lines are less affected by dust extinction than optical lines and, at high redshift (hereafter high-$z$), are shifted to the (sub-)mm wavelength range accessible to ground-based telescopes \citep{2020RSOS....700556H,2020ARA&A..58..661F}. The most important ISM emission cooling line is  [\ion{C}{II}] at 158\,$\mu$m. This line traces different phases of the ISM: photo-dissociation regions (PDRs), \textsc{Hii} regions, diffuse ionised gas \citep[DIG; also known as the warm ionised medium (WIM), e.g.][]{2009RvMP...81..969H,2019ARA&A..57..511K}, molecular clouds, and the cold and warm neutral media (CNM and WNM, respectively) \citep[e.g.][]{2012A&A...548A..20C,2017ApJ...845...96C,2017ApJ...842....4A,2020A&A...639A.110A}. [\ion{C}{II}] is thus a very important cooling line in the range of 20--8000\,K due to its low ionisation potential \citep[11.3\,eV compared to 13.6\,eV for hydrogen:][]{2012ApJ...745...49G,2012ApJS..203...13G}. Furthermore, it is easily observable, as its luminosity is around 1\% of the FIR luminosity of galaxies \citep[e.g.][]{1991ApJ...373..423S,2008ApJS..178..280B}. Other important FIR lines include: i) atomic oxygen ([\ion{O}{I}]) at 63 and 145\,$\mu$m, which traces the denser and warmer neutral ISM environments important for star formation \citep{2001ApJ...561..766M,2019ApJ...887...54G}; ii) ionised nitrogen ([\ion{N}{II}]) at 122 and 205\,$\mu$m, which traces the ionised medium from DIG and \textsc{Hii} regions \citep[in the local Universe,][and at high-$z$, e.g. \citealt{2016ApJ...832..151P}]{2012A&A...548A..20C, 2015ApJ...814..133G,2016ApJ...819...69Z,2017ApJ...845...96C,2021A&A...651A..59L}; iii) [\ion{O}{iii}] at 52 and 88\,$\mu$m, which traces \textsc{Hii} regions around young stars \citep[at high-$z$][]{2010ApJ...714L.147F,2014ApJ...780L..18I}; and iv) [\ion{N}{iii}] at 57\,$\mu$m, which also traces \textsc{Hii} regions \citepads{2011A&A...526A.149N}. Using these lines to understand the ISM evolution of galaxies requires a self-consistent model for all of these FIR lines over the $z=0$--$6$ range.

\begin{table}
    \centering
    \caption{Important FIR fine-structure emission lines. Data from \citet{NIST_ASD} and \citetads{2017PASA...34...57S}.}
    \begin{tabular}{ccccc}
\hline
\hline
Line  & $\lambda$ & Transition & IP & $n_{\rm{crit}}$ \\
Species & [$\mu$m] &  & [eV] & [cm$^{-3}$]\\
\hline 
{[\ion{O}{III}]} & 51.81 & $^3$P$_2$ -- $^3$P$_1$ & 35.12 & $3.6 \times 10^3$ \\
{[\ion{N}{III}]} & 57.34 & $^2$P$_{3/2}$ -- $^2$P$_{1/2}$ & 29.60 & $3.0 \times 10^3$\\
{[\ion{O}{I}]} & 63.18 & $^3$P$_2$ -- $^3$P$_1$ & $\cdots$ & $4.7 \times 10^5$\\
{[\ion{O}{III}]} & 88.36 & $^3$P$_1$ - $^3$P$_0$ & 35.12 & $5.1 \times 10^2$ \\
{[\ion{N}{II}]} & 121.80 & $^3$P$_1$ -- $^3$P$_0$ & 14.53 & $3.1 \times 10^2$ \\
{[\ion{O}{I}]} & 145.53 & $^3$P$_1$ -- $^3$P$_0$ & $\cdots$ & $9.5 \times 10^4$ \\
{[\ion{C}{II}]} & 157.68 & $^2$P$_{3/2}$ -- $^2$P$_{1/2}$ & 11.26 & $2.8 \times 10^3$\\
{[\ion{N}{II}]} & 205.30 & $^3$P$_2$ -- $^3$P$_1$ & 14.53 & $4.8 \times 10^1$\\
\hline
\end{tabular}
    \label{tab:lines}
\end{table}

For many decades, the lack of suitable instruments has hampered observations of these FIR lines in high-$z$ galaxies \citepads{2014ApJ...780L..18I}. Fortunately, observations taken with telescopes such as IRAM, CSO, \textit{Herschel} and ALMA have provided the first high-$z$ detections of lines like [\ion{C}{II}], [\ion{N}{II}], [\ion{O}{I}] and [\ion{O}{III}] \citep[e.g.][]{2005A&A...440L..51M,2010ApJ...714L.147F,2011ApJ...740L..29F,2016Sci...352.1559I,2016ApJ...832..209U}. Moreover, recent emission line surveys like ALPINE \citepads{2020A&A...643A...1L} and REBELS \citepads{2021arXiv210613719B} are gathering data for larger samples of high-$z$ galaxies, which are ideal to diagnose the ISM of galaxies over cosmic time.  

With these new observations, different tools can be used to describe the different physical conditions in the ISM of high-$z$ galaxies. The most common and accessible tool is the use of emission line ratios that reflect the physical conditions of the ISM. For example, the [\ion{C}{II}]/[\ion{N}{II}] ratio has been used to describe and constrain whether \textsc{Hii} regions and/or PDRs contribute to the ISM phases of galaxies \citepads{2014ApJ...782L..17D}, and can be used to estimate the amount of ionised gas in [\ion{C}{II}] \citepads{2017ApJ...845...96C}. Another useful line ratio is [\ion{O}{III}]/[\ion{C}{II}], used to understand ionised and neutral gas in high-$z$ galaxies \citep[e.g.][]{2020ApJ...896...93H, 2020MNRAS.499.5136C}. This ratio has the advantage that [\ion{O}{III}] can be brighter than [\ion{C}{II}] at redshifts around the reionisation epoch \citep[$z\gtrsim7$, ][]{2014ApJ...780L..18I, 2016Sci...352.1559I} and can be observed efficiently with ALMA \citepads{2021arXiv210613719B}. Finally, the [\ion{O}{III}]/[\ion{N}{III}] ratio is used to estimate the gas metallicities \citep{2011A&A...526A.149N, 2018ApJ...861...94H}, although other line ratios can also be used for this purpose \citepads{2021A&A...652A..23F}.
 
A more sophisticated tool to describe the physical processes of the ISM is the use of line luminosity predictions from simulations or analytical models. Simple models involve ratios between FIR emission lines and/or the total luminosity in the IR to obtain physical conditions such as hydrogen density, FUV radiation flux or stellar temperatures \citep[e.g.][]{2001ApJ...561..766M,2011ApJ...740L..29F}, while more complex models use magneto-radiation hydrodynamics simulations of the Universe with different radiative transfer codes to predict emission line luminosities \citep[e.g.][and references therein]{2021ApJ...922...88O,2022MNRAS.510.5603K,2022arXiv220102636P}. Some of these studies focus on specific emission lines such as [\ion{C}{II}], [\ion{O}{I}] or [\ion{O}{III}] in analytic models \citep[e.g.][]{2019ApJ...887...54G,2019MNRAS.489....1F,2020MNRAS.499.3417Y} and simulations \citep[e.g.][]{2018MNRAS.481L..84M,2020ApJ...905..102L,2021A&A...645A.133R}, while others study interesting line ratios like [\ion{O}{III}]/[\ion{C}{II}] \citep{2020MNRAS.498.5541A,2021MNRAS.505.5543V}. A large effort has also been made to model various FIR emission lines in models at different cosmic times \citep[e.g.][]{2013MNRAS.433.1567V,2015ApJ...813...36V,2015ApJ...814...76O,2017ApJ...846..105O,2021ApJ...922...88O,2019MNRAS.482.4906P,2019ApJ...887..142S,2019MNRAS.487.5902K,2022MNRAS.510.5603K,2019MNRAS.487.1689P,2021ApJ...911..132Y,2022ApJ...929..140Y}. However, these studies do not examine the emission of these FIR lines in a consistent way in terms of their redshift evolution, from the local Universe to the epoch of reionisation, due to computational constraints or focus on certain cosmic epochs (with the exception of \citet{2019MNRAS.482.4906P} which modelled the [\ion{C}{II}] line). Therefore, we need better models to understand current observations consistently.

With this in mind, we aim to predict luminosities of the main FIR lines in a cosmological context through the use of cosmological hydrodynamical simulations to infer the physical conditions of galaxies across a wide range of redshifts. The goal of this paper is to test the impact of physical parameters on the FIR emission lines tracing different ISM phases in galaxies. We will use these predictions as diagnostic tools, which will be useful for both current and future observations. To do this, we model the emission of FIR lines by post-processing the hydrodynamical simulations of the Evolution and Assembly of GaLaxies and their Environments (\textsc{EAGLE}) project \citep{2015MNRAS.446..521S,2015MNRAS.450.1937C} with a physically motivated model of the ISM presented in \citetads[][hereafter \citetalias{2021A&A...645A.133R}]{2021A&A...645A.133R}. We use \textsc{Cloudy} \citepads{2017RMxAA..53..385F} cooling tables \citepads{2020MNRAS.497.4857P} to predict the emission from different ISM phases in galactic environments. Throughout this paper, we assume the $\Lambda$CDM cosmology from the \citetads{2014A&A...571A..16P} results ($\Omega= 0.307$, $\Omega_\Lambda$=0.693, $H_0 =67.7$ km s$^{-1}$ Mpc$^{-1}$ and $\sigma_8=0.8288$).

In this paper, we first briefly describe the simulation data and the ISM model that we use to predict FIR emission line luminosities (Sect.~\ref{sec:meth}). Next, we present the results of the FIR emission line predictions from the simulations and how they compare with the observations from the local Universe all the way out to $z{\sim}6$ (Sect.~\ref{sec:lines}). In Sect.~\ref{sec:diag} we evaluate some FIR diagnostic diagrams used in various high-$z$ studies. After that, we discuss the potential systematic uncertainties that can affect the predictions and the comparisons with observations (Sect.~\ref{sec:systematics}). Finally, we present our conclusions in Sect.~\ref{sec:concl}.

\section{Methodology}\label{sec:meth}

In this section, we first describe the sets of simulations that we use in this work (Sect.~\ref{sec:EAGLE}), then we briefly explain the initial model used to characterise the structure of the ISM (Sect.~\ref{sec:ISMmodel}). Finally, we present in detail the addition of \textsc{Hii} regions as a new ISM phase (Sect.~\ref{sec:HIIregions}) in our model.  

\subsection{The EAGLE simulations}\label{sec:EAGLE}

\textsc{EAGLE} \citep{2015MNRAS.446..521S,2015MNRAS.450.1937C} is a suite of cosmological hydrodynamical simulations which were run using a modified version of \textsc{GADGET-3} \citepads[last described by][]{2005MNRAS.364.1105S}, a smoothed-particle hydrodynamics (SPH) code. Briefly, \textsc{EAGLE} adopts an SPH pressure-entropy parameterisation 
following \citetads{2013MNRAS.428.2840H}. The simulations include radiative cooling and photo-electric heating \citepads{2009MNRAS.393...99W}, star formation \citepads{2008MNRAS.383.1210S}, stellar evolution and
mass loss \citepads{2009MNRAS.399..574W}, black hole growth \citep{2005MNRAS.361..776S,2015MNRAS.454.1038R}, and feedback from star formation and active galactic nuclei (AGN) \citepads{2012MNRAS.426..140D}. The simulations provide the properties for gas, dark matter, stellar and supermassive black hole SPH particles. 

For this work, based on the results from \citetalias{2021A&A...645A.133R}, we used two simulations from the EAGLE suite: \textsc{Ref-L0100N1504} and \textsc{Recal-L0025N0752}, as described in Table~\ref{tab:EAGLESims}. The main differences between the two simulations are the box-size of the simulation (100 and 25 cMpc (comoving Mpc), respectively), the mass resolution (${\sim} 10^6\, \mathrm{M_{\sun}}$ and $ {\sim} 10^5\, \mathrm{M_{\sun}}$, respectively), and the calibration of the physical parameters of the subgrid routines to reproduce the galaxy stellar mass function \citepads[GSMF;][]{2015MNRAS.446..521S}. Both simulations are similar in terms of ``weak convergence'', which means numerical results converge in different simulations after re-calibrating the sub-grid parameters \citep{2015MNRAS.450.4486F,2015MNRAS.446..521S}.

\begin{table*}
    \centering
    \caption{\textsc{EAGLE} simulations used in this work. The box-size, number of particles and gas particle mass define the resolution of the simulation. The right columns show the number of galaxies used in this work for a given simulation at each redshift.}
    \label{tab:EAGLESims}
    \begin{tabular}{lcccccccccc}
    \hline
    \hline
    Name in & Box-size & Number of & SPH Gas mass & \multicolumn{7}{c}{Number of galaxies}\\
    \cline{5-11}
    \citetads{2015MNRAS.446..521S}&(cMpc)& particles& ($\mathrm{M_{\sun}}$) & $z=0$& $z=1$& $z=2$& $z=3$& $z=4$& $z=5$& $z=6$\\
    \hline
    \textsc{Recal-L0025N0752}&25& $752^{3}$ & $ 2.26\times 10^5$ &415 & 426 & 339 & 252 & 154 & 75 & 37\\
    \textsc{Ref-L0100N1504} & 100 & $1\,504^{3}$&$1.81 \times 10^6$&1\,000&1\,000&1\,000&1\,000&1\,000&1\,000& 579\\
    \hline
    \end{tabular} 
\end{table*}

In this study, we retrieve simulated galaxies from the SPH data \citepads{2017arXiv170609899T} by using textsc{FoF} (Friends-of-Friends) and \textsc{SUBFIND} algorithms \citep{2001MNRAS.328..726S,2009MNRAS.399..497D} in the dark matter halos. With these algorithms, the sub-halos containing the particle with the lowest value of the gravitational potential are called ``central'' galaxies. We focus on these ``central'' galaxies to estimate line luminosities. We use galaxies with at least 300 star particles (i.e.\ stellar masses higher than ${\sim} 10^8\, \mathrm{M_{\sun}}$ and ${\sim} 10^{8.5}\, \mathrm{M_{\sun}}$ for \textsc{Recal-L0025N0752} and \textsc{Ref-L0100N1504}, respectively) within 30\,pkpc (proper kpc) from the centre of the potential. We selected our sample of galaxies from the \textsc{EAGLE} database \citepads{2016A&C....15...72M} in redshifts (the closest snapshot) between $z=0$ and $z=6$ in steps of $\Delta z=1$, using a single snapshot at each redshift (the snapshot closest in time; e.g. at $z=6$, we use the $z=5.97$ snapshot from \textsc{EAGLE}), where the total number of galaxies depends on the simulation used. The $z=6$ cutoff was selected due to the availability of observational data and the number of galaxies recovered in \textsc{EAGLE} required to compare them statistically. For \textsc{Recal-L0025N0752}, we select all retrieved galaxies, while for \textsc{Ref-L0100N1504}, we randomly select up to 1\,000 galaxies that fulfil the previous conditions per redshift. In the last columns of Table~\ref{tab:EAGLESims}, we present the total number of galaxies used per redshift slice in each of the simulations.

We selected a total sample of 8\,277 galaxies simulated with EAGLE at redshifts between $z=0$ and $z=6$. In each of those galaxies, we model the emission coming from the eight FIR emission lines (Table~\ref{tab:lines}) that trace different phases of the ISM.

\subsection{The multi-phase ISM model}\label{sec:ISMmodel}

\begin{figure*}
    \centering
    \includegraphics[width=\textwidth]{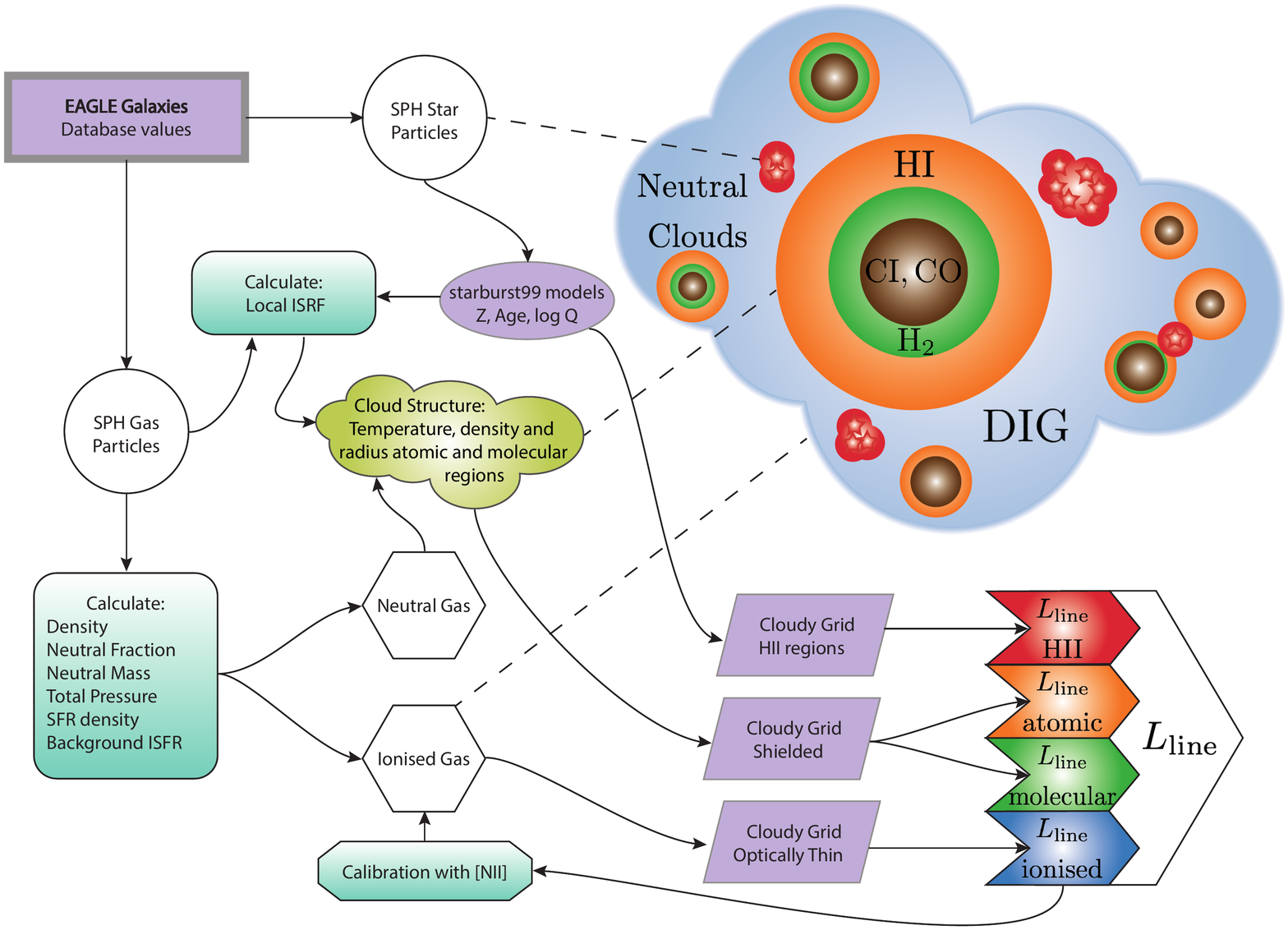}
    \caption{Flowchart of the sub-grid procedures applied to the SPH to simulate FIR line emission in post-processing. This flowchart is similar to the one presented in \citetalias{2021A&A...645A.133R} with the main difference being the added \textsc{Hii} regions as a new ISM phase. The dashed lines connect the gas and star environments to the ingredients of the model.}
    \label{fig:F1_HIImodel}
\end{figure*}

To predict these emission lines, we used the ISM model presented in \citetalias{2021A&A...645A.133R}, where the luminosity estimations from the [\ion{C}{II}] emission line at 158\,$\mu$m  in the local Universe ($z=0$) showed good agreement with observations. However, this model must be improved if we want to properly account for other FIR lines such as [\ion{O}{iii}] and [\ion{N}{iii}], which probe denser ionised regimes. Therefore, we add \textsc{Hii} regions as a new phase in our ISM model. In Fig.~\ref{fig:F1_HIImodel}, we illustrate the path from the EAGLE simulations (Sect.~\ref{sec:EAGLE}) to the total luminosity of the lines in the current model. The main difference between \citetalias{2021A&A...645A.133R} and this work is the addition of the contribution to the line luminosities from \textsc{Hii} regions (Sect.~\ref{sec:HIIregions}). In this subsection, we briefly explain the model presented in \citetalias{2021A&A...645A.133R}.

After selecting the sample of galaxies for which we want to calculate the line luminosity, we retrieve and post-process the gas and star particle data of the galaxies in the simulation. Physical properties such as total hydrogen number density were estimated for all gas particles using the information available in the SPH particle data. For example, we calculated the total hydrogen number density as $n(\mathrm{H})=\frac{\rho X_H }{m_{H}}$, with $m_{H}$ the hydrogen mass, $\rho$ the density and $X_{H}$ the SPH weighted hydrogen abundance. In addition, the fraction of neutral hydrogen was estimated for all gas particles according to \citetads[][]{2013MNRAS.430.2427R}, following ionisation equilibrium as described in Sect. 2.2.1 in \citetalias{2021A&A...645A.133R}. We calculated the background interstellar radiation field (ISRF) from the star formation rate (SFR) surface density of the gas particles and the local ISRF from the star particles as described by \citetads{2017ApJ...846..105O}. The local ISRF is estimated with \texttt{starburst99} models \citepads{2014ApJS..212...14L} for star particles with a distance below the smoothing length of a given gas particle (as described in Sect. 2.2.1 in \citetalias{2021A&A...645A.133R}). The sum of the background ISRF and the local ISRF defines the total ISRF impinging on the neutral cloud. With the fraction of neutral hydrogen, we split the gas into neutral and ionised components. The neutral gas mass and the ISRF are used to estimate the sizes of the neutral clouds, while the ionised mass is used to describe the diffuse ionised gas (DIG).

The neutral clouds are defined as concentric spheres with densities following a Plummer profile. Each ISM phase in the neutral cloud is defined by a different radius as described in Sect. 2. in \citetalias{2021A&A...645A.133R}. The transition between atomic and molecular hydrogen defines the limit between the neutral atomic gas and the dense molecular gas, while the transition between ionised to neutral carbon defines the limit for the inner core region of the cloud that is completely shielded from FUV radiation. This definition assumes that the inner core can only be traced by neutral species such as CO, and the inner core is therefore ignored in the estimation of the total luminosity of the FIR emission lines\footnote{[\ion{O}{I}] emission can come also from the inner core. However, as our results show, the contribution of this inner core to the [\ion{O}{I}] luminosity is usually very small.}. 

On the other hand, the volumetric structure of the DIG is assumed to be spherical with a radius drawn from a smoothed broken power-law distribution. The parameters of the power-law distribution are derived based on calibrating the [\ion{N}{ii}] line (at 122 and 205\,$\mu$m) predictions of 492 galaxies from the \textsc{Ref-L0100N1504} and 200 galaxies from \textsc{Recal-L0025N0752} simulations to the observational dataset from \citetads{2015ApJ...799...21S}, \citetads{2016ApJS..226...19F} and \citetads{2019A&A...626A..23C} (containing 8, 53 and 2 galaxies, respectively, of a sample of 63 galaxies). This led us to DIG clouds with average sizes of ${\sim}$900 pc for \textsc{Recal-L0025N0752} galaxies and ${\sim}$2 kpc for \textsc{Ref-L0100N1504}, which is around 3 times the  maximum softening length of the original simulations. 

These structures define the phases of the ISM that contribute to the total luminosity of the emission line: DIG, neutral atomic and dense molecular gas. The estimation of the luminosities for each phase is obtained by using \textsc{Cloudy}  \citepads[v17.01, ]{2017RMxAA..53..385F} cooling tables of shielded and optically thin gas \citepads{2020MNRAS.497.4857P}. The sum of those phases then gives us the total luminosity for a FIR emission line. For a complete description of the model, we refer the reader to Sect. 2.2 and 2.3 in \citetalias{2021A&A...645A.133R}.

\subsection{\textsc{Hii} regions as a new ISM phase}\label{sec:HIIregions}

The luminosity estimations for FIR lines such as [\ion{O}{iii}] and [\ion{N}{iii}], which require high ionisation potentials (35.1 and 29.6\,eV, respectively), can only be found in \textsc{Hii} regions and other dense, ionised regimes. The addition of this phase as an ISM component to the model allows us to compare the predicted luminosities of these lines with observations. Therefore, we update the model described in \citetalias{2021A&A...645A.133R} to simulate the \textsc{Hii} regions production of the most prominent FIR lines, including [\ion{C}{II}], [\ion{O}{III}], [\ion{O}{I}], as shown in Table~\ref{tab:lines}, in order to infer the ISM conditions in galaxies from the local Universe out to $z\approx6$.

In \citetalias{2021A&A...645A.133R}, we use \texttt{starburst99} to calculate the ISRF from the stars close to gas particles (the distance between the particles is less than one smoothing length). Now, we also use the spectrum from \texttt{starburst99} to calculate the emission coming from \textsc{Hii} regions. To generate this spectrum, we adopted the Geneva stellar models \citepads{1992A&AS...96..269S} with standard mass loss for five metallicities ($Z=0.001, 0.004, 0.008, 0.020, \mathrm{and}\, 0.040$). We split \texttt{starburst99} grid values in young ($\leq100$Myr) and old ages ($>100$Myr), to improve our estimations at younger ages. For young ages, we estimate the parameters every 1 Myr, while for old ages we calculate 100 steps on a logarithmic scale up to 10 Gyr. We assume a total stellar mass of $10^4\,\mathrm{M_{\sun}}$, with a Kroupa initial mass function (IMF). The SPH star particles are divided to match the stellar mass from the \texttt{starburst99} models, assuming a random exponential distribution of the original ages of the SPH particles. By doing this we try to avoid the poor sampling of star formation that can affect luminosity estimates, especially in \textsc{Hii} regions.

We use a a photoionisation model from \textsc{Cloudy} to simulate the line emissitivites of \textsc{Hii} regions based on \texttt{starburst99} spectra of their underlying stellar populations. We calculate the stellar atmospheres, or spectral energy distributions (SEDs), in the photoionisation models for a given age, metallicity and the ionising photon flux ($Q$). In this way, the stellar mass and metallicity from the star SPH particles are used to obtain $Q$ coming from a stellar cluster from the \texttt{starburst99} grids. Thus, we use these three physical parameters (age, metallicity and $Q$) to construct spherical clouds where the emissivities depend only on them. The range of values for these parameters is presented in Table~\ref{tab:HIIGridValues}, totalling 25\,200 grid points for all redshifts in this work. 

\begin{table}
    \caption{Sampling of the properties of \textsc{Hii} regions in the \textsc{Cloudy} grid used in this work. The resulting number of grid points is 3600 per redshift.}
    \label{tab:HIIGridValues}
    \centering
    \begin{tabular}{lcccc}
    \hline
    \hline
    Parameter & Unit & Min. & Max. & Interval\\
    \hline
    Metallicity &[$Z_{\sun}$]& 0.075 & 2.991 & 0.324\\
    $\log(Q)$ &[s$^{-1}$] &52&44&1\\
    $\log(\mathrm{Age})$ &[Gyr] &$-$3.0&0.9&0.1\\
    \hline
    \end{tabular}    
\end{table}

In terms of the structure, we assume a fixed density of ${\sim} 30\, \mathrm{cm}^{-3}$ to resemble classical \textsc{Hii} regions similar to the Str\"{o}mgren sphere, where densities are in the range of 10--100\,$\mathrm{cm}^{-3}$ \citepads{2011piim.book.....D}. Choosing a higher or lower density affects the \textsc{Hii} luminosity of some lines as we describe in Sect.~\ref{sec:systematics}. In ionisation balance the Str\"{o}mgren radius is
\begin{equation}
    R_{\mathrm{Str}} = \left( \frac{3 Q}{4\pi\, n(\mathrm{H})^{2}\, \alpha_{B}} \right)^{1/3},
    \label{eq:RStr}
\end{equation}
with $Q$ the ionising photon flux in units of $s^{-1}$, $n(\mathrm{H})$ the total hydrogen number density in units of cm$^{-3}$ and $\alpha_{B}$ is the Case B recombination rate coefficient. 

These \textsc{Hii} regions are radiation bound in terms of ionised hydrogen. Once the ratio of ionised hydrogen to the total hydrogen density drops below 5\%, the calculation stops, which defines the outer radius. The inner radius is set to 1\% of the expected Str\"{o}mgren sphere radius (Eq.~\ref{eq:RStr}), as in \citetads{2020MNRAS.499.3417Y}, allowing us to obtain ``ultracompact'' \textsc{Hii} regions, which can have sizes of ${\sim} 0.03\, \mathrm{pc}$ \citepads{2005IAUS..227..111K}. We calculate the emissivities without further iteration on the assumed density, which is adequate for clouds with densities typical of \textsc{Hii} regions \citepads{1996hbic.book.....F}. For the rest of the parameters in \textsc{Cloudy} we choose a similar configuration as the one used by \citetads{2020MNRAS.497.4857P}. 

The \texttt{starburst99} outputs are also used to calculate the $L_{\mathrm{FUV}}$ and $Q$ for each star SPH particle. Thus, the look-up luminosity tables from \textsc{Cloudy} are used to interpolate the star SPH particles in terms of age, metallicity and $Q$. As a result, we obtain the respective \textsc{Hii} luminosity for each star particle for the FIR emission lines. This luminosity is then used as part of the total contribution of the ISM phases (DIG + neutral atomic gas + dense molecular gas + \textsc{Hii} regions) for a given simulated galaxy. 

\section{Individual line luminosities}\label{sec:lines}

In this section, we present the predictions of individual lines from our model and compare them with observations. For each line, we first examine the relation between the line luminosity and the SFR of the galaxy at different cosmic epochs. Our predictions of the line luminosity--SFR relation are also compared with observational measurements collected and homogenised from published work (see Appendix~\ref{app:obs_sample}). Then we check the contribution to the line luminosities from each of the ISM phases, i.e., the DIG, \textsc{Hii} regions, neutral atomic and dense molecular gas. We mainly discuss the results of the five FIR emission lines: [\ion{C}{ii}] at 158\,$\mu$m, [\ion{O}{i}] at 63\,$\mu$m, [\ion{N}{ii}] at 205\,$\mu$m, [\ion{O}{iii}] at 88\,$\mu$m, and [\ion{N}{iii}] at 57\,$\mu$m. The other three emission lines listed in Table~\ref{tab:lines} behave similarly to their aforementioned pairs (e.g.\ [\ion{O}{iii}] at 52\,$\mu$m is similar to [\ion{O}{iii}] at 88\,$\mu$m). We release all the data in a Zenodo repository as described in Appendix~\ref{app:estimations_dataset}. 

\subsection{\texorpdfstring{[\ion{C}{II}]}{} 158 \texorpdfstring{$\mu$m}{}} \label{sec:CII}

\subsubsection{The SFR--\texorpdfstring{$L_{\mathrm[\ion{C}{II}]}$}{} relationship}

\begin{figure*}
    \centering
    \includegraphics[width=\textwidth]{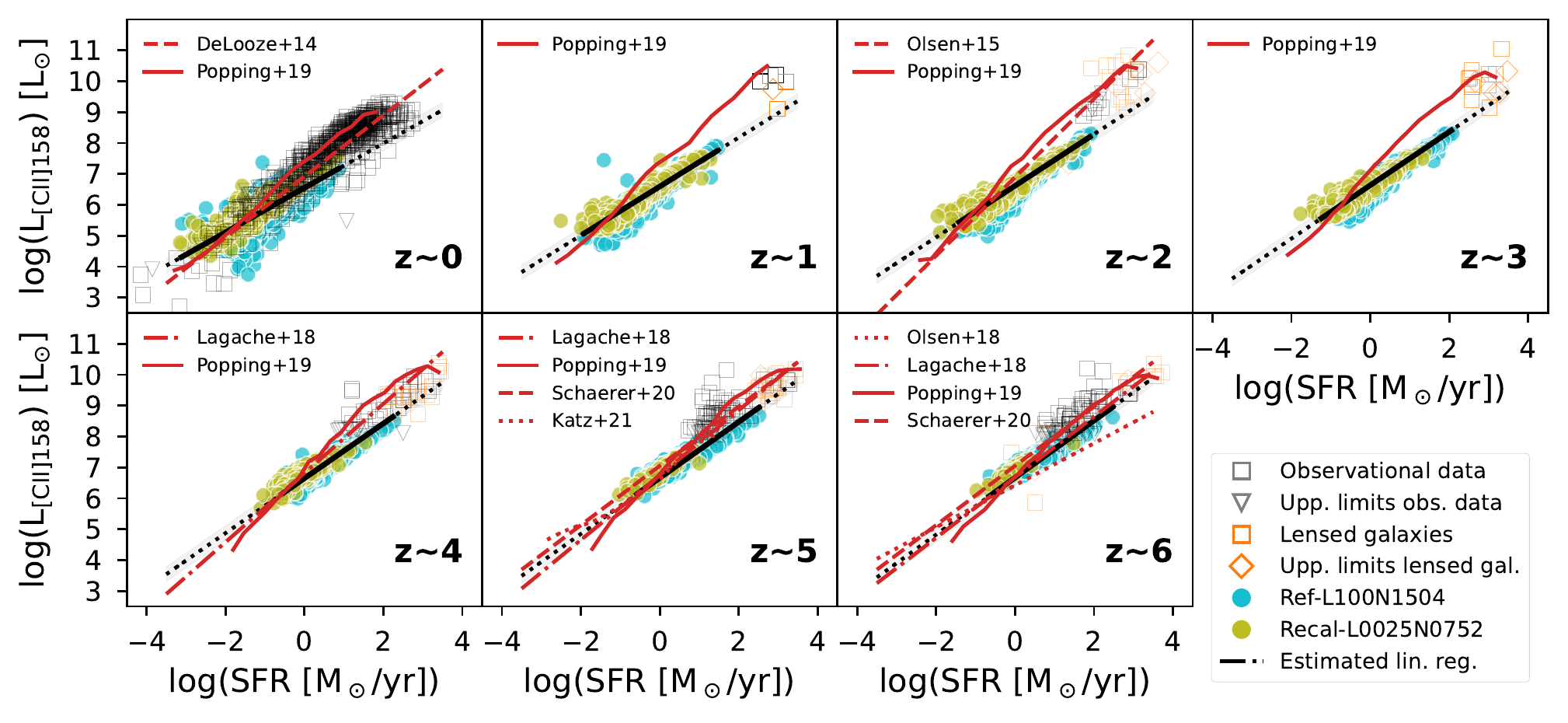}
    \caption{SFR--$L_{\mathrm[\ion{C}{II}]}$ relation for all redshift slices used in this work. We compare the obtained relations from the EAGLE simulations \textsc{Ref-L100N1504} and \textsc{Recal-L0025N0752} with predictions from other simulations \citep{2015ApJ...814...76O,2018ApJ...857..148O, 2022MNRAS.510.5603K}, semi-analytic models \citep{2018A&A...609A.130L,2019MNRAS.482.4906P} and linear relations derived from observations \citep{2014A&A...568A..62D,2020A&A...643A...3S}. Linear relations inferred from our models are shown as black solid lines over the dynamic range covered by the simulations and extrapolated to lower and higher SFRs as black dotted lines, with the grey shaded area  representing the 1\,$\sigma$ error. Collections of observational data (Appendix~\ref{app:obs_sample}) are plotted as grey squares for detections and as grey triangles for upper limits. For lensed galaxies, the markers are plotted in  and the upper limits that affect both the SFR and luminosity are plotted as diamonds.}
    \label{fig:LSRCII}
\end{figure*}

The most important and brightest of these FIR lines is [\ion{C}{ii}] at 158\,$\mu$m. This line follows a clear trend with SFR \citep[e.g.][]{1991ApJ...373..423S,2008ApJS..178..280B,2010ApJ...724..957S}, and it is used as a SFR tracer at different redshifts \citep[e.g.][]{2014A&A...568A..62D,2015ApJ...800....1H, 2020A&A...643A...3S}. In Fig.~\ref{fig:LSRCII}, we show the relationships between SFR and [\ion{C}{ii}] luminosity ($L_{\mathrm[\ion{C}{II}]}$) for the seven redshift slices analysed in this work. We compare the predictions from the EAGLE simulations \textsc{Ref-L100N1504} and \textsc{Recal-L0025N0752} with predictions from other simulations \citep{2015ApJ...814...76O,2018ApJ...857..148O,2022MNRAS.510.5603K}, semi-analytic models \citep{2018A&A...609A.130L,2019MNRAS.482.4906P} and linear relationships derived from observations \citep{2014A&A...568A..62D,2020A&A...643A...3S}. We also plot the linear relationships that we infer from our model (see Appendix~\ref{app:linregcoef}) and extrapolate them outside the dynamic range covered by the simulations but within $10^{-3.5} \mathrm{M_{\sun}} \, \mathrm{yr}^{-1} < \textrm{SFR} < 10^{3.5}\,\mathrm{M_{\sun}} \, \mathrm{yr}^{-1}$. With this extrapolation, we can compare the observations with high SFR values that the simulations do not cover. This is necessary as high-$z$ observations generally only include galaxies with very high SFR due to sensitivity limits. In general, the agreement between our models, observations and other models is good within the typical scatter ${\sim} 0.4$\, dex \citepads{2014A&A...568A..62D}, especially at $z=0$ and $z=6$, where there are more observational constraints.

In \citetads{2021A&A...645A.133R} we showed that the SFR--$L_{\mathrm[\ion{C}{II}]}$ relationship at $z=0$ could be reproduced with a model similar to that implemented in this work. Therefore, it is not surprising that the current model still reproduces this SFR--$L_{\mathrm[\ion{C}{II}]}$ relationship. Compared to \citetads{2019MNRAS.482.4906P}, the only other model that covers the same redshift range as this work for [\ion{C}{ii}], we find a flatter slope in the SFR--$L_{\mathrm[\ion{C}{II}]}$ relationship. Especially at $z=1$--$3$, the differences in the slopes can lead to up to ${\sim}$1.8\, dex change in $L_{\mathrm[\ion{C}{II}]}$ at SFRs around $1000\,\mathrm{M_{\sun}} \, \mathrm{yr}^{-1}$, but the relationships are more similar at other redshifts. The reason for these discrepancies may reside in the different galaxy formation physics in \textsc{EAGLE} and the \textssc{Santa Cruz SAM} used in \citetads{2019MNRAS.482.4906P}. Unfortunately, the comparison between the two galaxy formation models is out of the scope of this work.

At $z=2$, the linear relationship of \citetads{2015ApJ...814...76O} predicted from seven simulated galaxies shows a behaviour similar to that of \citetads{2019MNRAS.482.4906P} and agrees with the estimated scatter of our linear regression (0.2\,dex) at SFRs around $1\,\mathrm{M_{\sun}} \, \mathrm{yr}^{-1}$. Over the redshift range $z=1$--$3$, the extrapolation of our relation show a potential small offset compared with observations. However, this discrepancy is not significant, taking into account the small sample size, large scatter (around 0.5\,dex), potential bias towards galaxies with high line luminosities, and systematics in deriving luminosities and SFR. Furthermore, altering the assumptions in our modelling process could affect our predictions, as we show in Sect.~\ref{sec:systematics}. 

At $z=4$--$6$, most of the models and observations match well with the linear relationships derived from our predictions. Similar predictions from \citetads{2015ApJ...813...36V}, \citetads{2020ApJ...905..102L} and \citetads{2020MNRAS.499.5136C}, which are not shown in the plot, also agree at $z=6$, indicating that the SFR--$L_{\mathrm[\ion{C}{II}]}$ relationship can be tight at higher redshift, even if there is some scatter in the data related to observational errors \citepads{2020A&A...643A...3S}. This demonstrates that our physically motivated model of the ISM is valid not only for estimating the luminosity of [\ion{C}{ii}] in the local Universe but also up to $z=6$.

\subsubsection{\texorpdfstring{[\ion{C}{II}]}{} deficit}

\begin{figure}
    \centering
    \includegraphics[width=\columnwidth]{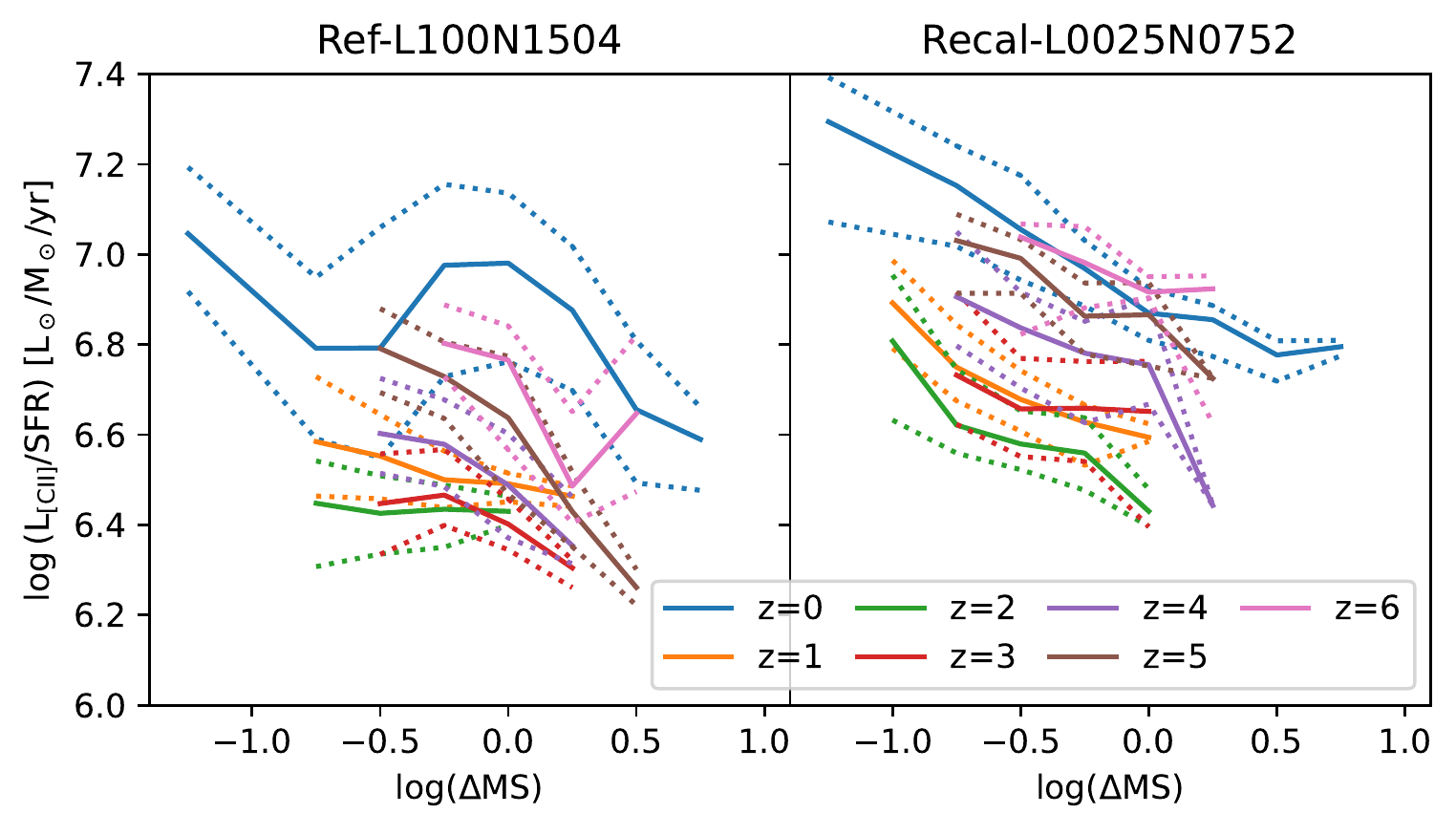}
    \caption{The ratio [\ion{C}{II}]/SFR as a function of the offset from the star-forming main-sequence $\Delta$MS for the simulations \textsc{Ref-L100N1504} (left) and \textsc{Recal-L0025N752} (right). We show the median values of the different redshifts (solid lines) and their 25th and 75th percentiles (dotted lines). We only show the bins with more than 3\% of the total sample for the respective simulation.}
    \label{fig:DMS}
\end{figure}

Although a linear SFR--$L_{\mathrm[\ion{C}{II}]}$ relation is commonly used to assume that $L_{\mathrm[\ion{C}{II}]}$ is a good SFR tracer, observational data of local galaxies show a decrease of $L_{\mathrm[\ion{C}{II}]}$ at IR luminosities above $10^{12}\, \mathrm{L_{\sun}}$, known as the ``[\ion{C}{II}] deficit'' \citep[e.g.][]{2017ApJ...846...32D,2018ApJ...861...95H}. However, recent observations of $z>4$ galaxies show no evidence of such a deficit \citep[e.g.]{2019ApJ...881..124M,2020MNRAS.499.5136C,2020A&A...643A...3S}. In \citetads{2021A&A...645A.133R}, we examined this deficit by comparing the SFR--$L_{\mathrm[\ion{C}{II}]}$ relationship with deviations from the star-forming main-sequence (MS), following the suggestion of \citetads{2018ApJ...861...95H}. In this work, we compare the $L_{\mathrm[\ion{C}{II}]}$/SFR ratio with the offset from the MS ($\Delta\mathrm{MS}$) in Fig.~\ref{fig:DMS}. We estimate the specific SFR ($\mathrm{sSFR}=\mathrm{SFR}/\mathrm{M}_{\star}$) of our galaxies and normalise this by the derived sSFR for the MS from \citetads{2014ApJS..214...15S}, which gives us the $\Delta\mathrm{MS}$.

We find that the $L_{\mathrm[\ion{C}{II}]}$/SFR ratio almost always decreases with $\Delta\mathrm{MS}$. If the decrease in the $L_{\mathrm[\ion{C}{II}]}$/SFR ratio extends to a higher $\Delta\mathrm{MS}$, it supports the idea that starburst galaxies may not follow the SFR--$L_{\mathrm[\ion{C}{II}]}$ relationship shown in Fig.~\ref{fig:LSRCII} \citep{2018ApJ...861...95H,2019MNRAS.489....1F}. Therefore, $L_{\mathrm[\ion{C}{II}]}$ may not be a good tracer of SFR for starburst galaxies, as we discuss in Sect.~\ref{subsec:summFIR}. 

\subsubsection{Contribution to \texorpdfstring{$L_{\mathrm[\ion{C}{II}]}$}{} from each ISM phase}

\begin{figure}
    \centering
    \includegraphics[width=\columnwidth]{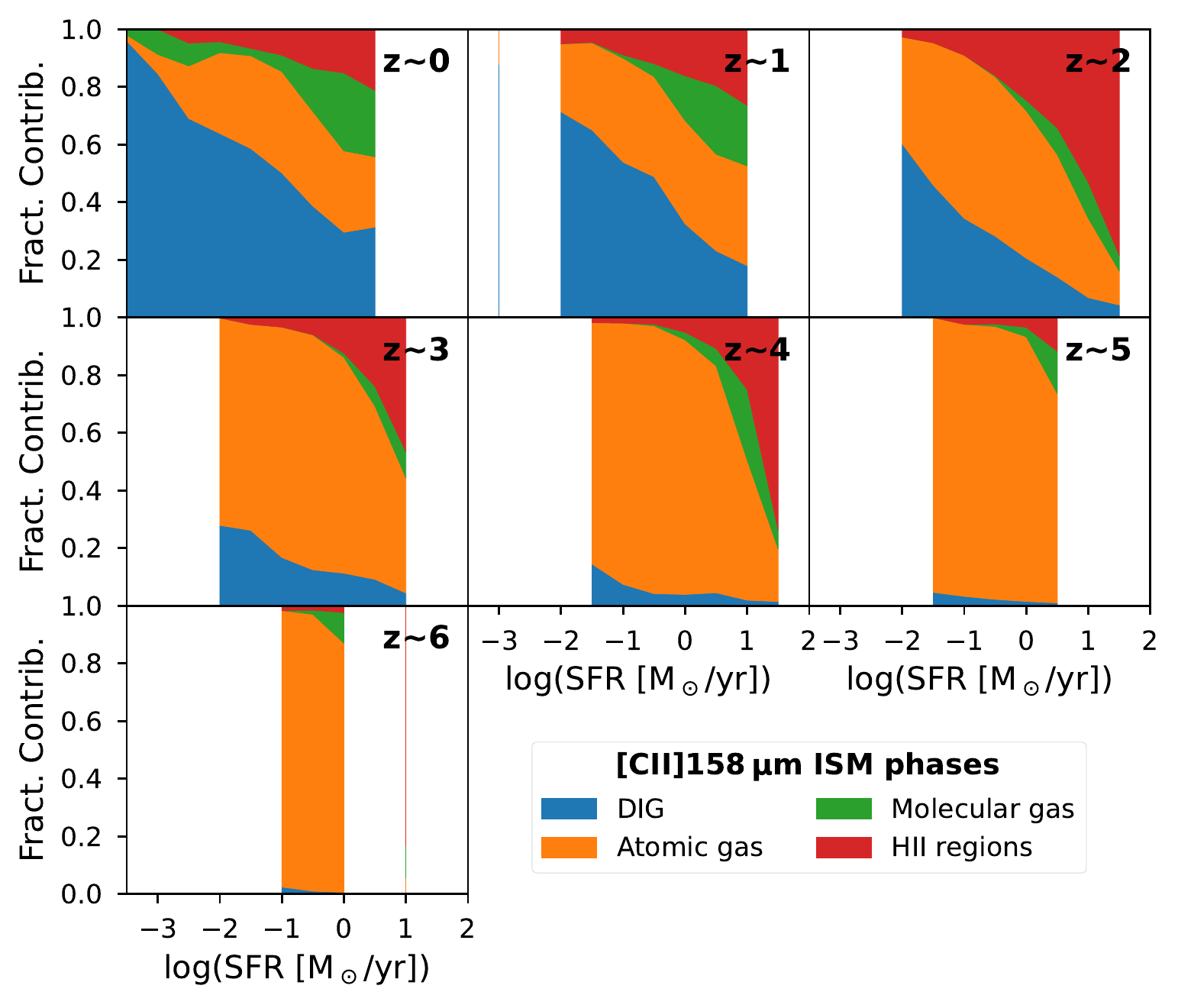}
    \caption{Contribution from the different ISM phases for the [\ion{C}{II}] emission line in \textsc{Recal-L0025N0752}. The regions define the DIG (blue), neutral atomic gas (orange), dense molecular gas (green) and \textsc{Hii} regions (red).}
    \label{fig:Cont_CII}
\end{figure}

We present the average contribution of the ISM phases to $L_{\mathrm[\ion{C}{II}]}$ at each redshift as a function of SFR for the galaxies in the \textsc{Recal-L0025N0752} simulation in Fig.~\ref{fig:Cont_CII}. We examine the ISM phase contributions of the \textsc{Recal-L0025N0752} simulation because we expect this simulation to behave more like observed local galaxies, as shown in \citetalias{2021A&A...645A.133R}. In general, we find that most of the [\ion{C}{II}] emission comes mainly from the atomic phase, especially at $z>2$, in agreement with the general assumption that the neutral gas is the dominant ISM phase contributing to $L_{\mathrm[\ion{C}{II}]}$ as estimated from observations \citep[e.g.][]{2017ApJ...845...96C,2019A&A...626A..23C} and suggested by models \citep[e.g.][]{2015ApJ...814...76O,2018ApJ...857..148O,2018A&A...609A.130L}. However, the contribution by neutral atomic gas to $L_{\mathrm[\ion{C}{II}]}$ changes with redshift, from ${\sim}$20--40\% at $z\leq1$, to ${\sim}$70--90\% at $z\geq3$. The most important reason for these differences is the contribution from the DIG. At $z=0$, the DIG dominates (the contribution is greater than 50\%) in most of the galaxies with a SFR$<0.1\,\mathrm{M}_\sun \, \mathrm{yr}^{-1}$, then the DIG contribution reduces to 30\% at $z=2$, and finally it is negligible at $z=6$. This negligible contribution of the DIG at $z=6$ does not agree with the estimated contribution of 44\% by \citetads{2018ApJ...857..148O}. This result is expected as we estimate the size of the DIG using a physical assumption instead of using the smoothing length as in \citetads{2018ApJ...857..148O}, which leads to a more compact DIG. However, as our modelled SFR--$L_{\mathrm[\ion{C}{II}]}$ relation shows a better agreement with observations, this may imply that a higher contribution from the atomic gas is needed to match the observed galaxies at $z=6$. Therefore, our estimations seem to favour the atomic gas as the main responsible for the $L_{\mathrm[\ion{C}{II}]}$ at $z=6$.

On the other hand, the fractional contribution of \textsc{Hii} regions to $L_{\mathrm[\ion{C}{II}]}$ reaches its maximum at $z=2$, where they contribute up to 80\% of the luminosity. This trend is expected as \textsc{Hii} regions trace young stars, and it is well known that the co-moving star formation rate density reaches its peak value at $z\approx2$ \citepads{2014ARA&A..52..415M}. In general, the average contribution by \textsc{Hii} regions is ${\sim}$20\% of $L_{\mathrm[\ion{C}{II}]}$ over all $z<4$. For molecular gas, the contribution is on average $\lesssim$10\% with a maximum at $z=0$. In both the molecular and \textsc{Hii} regions phases we obtain higher fractional contributions to $L_{\mathrm[\ion{C}{II}]}$ at higher SFR. This confirms the results found by \citetads{2015ApJ...814...76O} and in \citetalias{2021A&A...645A.133R} that the contribution of a given phase to $L_{\mathrm[\ion{C}{II}]}$ depends on the global SFR of the galaxy.

These predictions are not in agreement with recent results from \citetads{2021ApJ...915...92T} in resolved regions of two local galaxies (\object{M\,101} and \object{NGC\,6946}) where the ionised gas contribution to $L_{\mathrm[\ion{C}{II}]}$ is negligible (average upper limit of 12\%). This disagreement in the ISM phases may be related to the spatial constraints of the observations, as they focus on the arms regions of galaxies where denser neutral gas is expected. For example, \citetads{2018ApJ...869L..30P} estimated the $L_{\mathrm[\ion{C}{II}]}$ contribution coming from different regions inside \object{M\,51}. They found that the region between the arms in \object{M\,51}, where the diffuse ionised gas is expected be located, is approximately 20\% of the total $L_{\mathrm[\ion{C}{II}]}$, compared with ${\sim}$80\% that comes from the arms and the nucleus. In our model, we calculate the global properties of galaxies (in a 30\,pkpc aperture), and thus we expect a higher contribution from the diffuse gas, as this ISM phase can cover more extended regions throughout a galaxy. The information from the diffuse phases may therefore be missing when regions close to the arms of disk galaxies are observed. In general, contributions from different phases will depend on the scales over which we observe the galaxies \citepads{2021ApJ...915...92T}.

\begin{figure}
    \centering
    \includegraphics[width=\columnwidth]{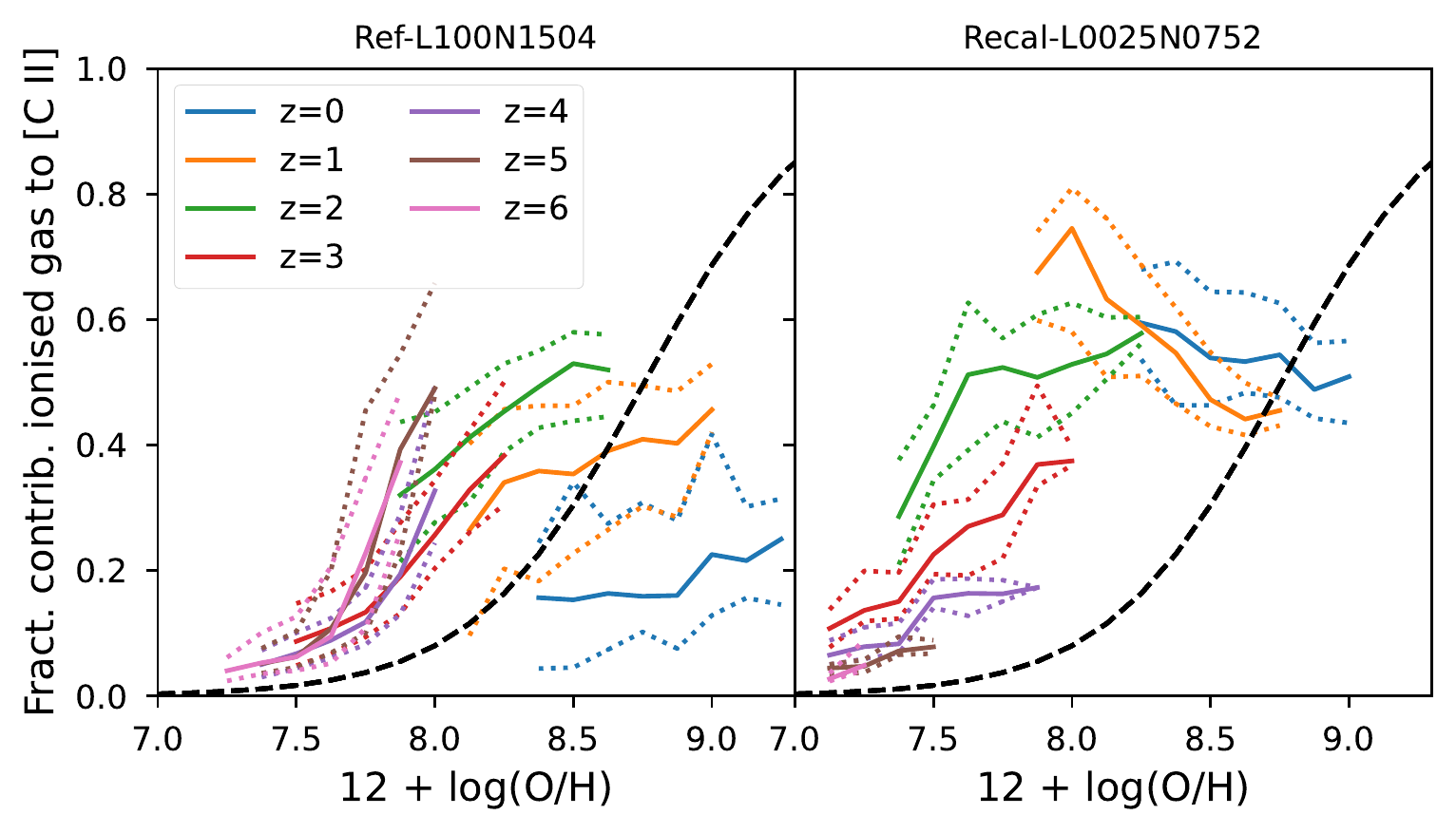}
    \caption{Contribution of the ionised gas phase to [\ion{C}{II}]. We show the median values of the different redshifts (solid lines) and their 25th and 75th percentiles (dotted lines) with respect to the fit to the observational data made by \citetads[][black dashed line]{2019A&A...626A..23C}. In the left panel, we show the predictions of \textsc{Ref-L100N1504} while in the right panel we show the predictions of \textsc{Recal-L0025N0752}. We only show the bins with more than 5\% of the total sample of galaxies for the respective simulation.}
    \label{fig:fDIG_metal}
\end{figure}

It is important to note that our model seems to overpredict the contribution of the ionised phases (DIG $+$ \textsc{Hii} regions) for the \textsc{Recal-L0025N0752} simulation. In Fig.~\ref{fig:fDIG_metal}, we examine the relation between the ionised phases to [\ion{C}{II}] as a function of metallicity ($12 + \log{(\mathrm{O/H})}$). We compare the median predictions of  \textsc{Ref-L100N1504} and \textsc{Recal-L0025N0752} with the relationship obtained by \citetads{2019A&A...626A..23C}, based on the data from their work and \citetads{2017ApJ...845...96C}, where at a higher metallicity there is a higher contribution to $L_{\mathrm[\ion{C}{II}]}$ from the ionised phase. For \textsc{Ref-L100N1504}, we estimate fractional contributions ${\sim}$10\%  higher than those inferred from the \citetads{2019A&A...626A..23C} fitting function, although at $z=0$ our predictions can be 40\% lower for higher metallicities. In \textsc{Recal-L0025N0752} we see that the fractional contributions are always above the empirical fitting function, especially at $z=1$ and $z=2$, by around 40\%. This means that there may be an overestimation in the ionised component of the \textsc{Recal-L0025N0752} luminosities.

The simulations and observations presented in Fig.~\ref{fig:fDIG_metal} are in any case difficult to compare. The simulated galaxies may not have the same metallicity calibration as the observed metallicities in \citetads{2017ApJ...845...96C} and \citetads{2019A&A...626A..23C}, and the method by which fractional contributions of the ionised gas are calculated may be different. Nonetheless, in general, we find that the contribution of the ionised phase to [\ion{C}{II}] increases with increasing metallicity, as observed by \citetads{2019A&A...626A..23C}.

\subsection{\texorpdfstring{[\ion{N}{II}]}{} 122 and 205 \texorpdfstring{$\mu$m}{}}

\subsubsection{The SFR--\texorpdfstring{$L_{\mathrm[\ion{N}{II}]}$}{} relationship}

\begin{figure*}
    \centering
    \includegraphics[width=\textwidth]{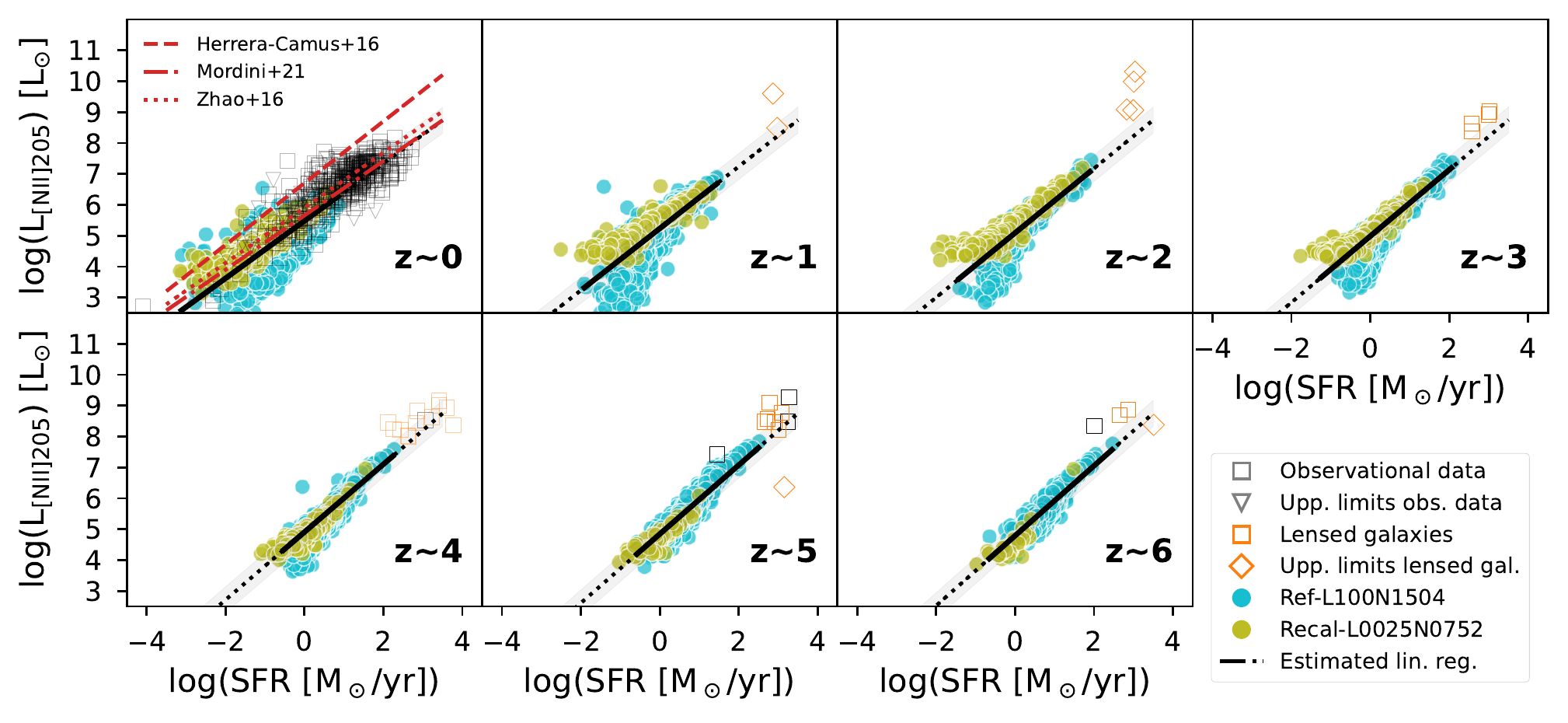}
    \caption{As Fig.~\ref{fig:LSRCII} for the [\ion{N}{II}] 205 \,$\mu$m line. We present the linear relations estimated by \citetads{2016ApJ...819...69Z}, \citetads{2016ApJ...826..175H} and \citetads{2021A&A...653A..36M} at $z=0$.}
    \label{fig:LSRNII}
\end{figure*}

The emission lines at 122 and 205\,$\mu$m of [\ion{N}{II}] are commonly used to trace ionised gas phases around neutral clouds, because its ionisation potential is only slightly above that of hydrogen. These lines are also used to disentangle the ionised gas contribution to the [\ion{C}{II}] luminosity \citep[e.g.][]{2015ApJ...814..133G,2015ApJ...806..260F,2016ApJ...832..151P,2017ApJ...845...96C,2019A&A...626A..23C,2021A&A...651A..59L}, as we do in this work. The relationship of these lines with SFR has been explored in the local Universe \citep[e.g.][]{2016ApJ...826..175H,2016ApJ...819...69Z}; however, at higher redshifts we have very little observational data, especially for the 122\,$\mu$m line. Due to this lack of data at $z>0$, we focus here on the [\ion{N}{II}] 205\,$\mu$m line. We present [\ion{N}{II}] luminosities at 205\,$\mu$m as a function of SFR in Fig.~\ref{fig:LSRNII}. We compare our results with the linear relationships in the local Universe estimated by \citetads{2016ApJ...819...69Z}, \citetads{2016ApJ...826..175H} and \citetads{2021A&A...653A..36M}. For the relation of \citet[][eq. 10]{2016ApJ...826..175H}, we assume the values of the collisional excitation coefficients from \citetads{2011ApJS..195...12T}, and abundances close to solar. For the relation of \citetads{2021A&A...653A..36M}, we use the sample of AGN galaxies assuming the conversion from infrared luminosities ($\mathrm{L}_{\mathrm{IR}}$) to SFR of \citetads{2012ARA&A..50..531K}. We use the AGN sample in \citetads{2021A&A...653A..36M} as those galaxies also follow the relation between SFR and the luminosity of the PAH feature at 6.2 $\mu$m (see their Fig.~6).

At $z=0$, the luminosity predictions of [\ion{N}{II}] follow a similar relationship to the observational relations of \citetads{2016ApJ...819...69Z} and \citetads{2021A&A...653A..36M}. A potential reason for the difference of one to two orders of magnitude with respect to the \citetads{2016ApJ...826..175H} relation is that the assumption of solar abundances is incorrect\footnote{Compared to the relations of \citetads{2016ApJ...819...69Z} and \citetads{2021A&A...653A..36M}, the N/H abundance that best matches the data is between the values of 0.03 and 0.16 with respect to solar abundance.}. If we assume an abundance below solar in the \citetads{2016ApJ...826..175H} relation, the relation is closer to our model.  At $z=1$ and $z=2$ our model is consistent with the upper limits of the observational data. At higher redshifts, $3<z<6$, our models agree with the observations, although the range of $L_{\mathrm[\ion{N}{II}]}$/SFR in the observations is larger than the estimations (almost 0.5\,dex at $z=4$). Most of the data at these redshifts come from the work of \citetads{2020MNRAS.494.4090C}, who observed 40 gravitationally lensed galaxies with the Morita Atacama Compact Array (ACA) of ALMA. The luminosities of these galaxies depend strongly on the lensing magnification factor, which could create a large scatter in the inferred luminosities (due to uncertainties in the assumed lensing model), as observed in the orange data points. Our model shows a good agreement with the observations, including the lensed galaxies after correction for magnification.

\subsubsection{Contribution to \texorpdfstring{$L_{\mathrm[\ion{N}{II}]}$}{} from each ISM phase}

\begin{figure}
    \centering
    \includegraphics[width=\columnwidth]{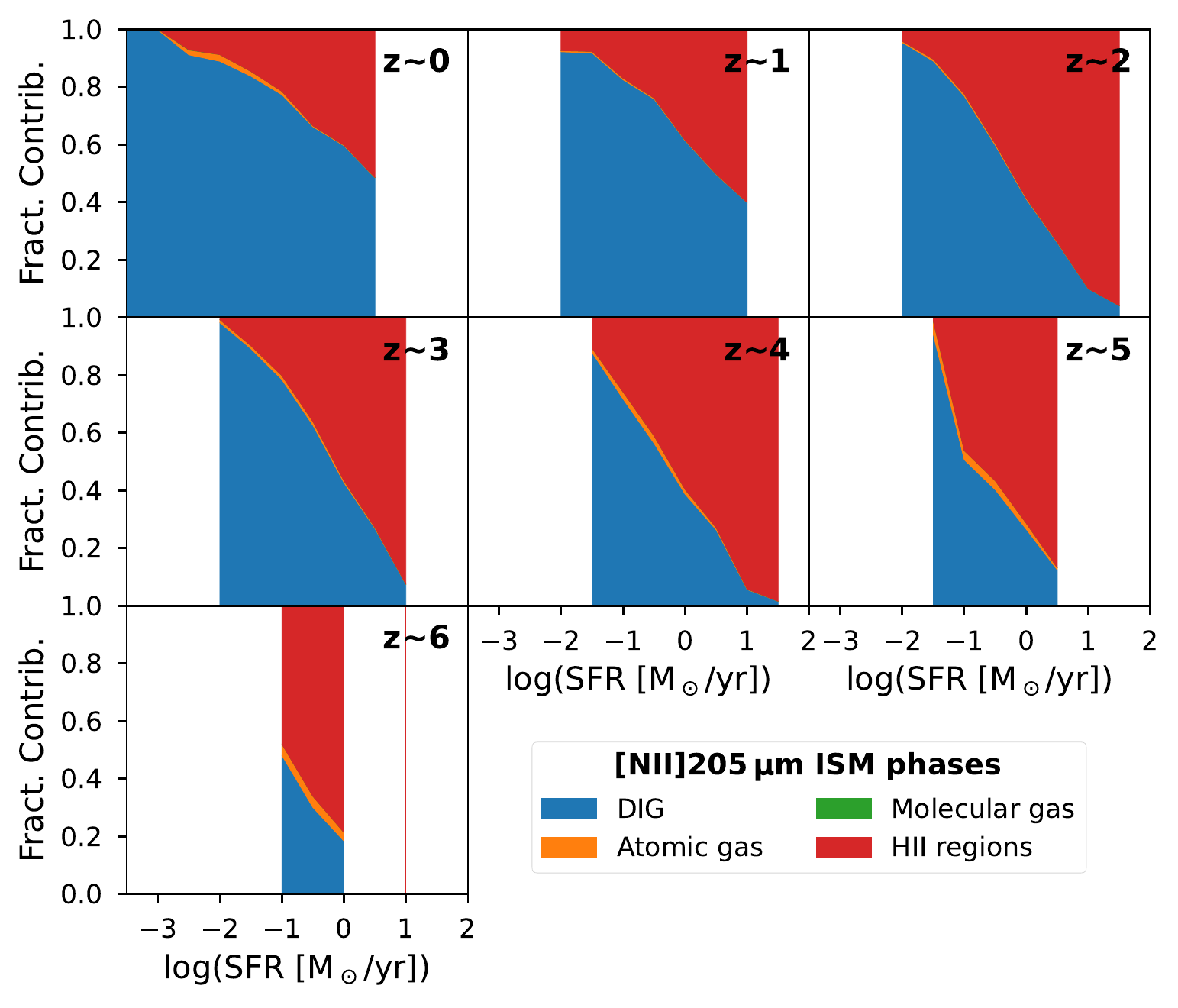}
    \caption{Contribution from the different ISM phases to the [\ion{N}{II}] emission line at 205\,$\mu$m in \textsc{Recal-L0025N0752}. Colour-coding is the same as in Fig.~\ref{fig:Cont_CII}}
    \label{fig:Cont_NII}
\end{figure}

In Fig.~\ref{fig:Cont_NII}, we show the contribution of the different phases of the ISM to the [\ion{N}{II}] line at 205\,$\mu$m. The luminosity of this line comes mainly from the two ionised phases (DIG and \textsc{Hii} regions), as expected from observations \citepads[e.g.][]{2021A&A...651A..59L}. Interestingly, the relative dominance of these two phases seems to change with redshift and SFR. At higher SFR most of the luminosity comes from \textsc{Hii} regions, while at low SFR, most of the luminosity comes from the DIG phase. The DIG can contribute significantly to the luminosity, which is clearer at lower SFR when there are not so many \textsc{Hii} regions. In the local Universe, the contribution from the DIG dominates (${\sim}$80\%) over the contribution from \textsc{Hii} regions. At higher redshifts, $1<z<4$, the contribution of the two phases is split relatively evenly. At $z>5$, \textsc{Hii} regions dominate the line emission. In this redshift range, at the highest SFRs, \textsc{Hii} regions contribute significantly more than the DIG. The transition point between the phases (the SFR where DIG contributes less than 50\% and \textsc{Hii} regions more than 50\%) is around $1\, \mathrm{M_{\sun}} \, \mathrm{yr}^{-1}$ at $z=0$ and decreases to $0.1\, \mathrm{M_{\sun}} \, \mathrm{yr}^{-1}$ at $z=6$. There is also a very small contribution from the atomic phase (<4\%) at some redshifts, but this is negligible compared to the ionised phases. 

\subsection{\texorpdfstring{[\ion{O}{I}]}{} 63 and 145 \texorpdfstring{$\mu$m}{}}

\subsubsection{The SFR--\texorpdfstring{$L_{\mathrm[\ion{O}{I}]}$}{} relationship}

The [\ion{O}{I}] emission lines at 63 and 145\,$\mu$m trace warm gas in neutral clouds and are commonly detected in galaxies in the local Universe \citepads{2001ApJ...561..766M}. However, the line at 145\,$\mu$m is fainter than the 63\,$\mu$m line due to its lower spontaneous decay rate and higher upper level energy \citepads{2019ApJ...887...54G}. As we did with [\ion{N}{II}], here we focus on the results of the brighter line of [\ion{O}{I}], i.e., the emission line at 63\,$\mu$m. 

\begin{figure*}
    \centering
    \includegraphics[width=\textwidth]{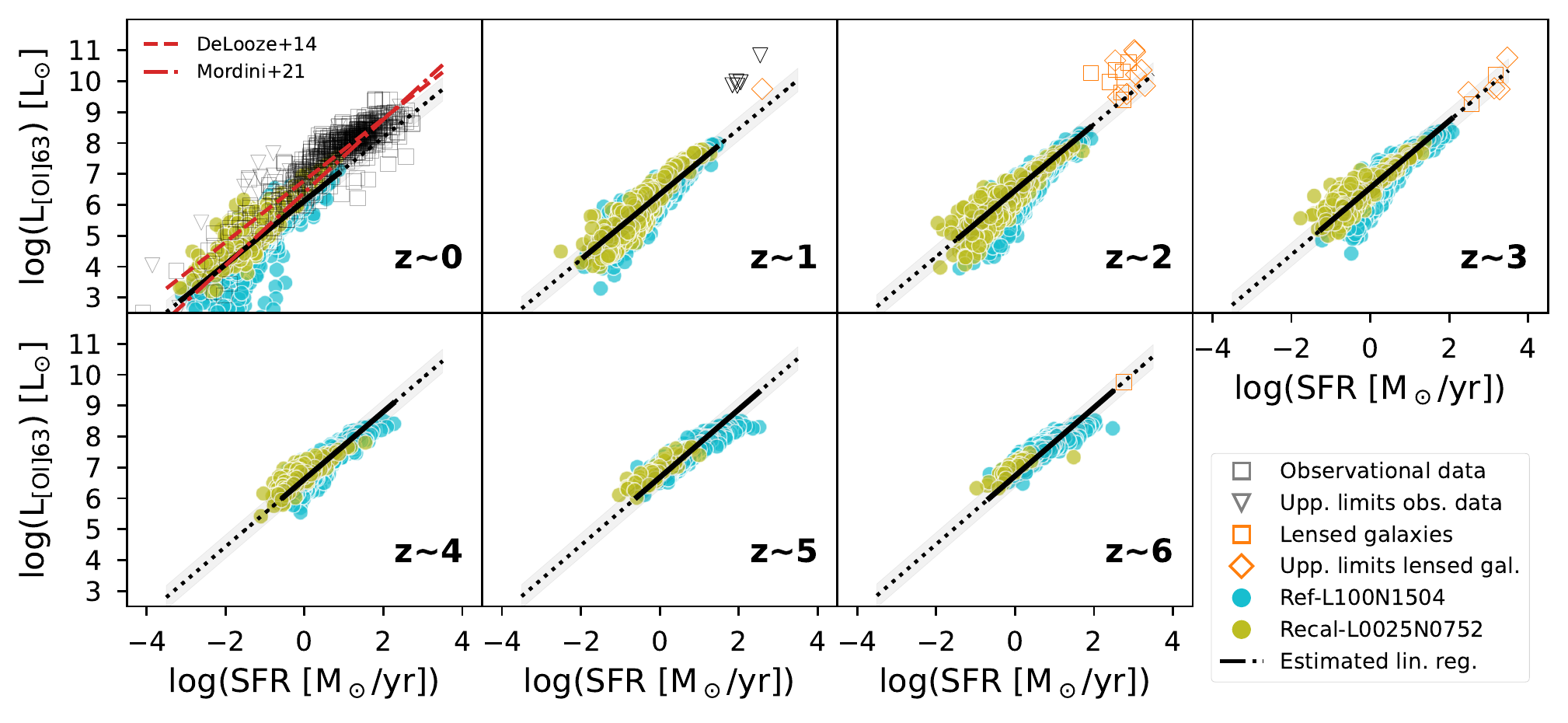}
    \caption{Similar to Fig. 2 but for the [\ion{O}{I}] 63\,$\mu$m line. We present the linear relations estimated by \citetads{2014A&A...568A..62D} and \citetads{2021A&A...653A..36M} at $z=0$.}
    \label{fig:LSROI}
\end{figure*} 

In Fig.~\ref{fig:LSROI} we present predictions of [\ion{O}{I}]  63\,$\mu$m luminosity as a function of SFR at different redshifts. Our model at $z=0$ agrees with the local Universe relationships of \citetads{2014A&A...568A..62D} and \citetads{2021A&A...653A..36M}. For the relation of \citetads{2021A&A...653A..36M}, we use the sample of star-forming galaxies assuming the conversion from $\mathrm{L}_{\mathrm{IR}}$ to SFR of \citetads{2012ARA&A..50..531K}. At $z>1$, extrapolation of the linear fits to our model predictions are consistent with the observational upper limits. There are only a few detections at $z=2$, $z=3$ and $z=6$ that can be directly compared with our predictions. There are seven detections in the $z=2$ bin, of which five are $>0.5$\,dex from the extrapolated  linear fit. However, all of these galaxies are gravitationally lensed \citepads{2015ApJ...799...13B}, and no corrections were applied to correct for lensing because the magnification factors were unknown \citepads[see also][]{2018MNRAS.481.1976Z}. Therefore, it is reasonable to assume that these five measurements should be treated as upper limits. In the other redshift bins, we observe a very good agreement with the linear relationship because the lensing magnification has been taken into account \citep{2018MNRAS.473...20R,2018MNRAS.481...59Z,2020ApJ...889L..11R}. 

Other numerical models, not shown in the plots, also predict the [\ion{O}{I}] 63\,$\mu$m line and show similar results. For example, \citetads{2018ApJ...857..148O} predict the emission of 30 simulated galaxies at $z=6$ in the \textsc{MUFASA} cosmological simulation. All of their galaxies have SFR${\sim} 10\, \mathrm{M_{\sun}} \, \mathrm{yr}^{-1}$ and $L_{\mathrm[\ion{O}{I}]}{\sim} 10^8\, \mathrm{L_{\sun}}$, very similar to the estimations of the \textsc{Ref-L100N1504} simulation. However in an updated model, \citetads{2021ApJ...922...88O} use the \textsc{SIMBA} cosmological simulation at $z=0$, and their predicted [\ion{O}{I}]  63\,$\mu$m luminosities are 1.2\,dex above the \citetads{2014A&A...568A..62D} relation shown in Fig.~\ref{fig:LSROI}. This contrasts with our model, which exhibits better agreement with the linear fits of \citetads{2014A&A...568A..62D} and \citetads{2021A&A...653A..36M} at $z=0$, with a difference of ${\sim}0.5$ dex at $\log(\mathrm{SFR}[\mathrm{M_{\sun}} \, \mathrm{yr}^{-1}]) = 1$. We conclude that the model predictions presented in this work provide a better match with observations across a wide range of redshifts than previous models that only focus on a single redshift slice.

\subsubsection{Contribution to \texorpdfstring{$L_{\mathrm[\ion{O}{I}]}$}{} from each ISM phase}

\begin{figure}
    \centering
    \includegraphics[width=\columnwidth]{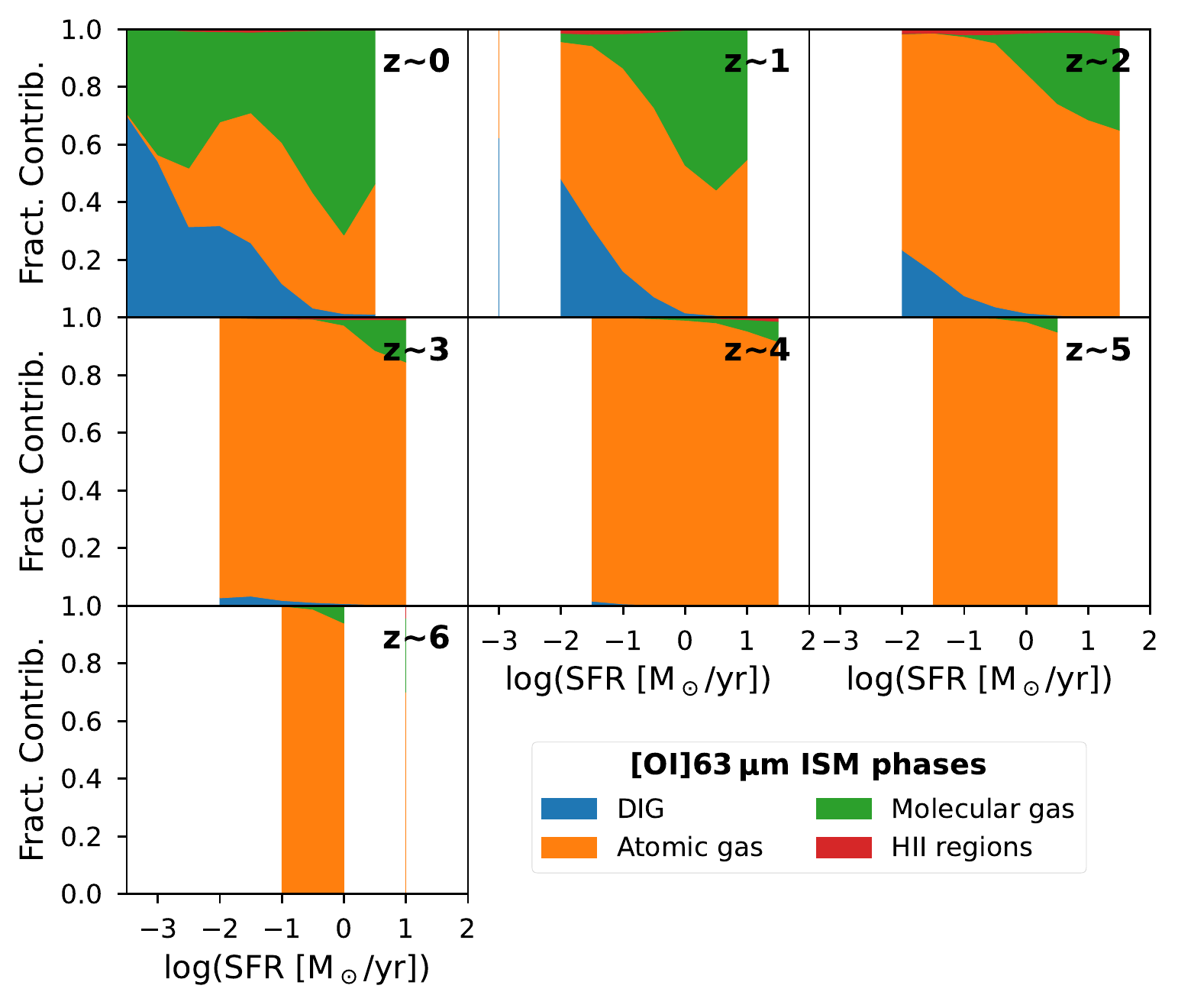}
    \caption{Contribution from the different ISM phases to the [\ion{O}{I}] emission line in \textsc{Recal-L0025N0752}. Colour-coding is the same as in Fig.~\ref{fig:Cont_CII}.}
    \label{fig:Cont_OI}
\end{figure}

Contributions to the [\ion{O}{I}] 63\,$\mu$m line come mainly from neutral clouds, i.e., neutral atomic gas and dense molecular gas, as shown in Fig.~\ref{fig:Cont_OI}. At $z=0$, the contribution to $L_{\mathrm[\ion{O}{I}]}$ from molecular gas is ${\sim} 40\%$ while for atomic gas it is ${\sim} 39 \%$. These percentages change with redshift: the contribution from molecular gas decreases and from atomic gas increases with increasing redshift. At $z=2$ the percentages are ${\sim} 7\%$ and ${\sim} 85\%$, respectively, while at $z=4$ they are ${\sim} 1\%$ and ${\sim} 98\%$. The contributions from the other phases are negligible, especially for the \textsc{Hii} regions. On average the contribution from the DIG is less than 10\%; however, it can be very high ($>80\%$) in galaxies with very low SFR ($<10^{-2} \mathrm{M_{\sun}} \, \mathrm{yr}^{-1}$) in the local Universe. At $z\geq3$, the molecular fraction, which is calculated from the line luminosity in the region defined by the radius at which the transition from atomic to molecular H occurs, in the Plummer profile (Equation 31 in \citetalias{2021A&A...645A.133R}) is very low. At those redshifts, even though the average density of the neutral cloud is higher than in the local Universe, the ISRF is also high, which causes the dominant emission to come from the atomic gas instead of the molecular gas. These results support the understanding that the [\ion{O}{I}] 63\,$\mu$m line originates in warm neutral environments \citep{2001ApJ...561..766M,2019ApJ...887...54G} even in high-$z$ galaxies ($z{\sim}6$).
 
\subsection{\texorpdfstring{[\ion{O}{III}]}{} 52 and 88 \texorpdfstring{$\mu$m}{}}

\subsubsection{The SFR--\texorpdfstring{$L_{\mathrm[\ion{O}{III}]}$}{} relationship}

\begin{figure*}
    \centering
    \includegraphics[width=\textwidth]{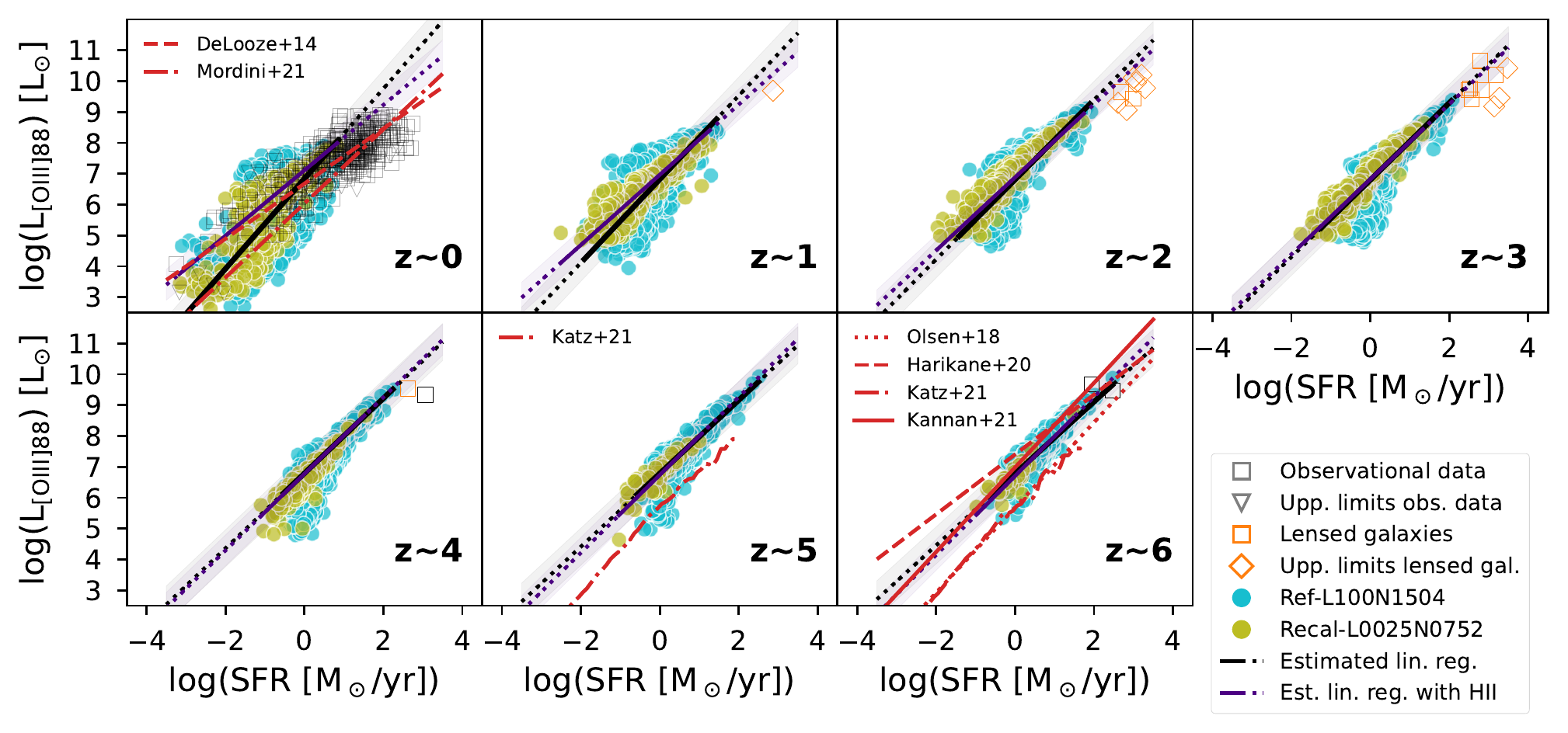}
    \caption{As Fig.~\ref{fig:LSRCII} for the [\ion{O}{III}] 88 
   \,$\mu$m line. We compare the obtained relations from the EAGLE simulations with relations derived from simulations \citep{2018ApJ...857..148O,2022MNRAS.510.5603K,2021arXiv211102411K} and observations \citep{2014A&A...568A..62D,2020ApJ...896...93H,2021A&A...653A..36M}. Linear relations inferred from our models using only \textsc{Hii} regions are shown as indigo solid lines over the dynamic range covered by the simulations and extrapolated to lower and higher SFRs as indigo dotted lines, with the shaded area representing the 1$\sigma$ error (see Appendix~\ref{app:linregcoef}).}
    \label{fig:LSROIII}
\end{figure*}

The [\ion{O}{III}] emission lines at 52 and 88\,$\mu$m are the best tracers of ionised gas in the FIR. These lines may be used as SFR tracers as they come mainly from young stars \citep[e.g][]{2010ApJ...714L.147F,2014A&A...568A..62D,2014ApJ...780L..18I,2020ApJ...896...93H, 2020MNRAS.499.3417Y,2021MNRAS.504..723Y}. The [\ion{O}{III}] 88\,$\mu$m line has become very important as it may be brighter than the [\ion{C}{II}] 158\,$\mu$m line in galaxies close to the reionisation epoch \citep[$z\gtrsim7$,][]{2014ApJ...780L..18I, 2016Sci...352.1559I,2021arXiv210613719B}. As in previous sections, we focus here on the results of only one of the [\ion{O}{III}] lines, the [\ion{O}{III}] 88\,$\mu$m line, as we have more observational data for this line. In Fig.~\ref{fig:LSROIII} we present the [\ion{O}{III}] 88\,$\mu$m luminosities as a function of SFR. We compare the predictions from the EAGLE simulations \textsc{Ref-L100N1504} and \textsc{Recal-L0025N0752} with observations of individual galaxies and linear relationships between [\ion{O}{III}] and SFR derived from observations \citep{2014A&A...568A..62D,2020ApJ...896...93H,2021A&A...653A..36M} and numerical models \citep{2018ApJ...857..148O,2022MNRAS.510.5603K,2021arXiv211102411K}.

Our predicted SFR--$L_{\mathrm[\ion{O}{III}]}$ relationships largely agree with observations, although it seems that our predicted relationship is steeper than some relations found in the literature. At $z=0$, we compare our model with the predictions of \citetads{2014A&A...568A..62D} and \citetads{2021A&A...653A..36M}. For the relation of \citetads{2021A&A...653A..36M}, we use the sample of star-forming galaxies assuming the conversion from $\mathrm{L}_{\mathrm{IR}}$ to SFR of \citetads{2012ARA&A..50..531K}. Both observational relationships appear to have slopes flatter than our model, separated by about an order of magnitude at SFR$ =10\,\mathrm{M_{\sun}} \, \mathrm{yr}^{-1}$. The reason for this difference is that at low SFRs, our luminosity predictions coming from \textsc{Hii} regions drop off sharply for both simulations (as we will discuss in the next subsection), leading to a steep slope even though our predicted $L_{\mathrm[\ion{O}{III}]}$ values agree with the observational data at SFR$ =1\, \mathrm{M_{\sun}} \, \mathrm{yr}^{-1}$. 

At $z=2$--$4$, due to the small sample size in the observational datasets, it is very difficult to properly assess the level of agreement between the observations and our model predictions. In some cases, the observations agree well with the extrapolation of our predicted SFR--$L_{\mathrm[\ion{O}{III}]}$ relations, with a few exceptions. For example, the galaxy located 1.1\,dex below the extrapolation at $z=4$ is \object{AzTEC\,1} \citepads{2019ApJ...876....1T}. The discrepancy in this galaxy is not related to gravitational lensing, but rather this galaxy has a very low [\ion{O}{III}]/[\ion{C}{II}] ratio compared with \object{SPT-S\,J041839-4751.8}, a galaxy that closely follows our relation \citepads{2019A&A...631A.167D}. This means that this galaxy may be an outlier and very different physical conditions can shift its position in the SFR--$L_{\mathrm[\ion{O}{III}]}$ relation, as we discuss in Sect.~\ref{sec:diag}. 

At $z=5$, we compare our results with those of \citetads{2022MNRAS.510.5603K}, which show a similar slope but lower $L_{\mathrm[\ion{O}{III}]}$ by ${\sim}1.3$\,dex. The same model was used to calculate $L_{\mathrm[\ion{O}{III}]}$ at $z=6$, showing that their predictions, together with the models of \citetads{2018ApJ...857..148O}, underestimate $L_{\mathrm[\ion{O}{III}]}$ compared with observations by 0.7\,dex at SFR${\sim} 100\,\mathrm{M_{\sun}} \, \mathrm{yr}^{-1}$. This is not the case with our model: it agrees very well with the observations and with the linear relation from \citetads{2020ApJ...896...93H}, with a difference of 0.2 dex at SFR${\sim} 100\,\mathrm{M_{\sun}} \, \mathrm{yr}^{-1}$. It is also interesting to note that the numerical results of \citetads{2021arXiv211102411K} are very similar to those presented in this work, with the difference of a slightly higher slope in their work, which gives a difference of 0.4 dex at SFR${\sim} 10\,\mathrm{M_{\sun}} \, \mathrm{yr}^{-1}$.

\subsubsection{Contribution to \texorpdfstring{$L_{\mathrm[\ion{O}{III}]}$}{} from each ISM phase}\label{subsec:OIII_ISM}

\begin{figure}
    \centering
    \includegraphics[width=\columnwidth]{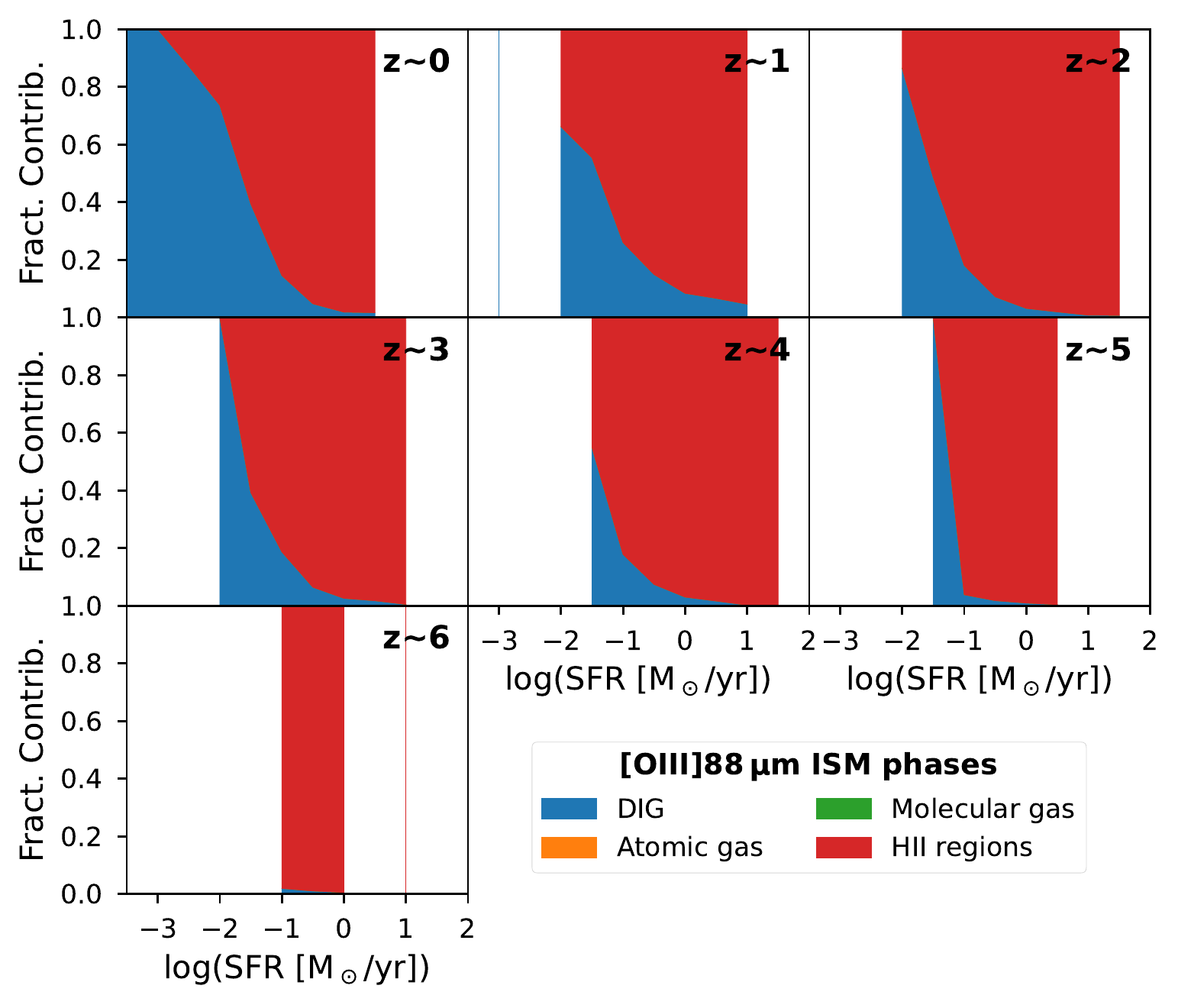}
    \caption{Contribution from the different ISM phases for the [\ion{O}{III}] emission line in \textsc{Recal-L0025N0752}. Colour-coding is the same as in Fig.~\ref{fig:Cont_CII}.}
    \label{fig:Cont_OIII}
\end{figure}

In Fig.~\ref{fig:Cont_OIII} we show the contribution of the ISM phases to $L_{\mathrm[\ion{O}{III}]}$, of which the ionised ISM is the only contributor, as expected. The dominant contributor to $L_{\mathrm[\ion{O}{III}]}$ are the \textsc{Hii} regions at most redshifts. However at $z=0$, the emission coming from the \textsc{Hii} regions drops sharply when the SFR decreases. The reason for this sharp drop is the low ionising photon flux coming from the star SPH particles in the simulated galaxies at low SFR. This sets the [\ion{O}{III}] line luminosities from the \textsc{Hii} regions to almost negligible values compared to the DIG, which explains the trend observed in $z=0$. At $z=1$, the lack of ionising photon flux affects galaxies less than at $z=0$, and at these redshifts the \textsc{Hii} regions dominate (72\%) over the DIG (28\%). For redshifts from $z=2$ to $z=6$ the contribution from the \textsc{Hii} regions changes from 85\% to 99\%. 

\subsection{\texorpdfstring{[\ion{N}{III}]}{} 57 \texorpdfstring{$\mu$m}{}} \label{sec:NIII}

\subsubsection{The SFR--\texorpdfstring{$L_{\mathrm[\ion{N}{III}]}$}{} relationship}

\begin{figure*}
    \centering
    \includegraphics[width=\textwidth]{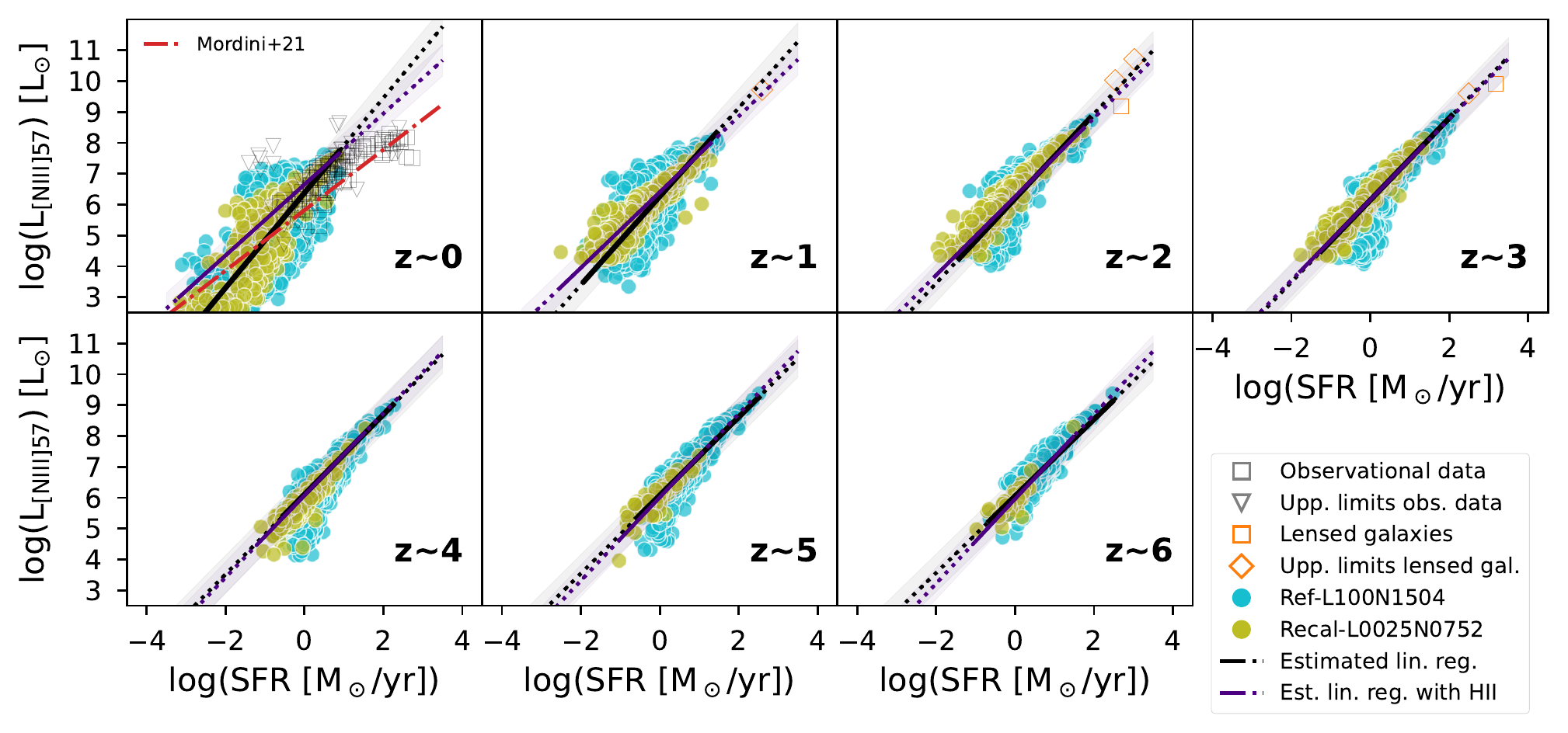}
    \caption{As Fig.~\ref{fig:LSROIII} for the [\ion{N}{III}] 57\,$\mu$m line. We present the linear relations estimated by \citetads{2021A&A...653A..36M} at $z=0$.}
    \label{fig:LSRNIII}
\end{figure*}

The [\ion{N}{III}] emission line at 57\,$\mu$m is very similar to the [\ion{O}{III}] 88\,$\mu$m  emission line, due to its excitation properties (Table~\ref{tab:lines}). Both trace \textsc{Hii} regions, with the difference that [\ion{N}{III}] is fainter in galaxies with metallicities below solar \citep[e.g.][]{2011A&A...526A.149N,2018MNRAS.473...20R}. The relationship between the SFR and $L_{\mathrm[\ion{N}{III}]}$ is less well known than the other FIR emission lines presented in this work. In Fig.~\ref{fig:LSRNIII}, we present the SFR-$L_{\mathrm[\ion{N}{III}]}$ relationship. We compare predictions from the EAGLE simulations with the observational sample and the linear relation at $z=0$ by \citetads{2021A&A...653A..36M} in AGN galaxies assuming the conversion from $\mathrm{L}_{\mathrm{IR}}$ to SFR of \citetads{2012ARA&A..50..531K}, as we did in the case of [\ion{N}{II}].

At $z=0$, our model predictions agree with the observations, with a mean offset of 0.2\,dex, as they are inside the observational scatter for the overlapping SFR range. However, the $L_{\mathrm[\ion{N}{III}]}$ predictions have the same problem as the $L_{\mathrm[\ion{O}{III}]}$ predictions: the low ionising photon flux coming from SPH star particles in simulated galaxies at low SFR. This leads to a steeper relation of $L_{\mathrm[\ion{N}{III}]}$ with SFR than that of \citetads{2021A&A...653A..36M}, which follows the other observational results. At $z\geq1$, even though there is not much information, we find that the extrapolation of our linear fit is consistent with the upper limits and detections in two galaxies at $z\approx2$ \citepads[\object{H-ATLAS\,J091043.1-000321},][]{2018ApJ...867..140L} and $z\approx3$ \citepads[\object{HERMES\,J105751.1+573027},][]{2018MNRAS.473...20R}. These two galaxies are magnified by around an order of magnitude (magnification factors of 11.5 and 10.9, respectively), therefore the agreement with our model, after applying the magnification correction, shows that the slope of the linear relationship may be correct for these redshifts.

\subsubsection{Contribution to \texorpdfstring{$L_{\mathrm[\ion{N}{III}]}$}{} from each ISM phase}

\begin{figure}
    \centering
    \includegraphics[width=\columnwidth]{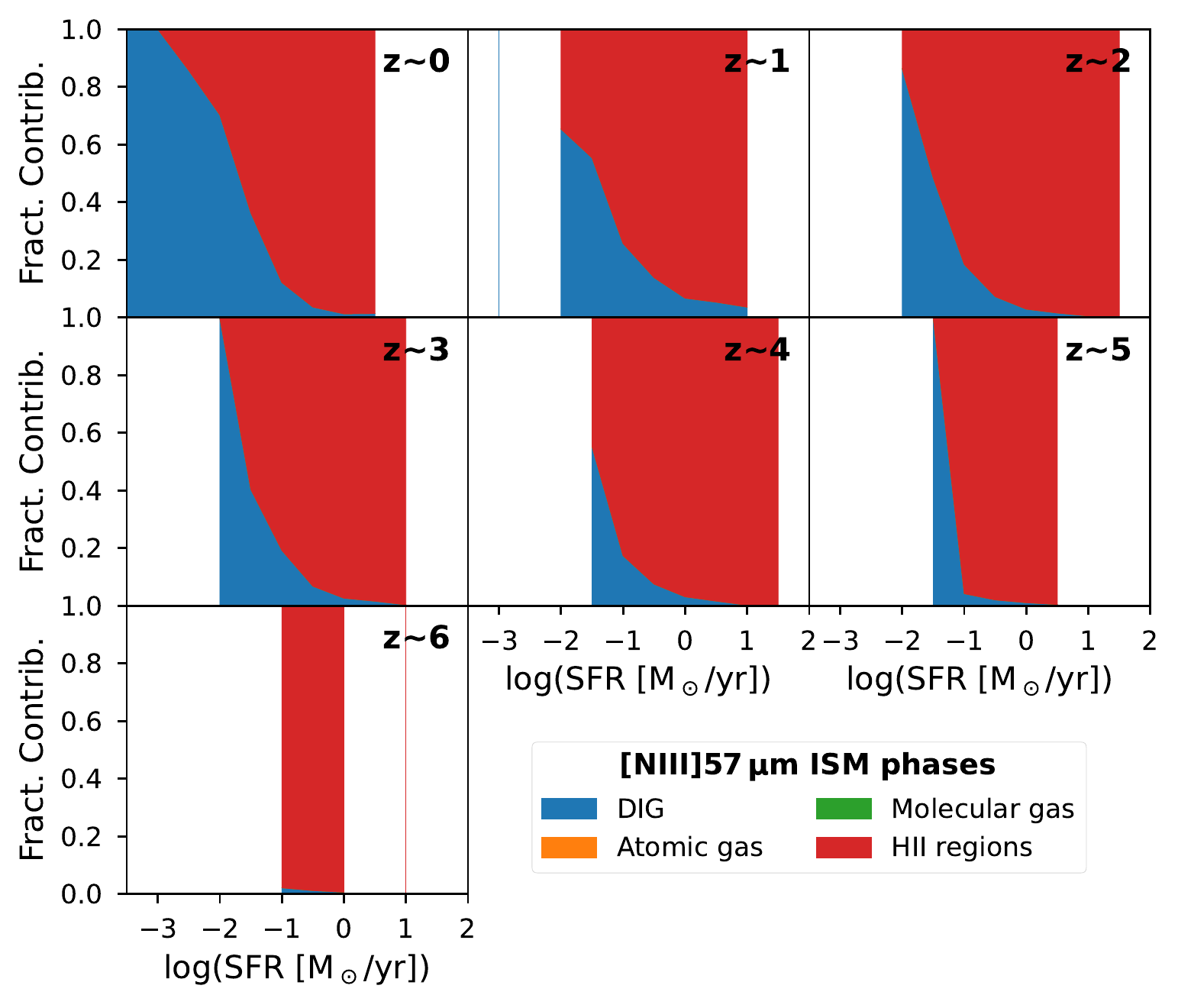}
    \caption{Contribution from the different ISM phases for the [\ion{N}{III}] emission line in \textsc{Recal-L0025N0752}. Colour-coding is the same as in Fig.~\ref{fig:Cont_CII}.}
    \label{fig:Cont_NIII}
\end{figure}

The contributions of the ISM phases to the [\ion{N}{III}] 57\,$\mu$m emission line are very similar to the contributions of the phases to the [\ion{O}{III}] 88\,$\mu$m line (Fig.~\ref{fig:Cont_OIII}), as shown in Fig.~\ref{fig:Cont_NIII}. The main difference resides in the exact percentages between the two dominant ionised ISM phases (DIG and \textsc{Hii} regions). At $z=0$, the contributions of the DIG and \textsc{Hii} regions to $L_{\mathrm[\ion{N}{III}]}$ are 38\% and 62\%, respectively. At higher redshifts these percentages change to 15\% and 85\% at $z=2$, 8\% and 91\% at $z=4$, respectively. Finally, at $z=6$ the \textsc{Hii} regions are responsible for almost all the $L_{\mathrm[\ion{N}{III}]}$.

\subsection{Summary of FIR line luminosities}\label{subsec:summFIR}

\begin{figure*}
    \centering
    \includegraphics[width=\textwidth]{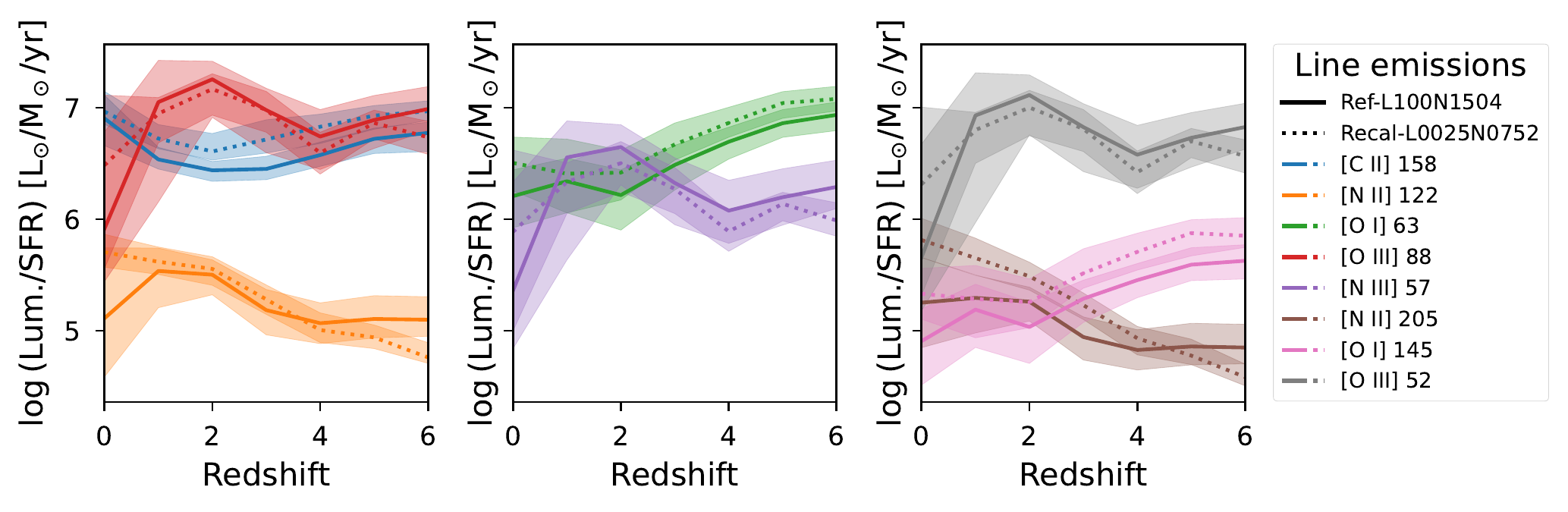}
    \caption{Evolution of the line luminosity--SFR ratio with redshift for the main FIR lines modelled in this work. We show the median values from \textsc{Recal-L0025N0752} (dotted lines) and \textsc{Ref-L100N1504} (solid lines). The shaded regions correspond to the 25th and 75th percentiles.}
    \label{fig:sfr_evol}
\end{figure*}

The relationships of the SFR with the different FIR line luminosities presented in Sects.~\ref{sec:CII}--\ref{sec:NIII} depend on redshift (Figs.~\ref{fig:LSRCII}, ~\ref{fig:LSRNII}, ~\ref{fig:LSROI}, ~\ref{fig:LSROIII} and ~\ref{fig:LSRNIII}). We assume a potential redshift dependency in the linear fits of our line predictions (Eq.~\ref{eq:linreg}), which shows a good agreement with the observations. Nonetheless, we want to quantify how these FIR lines and their dependence on SFR evolve with redshift. To do this, we plot in Fig.~\ref{fig:sfr_evol} the ratio between luminosity and SFR ($L$/SFR) versus redshift for the eight lines modelled in this work.

First, we compare the results of the two sets of simulations \textsc{Ref-L100N1504} and \textsc{Recal-L0025N0752}. Taking into account the scatter of the predictions, we find a good agreement between them, even though some of the physical properties of galaxies in the simulations are different, as we discuss in Sect.~\ref{sec:systematics}.

The evolution of the $L$/SFR ratio is almost flat for most FIR lines at $z\geq4$. At lower redshifts, $1<z<3$, however, there are drastic changes in the $L$/SFR ratio. For example, the [\ion{O}{III}] lines at 52 and 88\,$\mu$m have higher $L$/SFR values at $z=2$ and then decrease sharply towards $z=0$, by almost an order of magnitude. The opposite occurs for [\ion{C}{II}], where at $z=2$ the $L$/SFR ratio has lower values than at other redshifts, although the difference is less than 0.5\,dex.

These changes are related to the effects present in galaxies during ``cosmic noon'', where the cosmic star-formation history reaches its peak value at $z\approx2$ and around half of the stellar mass of the local Universe was formed, affecting the different phases of the ISM traced by the FIR lines in Fig.~\ref{fig:sfr_evol} \citep{2014ARA&A..52..415M,2020ARA&A..58..661F}. This result is expected because the \textsc{EAGLE} galaxies reproduce the observed trend in the SFR density and stellar mass assembly across cosmic time \citepads{2015MNRAS.450.4486F} and so the ISM phases evolve accordingly with redshift (Figs.~\ref{fig:Cont_CII}, \ref{fig:Cont_NII}, \ref{fig:Cont_OI}, \ref{fig:Cont_OIII} and \ref{fig:Cont_NIII}).

The two [\ion{O}{III}] lines have a similar shape in Fig.~\ref{fig:sfr_evol}, which explains why both lines can be used to constrain gas density and metallicity at high-$z$ \citepads{2021MNRAS.504..723Y}. The [\ion{N}{II}] pair shows a similar behaviour, where the main difference resides in the value of the $L$/SFR ratio at $z=0$, although the scatter is especially large at this redshift. This could be important when estimating electron densities from observations from the ratio of these two [\ion{N}{II}] lines \citep{2017ApJ...845...96C,2021A&A...651A..59L}. Finally, for both [\ion{O}{I}] lines, we see a clear difference between the estimated luminosities of the lines of around 1.15\,dex over most of the redshift range. This difference of more than an order of magnitude is expected from theoretical models. If the difference in [\ion{O}{I}] luminosities is more than 1.15\,dex, it may indicate higher kinetic temperatures ($>400$\,K) and/or lower gas densities ($\lesssim 10\,\mathrm{cm}^{-3}$) in observations \citepads{2019ApJ...887...54G}.

With this information, we can ask which FIR emission line is the best SFR tracer across cosmic time. The FIR line showing the least variation with $z$ is [\ion{C}{II}]. However, this tracer may not be ideal in some cases: observations, analytical models and simulations \citep[e.g.][]{2020A&A...643A...3S,2019MNRAS.482.4906P,2020MNRAS.499.5136C} suggest that there might be a weak evolution for the SFR--$L_{\mathrm[\ion{C}{II}]}$ relation with respect to redshift. Our model suggests that there is a slight evolution, although less than for the other lines, as shown by the coefficient $c_{2}$ in Table~\ref{tab:linreg} and Equation~\ref{eq:linreg}. The luminosity evolution may be related to the active star-formation processes that occur in starburst galaxies, as shown in Fig.~\ref{fig:DMS}, and in galaxies during ``cosmic noon'' at $z\approx2$. The regulation of these star-formation processes is reflected in the changes of the ISM phases (Fig.~\ref{fig:Cont_CII}), which tend to be more stable in the case of $L_{\mathrm[\ion{C}{II}]}$.

Another possibility is to use $L_{\mathrm[\ion{O}{III}]}$ as an SFR tracer, since the [\ion{O}{III}] lines tend to be equal to or brighter than $L_{\mathrm[\ion{C}{II}]}$ in some redshift ranges \citep[e.g.][]{2020MNRAS.498.5541A,2020MNRAS.499.5136C,2021MNRAS.505.5543V,2021arXiv210613719B,2022MNRAS.510.5603K}. However, our $L_{\mathrm[\ion{O}{III}]}$ predictions at $z=0$ may be underestimated for lower SFR (see Sect.~\ref{subsec:OIII_ISM}), which may explain the decrease in the median value of the $L$/SFR ratio for these lines. It is also possible to use $L_{\mathrm[\ion{O}{III}]}$ together with $L_{\mathrm[\ion{O}{I}]}$ at 63\,$\mu$m as an SFR tracer \citepads{2021A&A...653A..36M} to balance the neutral and ionised components of the ISM in these lines. Nevertheless, this use would require having access to both luminosities ($L_{\mathrm[\ion{O}{III}]}$ and $L_{\mathrm[\ion{O}{I}]}$) at $z<4$ to confirm the trends presented in Fig.~\ref{fig:sfr_evol}. The [\ion{O}{I}]  63\,$\mu$m line alone could also be used to trace the SFR, as this line has similar properties to the [\ion{C}{II}] line \citepads{2022MNRAS.510.5603K}. For example, in the predictions of \citetads{2018ApJ...857..148O}, [\ion{O}{I}] is brighter than [\ion{C}{II}] at $z=6$, and therefore is more easily observable. The trend of [\ion{O}{I}] being brighter than [\ion{C}{II}] is also predicted in this work, confirming the results from \citetads{2018ApJ...857..148O}. Therefore, $L_{\mathrm[\ion{C}{II}]}$ may be the best SFR tracer across the entire redshift range of this work ($z=0$--$6$), but at high-$z$ other FIR lines such as $[\ion{O}{III}]$ and $[\ion{O}{I}]$ are also very useful.

\section{Diagnostic diagrams using FIR lines}\label{sec:diag}

We now examine our predictions of FIR emission line strengths from the \textsc{EAGLE} simulations in the context of diagnostic diagrams. Typically, diagnostic diagrams use emission line ratios that reflect the physical conditions of the ISM. In this work, we focus on two diagnostic diagrams: one that normalises the emission line luminosity with SFR for [\ion{C}{II}] and [\ion{O}{III}] at 88 \,$\mu$m \citepads{2020ApJ...896...93H}, and the other that uses the ratios between [\ion{C}{II}]/[\ion{O}{III}] and [\ion{N}{II}]/[\ion{O}{I}], based on the [\ion{O}{III}], [\ion{N}{II}] and [\ion{O}{I}] lines at 88 \,$\mu$m, 205 \,$\mu$m and 63 \,$\mu$m, respectively. With these two diagnostic diagrams we investigate whether these ratios trace physical quantities related to the ISM, such as radiation field and density.

Other line ratios in the FIR are also of great interest for different types of studies \citepads[e.g.\ Sect.~3.3 of][]{2015A&A...578A..53C}. We therefore make our model predictions for different emission lines of these simulated \textsc{EAGLE} galaxies publicly available, as described in Appendix~\ref{app:estimations_dataset}. Similar FIR diagnostic diagrams, such as those presented by \citetads{2019A&A...631A.167D} and \citetads{2020ApJ...900..131L} are not discussed in this work, but plots are provided as supplementary material\footnote{Zenodo repository at \url{https://doi.org/10.5281/zenodo.6576202}}. 

\subsection{Comparison with observations}\label{subsec:comobs}

We compare our model predictions with the observational dataset of Appendix~\ref{app:obs_sample} and the results from the latest version of the \textsc{SIGAME} framework presented by \citetads{2021ApJ...922...88O}. The latest version of the \textsc{SIGAME} framework uses the \textsc{SIMBA} simulations \citepads{2019MNRAS.486.2827D} with the \textsc{SKIRT} \citepads{2020A&C....3100381C} radiative transfer code and \textsc{Cloudy} \citepads{2017RMxAA..53..385F} to predict line luminosities, similar to this work. The \textsc{SIGAME} predictions come from galaxies in two simulated boxes of 25\,cMpc and 100\,cMpc volumes in the local Universe, similar to the  \textsc{EAGLE} box sizes used in this work. \citetads{2021ApJ...922...88O} find that their $L_{\mathrm[\ion{C}{II}]}$ predictions appear to be an extension to higher SFRs of the model predictions presented in \citetalias{2021A&A...645A.133R}. However, \textsc{SIGAME} tends to have higher line luminosities relative to the SFR--$L_{\mathrm[\ion{C}{II}]}$ relationship (their Fig.~11). In addition, their 25\,cMpc box returns even higher line luminosities up to ${\sim}$0.5--1.5\,dex of the 100\,cMpc box.

\begin{figure*}
    \centering
    \includegraphics[width=\textwidth]{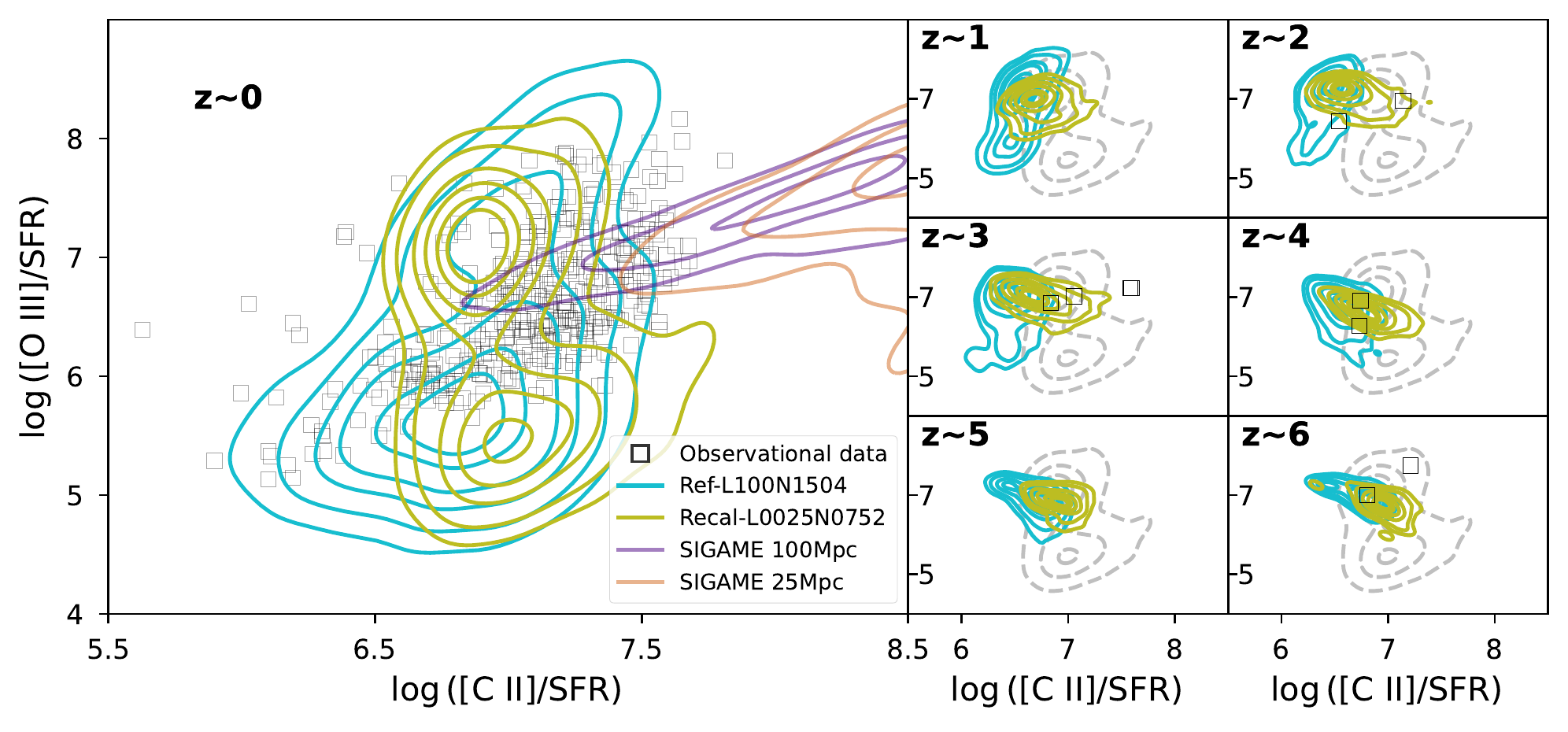}
    \caption{Diagnostic diagram for the $L_{\mathrm[\ion{O}{III}]}$/SFR and $L_{\mathrm[\ion{C}{II}]}$/SFR ratios, similar to that presented by \citetads{2020ApJ...896...93H}. Cyan and olive contours show the model predictions from \textsc{Ref-L100N1504} (cyan) and \textsc{Recal-L0025N0752} (olive). We compare with observational data (black squares) and \textsc{SIGAME} predictions \citepads{2021ApJ...922...88O} for the local Universe in 25\,Mpc and 100\,Mpc simulation boxes (purple and chocolate contours). All panels with redshifts above zero show the $z=0$ \textsc{Recal-L0025N0752} estimations as grey dashed contours.}
    \label{fig:Harikane}
\end{figure*}

In Fig.~\ref{fig:Harikane}, we compare the $L_{\mathrm[\ion{O}{III}]}$/SFR and $L_{\mathrm[\ion{C}{II}]}$/SFR ratios of our model predictions with the observational data and the \textsc{SIGAME} predictions. For the local Universe, our predictions share a similar range of values with the observational data, which tends to be in the range between $6.5 < \log(L_{\mathrm[\ion{C}{II}]}/\mathrm{SFR}) < 7.6$ and $5.8 < \log(L_{\mathrm[\ion{O}{III}]}/\mathrm{SFR}) < 7.6$. However, most of the simulated galaxies tend to be above ($\log(L_{\mathrm[\ion{O}{III}]}/\mathrm{SFR}) > 6.8$) or below ($\log(L_{\mathrm[\ion{O}{III}]}/\mathrm{SFR}) < 6.0$) the observational data. In contrast, \textsc{SIGAME} predictions tend to have very high [\ion{C}{II}] luminosities, which shifts most of the \textsc{SIGAME} simulated galaxies to $\log(L_{\mathrm[\ion{C}{II}]}/\mathrm{SFR}) > 7.5$ values, with the \textsc{SIGAME} values peaking an order of magnitude higher than the observed galaxies and our simulated ratios. This difference is expected from the comparisons presented by \citetads{2021ApJ...922...88O}. As noted above, the \textsc{SIGAME} predictions of $L_{\mathrm[\ion{C}{II}]}$/SFR are higher than those described in \citetalias{2021A&A...645A.133R}, which are similar to those in this work. The \textsc{SIGAME} $L_{\mathrm[\ion{O}{III}]}$/SFR values seem to be similar to our predictions, with values between $6.0 < \log(L_{\mathrm[\ion{O}{III}]}/\mathrm{SFR}) < 8$.  

At $z>0$, there are very few observed galaxies with which we can compare our results. At $z=2$, we only have two measurements for one galaxy: \object{H-ATLAS\,J091043.0-000322}. These two measurements come from \textit{Herschel} observations following different data reduction methods (i.e.\ pipeline versions) in \citetads{2018MNRAS.481...59Z} and \citetads{2018ApJ...867..140L}. ALMA observations presented by \citetads{2018ApJ...867..140L} show that the [\ion{C}{II}] luminosity is around half of the \textit{Herschel} measurement (right data point on the panel), but this difference cannot be fully explained. Therefore, we assume that \object{H-ATLAS\,J091043.0-000322} lies somewhere between the two measurements presented in the $z=2$ panel, agreeing with the predictions from \textsc{Recal-L0025N0752}. 

At $z=3$, we have three measurements for \object{SDP\,81} \citep{2011MNRAS.415.3473V,2014A&A...568A..62D,2018MNRAS.481...59Z} and one for \object{HLock01} \citepads{2018MNRAS.473...20R}. For \object{SDP\,81}, two of the measurements coincide (\citealtads{2014A&A...568A..62D} use the values from \citealtads{2011MNRAS.415.3473V}): the empty squares with $\log(L_{\mathrm[\ion{C}{II}]}/\mathrm{SFR}) {\sim} 7.6$. The other measurement for \object{SDP\,81} comes from \citetads{2018MNRAS.481...59Z}, which is the leftmost point in the $z=3$ panel. The main difference between the values of \citetads{2011MNRAS.415.3473V} and \citetads{2018MNRAS.481...59Z} is the different data reduction methods, and therefore we rely on the more recent results of \citetads{2018MNRAS.481...59Z}. Both \object{SDP\,81} and \object{HLock01} are close to the model predictions of \textsc{Recal-L0025N0752}.

At $z=4$, we have measurements for two galaxies: \object{SPT-S J041839-4751.8} \citepads{2019A&A...631A.167D} and \object{AzTEC 1} \citepads{2019ApJ...876....1T}. Our results agree with the location of both objects, a bit to the left of the results from the $z=0$ galaxies.


Finally, at $z=6$ we have measurements for two galaxies: \object{[DWV2017b] CFHQ J2100-1715 companion} \citepads{2018ApJ...869L..22W} and \object{[MOK2016b] HSC J121137.10-011816.4} \citepads{2020ApJ...896...93H}. \object{[DWV2017b] CFHQ J2100-1715 companion} is in the same region where most of our model predictions are at $z=6$. In contrast, \object{[MOK2016b] HSC J121137.10-011816.4} falls in the same upper region as \object{SPT-S J041839-4751.8} in the $z=4$ panel. The latter two galaxies have SFRs around $100\,\mathrm{M_{\sun}\,yr^{-1}}$ and have $L_{\mathrm[\ion{O}{III}]}$ higher than their respective companions in their panels. These high SFRs can explain their positions in the diagnostic diagrams. \citetads{2020ApJ...896...93H} presented two other galaxies in the same redshift range for the ratios presented in Fig.~\ref{fig:Harikane} besides \object{[MOK2016b] HSC J121137.10-011816.4}. Those two galaxies have been removed from the comparison because they have been identified as QSOs.

\begin{figure*}
    \centering
    \includegraphics[width=\textwidth]{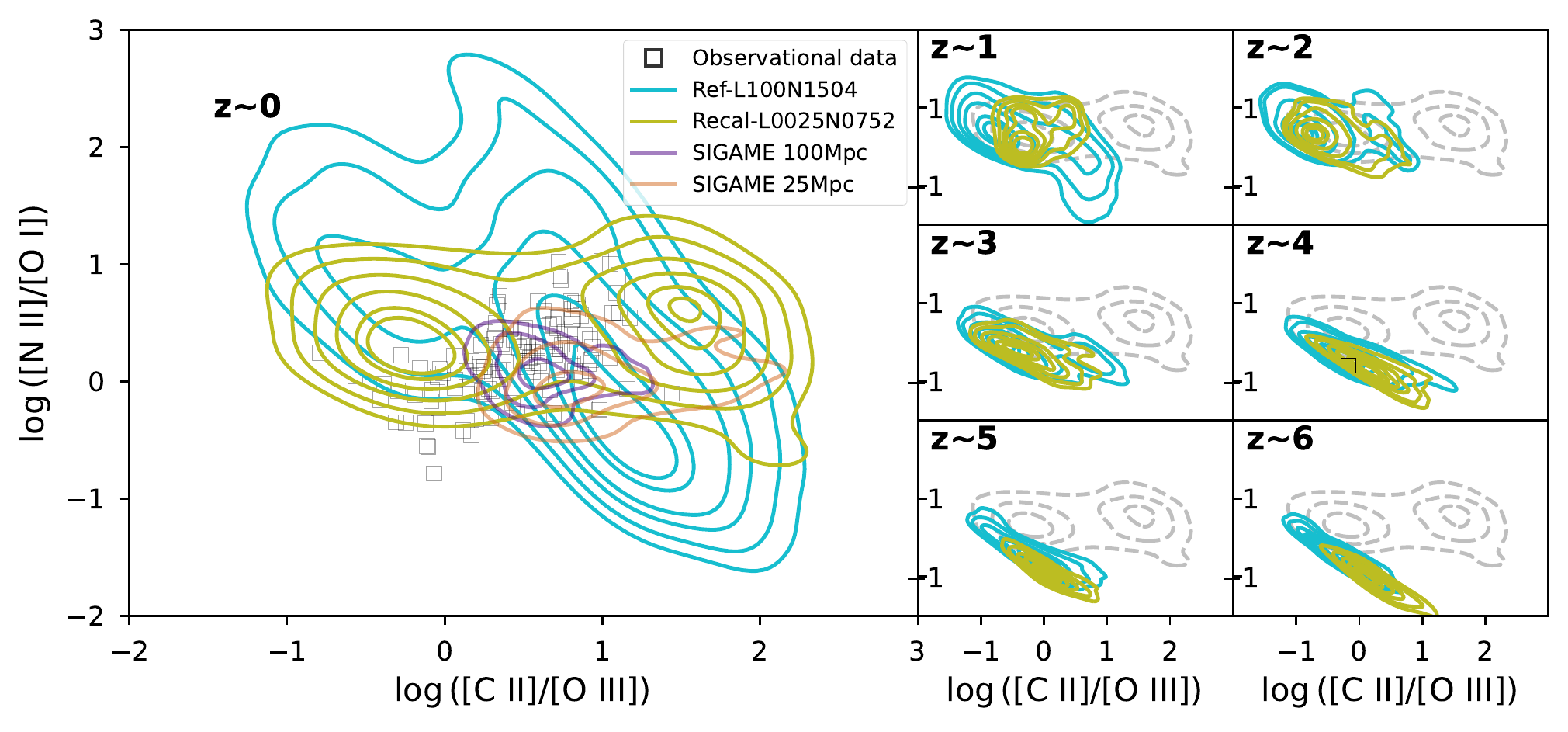}
    \caption{Diagnostic diagram using four different FIR emission lines comparing the [\ion{C}{II}]/[\ion{O}{III}] ratio against the [\ion{N}{II}]/[\ion{O}{I}] ratio. Colour-codes are the sames as in Fig.~\ref{fig:Harikane}.}
    \label{fig:DDFIR}
\end{figure*}

For the other diagnostic diagram, in Fig.~\ref{fig:DDFIR}, we compare the [\ion{C}{II}]/[\ion{O}{III}] and [\ion{N}{II}]/[\ion{O}{I}] ratios of our predictions with the observational data and \textsc{SIGAME} predictions. For the local Universe ($z=0$ panel), we observe that our predictions, observational data and \textsc{SIGAME} predictions tend to be grouped in the region between $-0.5 < \log(L_{\mathrm[\ion{C}{II}]}/L_{\mathrm[\ion{O}{III}]}) < 1.5$ and $-1 < \log(L_{\mathrm[\ion{N}{II}]}/L_{\mathrm[\ion{O}{I}]}) < 1$. The \textsc{SIGAME} predictions match most of the observational data, especially for the 100\,cMpc box, although they do not reach the values close to $\log(L_{\mathrm[\ion{C}{II}]}/L_{\mathrm[\ion{O}{III}]}) = 0$ and $\log(L_{\mathrm[\ion{N}{II}]}/L_{\mathrm[\ion{O}{I}]}) = 0$. The difference between the \textsc{SIGAME} 100\,cMpc and 25\,cMpc boxes comes from the high $L_{\mathrm[\ion{C}{II}]}$ that galaxies in the latter box can have, something that we also commented on at the beginning of this section. Our model behaves similarly to Fig.~\ref{fig:Harikane}, where most of the observations lie between the two concentration regions for \textsc{Recal-L0025N0752}, and the \textsc{Ref-L100N1504} agree with the observations. Our estimates with a simple model of the ISM are in the same parameter space as observations. Unfortunately, a completely fair comparison cannot be made because of selection bias in both the observational and simulated galaxies.

Unfortunately, as we are using four FIR emission lines, there are few observational data points to which we can compare at higher redshifts. The only galaxy with all four FIR lines is \object{SPT-S J041839-4751.8} at $z=4$ \citepads{2019A&A...631A.167D}. Our model matches the position of this galaxy in this diagram, similar to the results presented in Fig.~\ref{fig:Harikane}. The location of this galaxy and most of our predictions at high-$z$ in Fig.~\ref{fig:DDFIR} seem to coincide with some of the observational data at $z=0$, which may imply that some physical parameters (e.g. sSFR, metallicity and/or density) in these galaxies will be similar at different redshifts. 

The comparison presented in Figs.~\ref{fig:Harikane} and \ref{fig:DDFIR} show that our model predictions largely match the parameter space of the observational data in the diagnostic diagrams. We conclude that our model predictions can be used to interpret the physical parameters of  observed galaxies.

\subsection{Physical parameters in diagnostic diagrams}

Now, we use the diagnostic diagrams to infer the ISM physical conditions in galaxies. We use the simulation data of all modelled galaxies and compare the sensitivity of the line luminosity-to-SFR ratio to eight physical parameters as estimated in \citetalias{2021A&A...645A.133R}: gas mass ($\mathrm{M}_{\mathrm{gas}}$), stellar mass ($\mathrm{M}_{\ast}$), metallicity ($\mathrm{Z/Z}_{\odot}$), specific star formation rate (sSFR), interstellar radiation field (ISRF), total hydrogen number density in the neutral clouds ($\mathrm{n(H)}_{\mathrm{cloud}}$), external pressure ($\mathrm{P}_{\mathrm{ext}}$) and radius of the neutral clouds ($\mathrm{R}_{\mathrm{cloud}}$). We divide these physical parameters into seven ranges to compare them with the observational dataset of Appendix~\ref{app:obs_sample} and, as a reference, our mock data from \textsc{Recal-L0025N0752} at $z=0$.

\begin{figure*}
    \centering
    \includegraphics[width=\textwidth]{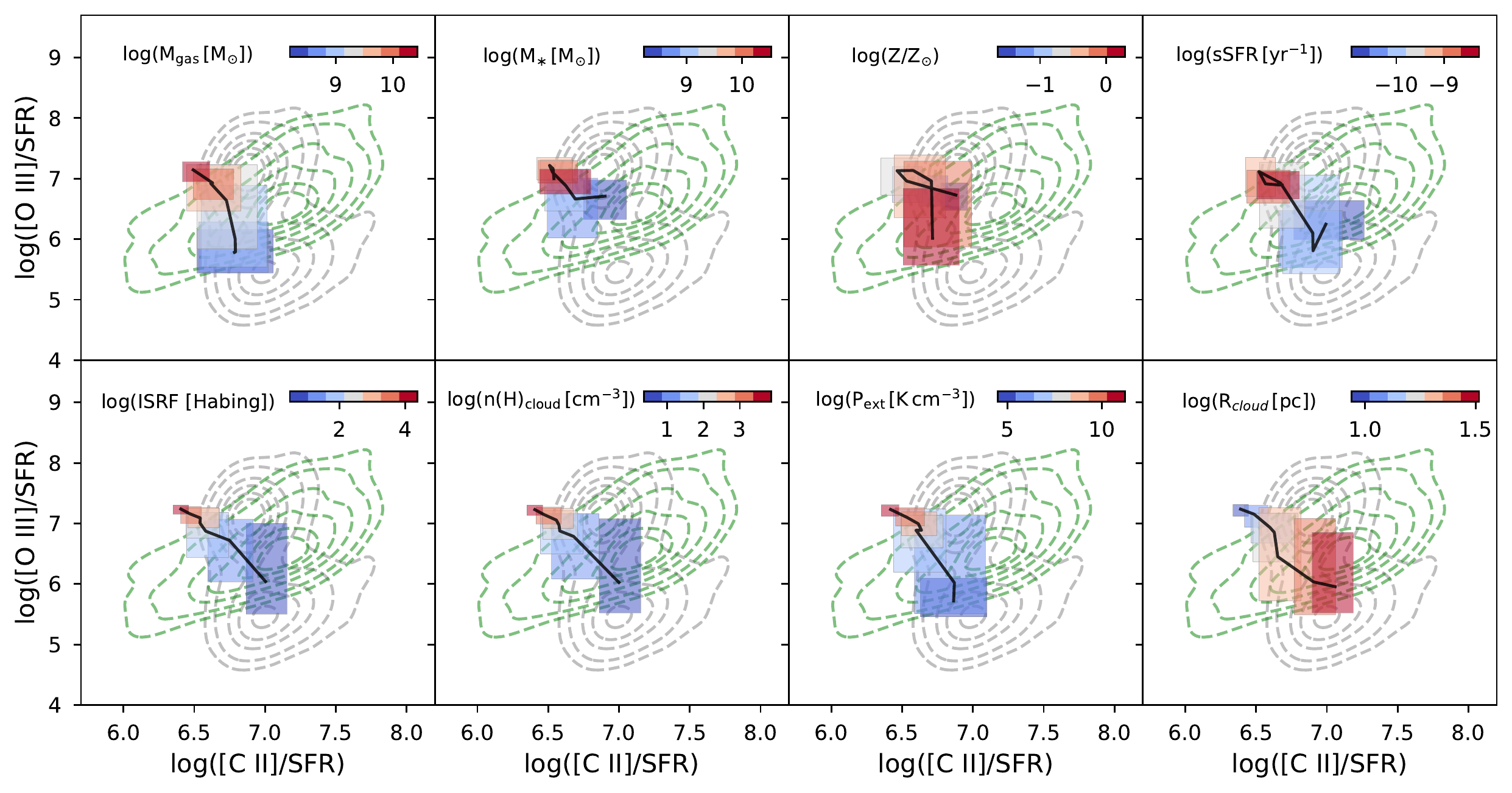}
    \caption{Physical parameters in the $L_{\mathrm[\ion{O}{III}]88}$/SFR--$L_{\mathrm[\ion{C}{II}]}$/SFR diagnostic plot (see Fig.~\ref{fig:Harikane}). All panels show the $z=0$ \textsc{Recal-L0025N0752} model predictions as grey dashed contours and the observational data as green dashed contours. In each panel, colour-coded rectangles represent the 25th and 75th percentiles for a given parameter from low (blue) to high (red) values. The solid black lines connect the median values of each rectangle. The scales for all physical parameters are logarithmic.}
    \label{fig:Harikane_pars}
\end{figure*}

We begin by presenting the physical parameters in the $L_{\mathrm[\ion{O}{III}]}$/SFR--$L_{\mathrm[\ion{C}{II}]}$/SFR diagnostic diagram in Fig.~\ref{fig:Harikane_pars}. We note that the impact of almost all the physical parameters is perpendicular to the observational sample. This effect arises because our mock data also tend to be perpendicular to the observational data, especially at $z=0$. From these physical parameters, we see that most of the predicted galaxies at the upper-left boundary of the observational contour tend to have higher $\mathrm{M}_{\mathrm{gas}}$, $\mathrm{M}_{\ast}$, sSFR, ISRF, $\mathrm{n(H)}_{\mathrm{cloud}}$) and $\mathrm{P}_{\mathrm{ext}}$, and lower $\mathrm{R}_{\mathrm{cloud}}$. In addition, low values of $\mathrm{M}_{\ast}$ do not reach the bottom-right limit of the observational contour, while the metallicity spans all over the observational contours and does not have a clear trend.

\citetads{2020ApJ...896...93H} used the $L_{\mathrm[\ion{O}{III}]}$/SFR--$L_{\mathrm[\ion{C}{II}]}$/SFR diagnostic diagram to explain the physical properties of galaxies at $z=6$--$9$ compared with the local Universe. Using simple \textsc{Cloudy} grids, they found that one of the probable reasons for the location of some of their galaxies in the upper-right region of the diagnostic diagrams was a high ionisation parameter, which is proportional to the ISRF. Their result is similar to what we find from Fig.~\ref{fig:Harikane_pars}; however, it is also important to compare the diagnostic diagram with our metallicity and density results. In terms of metallicity, \citetads{2020ApJ...896...93H} find that $L_{\mathrm[\ion{O}{III}]}$/SFR decreases with metallicity while $L_{\mathrm[\ion{C}{II}]}$/SFR does not change. We find the same result for $L_{\mathrm[\ion{C}{II}]}$/SFR, but the range of change of the $L_{\mathrm[\ion{O}{III}]}$/SFR ratio is not directly dependent on metallicity. In terms of density, \citetads{2020ApJ...896...93H} find that both ratios decrease with density, which we also find for our predictions of $L_{\mathrm[\ion{C}{II}]}$/SFR -- but not for $L_{\mathrm[\ion{O}{III}]}$/SFR. A simple reason for this discrepancy is that $L_{\mathrm[\ion{C}{II}]}$/SFR depends mainly on density while $L_{\mathrm[\ion{O}{III}]}$/SFR is more dependent on other physical parameters: namely metallicity, sSFR and gas mass.

\begin{figure*}
    \centering
    \includegraphics[width=\textwidth]{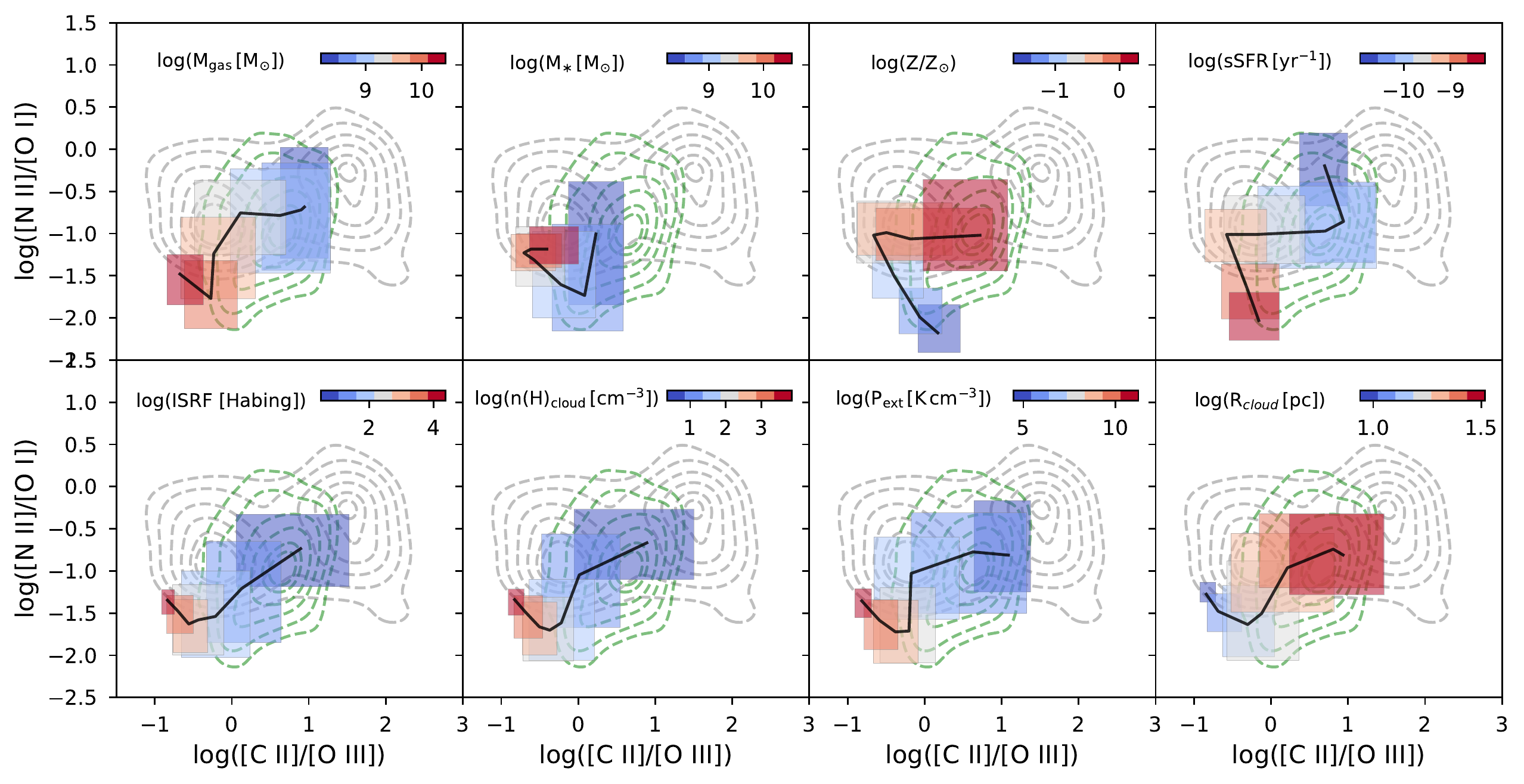}
    \caption{Physical parameters in the diagnostic plot comparing the [\ion{C}{II}]/[\ion{O}{III}] ratio against the [\ion{N}{II}]/[\ion{O}{I}] ratio. Colour-codes are the sames as in Fig.~\ref{fig:Harikane_pars}.}
    \label{fig:DDFIR_pars}
\end{figure*}

Figs.~\ref{fig:Harikane} and \ref{fig:Harikane_pars} highlight the importance of the [\ion{C}{II}] and [\ion{O}{III}] emission lines in understanding the physical conditions of gas of galaxies, especially by using their ratio, as other recent studies have done \citep[e.g.][]{2020ApJ...896...93H,2020MNRAS.498.5541A,2020MNRAS.499.5136C,2021MNRAS.505.5543V,2021arXiv210613719B} and as we show below. We check its correlation with other FIR lines that trace the neutral and ionised gas components, such as the [\ion{N}{II}]/[\ion{O}{I}] ratio. We present the physical parameters of the diagnostic diagram of the [\ion{C}{II}]/[\ion{O}{III}] vs [\ion{N}{II}]/[\ion{O}{I}] ratios in Fig.~\ref{fig:DDFIR_pars}. In this diagnostic diagram, we note how the physical parameters cross the observational data contours in different ways. Interestingly, the physical parameters in the bottom row panels show a ``spoon-like'' shape showing that most of the observed galaxies have low to moderate values for the ISRF, $\mathrm{n(H)}_{\mathrm{cloud}}$), and $\mathrm{P}_{\mathrm{ext}}$, and moderate to high values for the $\mathrm{R}_{\mathrm{cloud}}$. The opposite occurs in the region with low  [\ion{C}{II}]/[\ion{O}{III}],  (i.e.\ high ISRF, $\mathrm{n(H)}_{\mathrm{cloud}}$ and $\mathrm{P}_{\mathrm{ext}}$, and low $\mathrm{R}_{\mathrm{cloud}}$), which coincides with our predictions for high-$z$ galaxies ($z>3$, in Fig.~\ref{fig:DDFIR}). In terms of mass, $\mathrm{M}_{\mathrm{gas}}$ is more extended but follows a similar trend to the parameters mentioned above, while $\mathrm{M}_{\ast}$ does not show a clear trend. The most intriguing trends in the [\ion{C}{II}]/[\ion{O}{III}]--[\ion{N}{II}]/[\ion{O}{I}] diagnostic diagram come from metallicity and sSFR. In terms of metallicity, the [\ion{C}{II}]/[\ion{O}{III}] ratio seems to be a good tracer for metallicities close to solar, while the [\ion{N}{II}]/[\ion{O}{I}] ratio is a good tracer for metallicities below $\log(\mathrm{Z}/\mathrm{Z}_{\odot}) \lesssim 0.5$, which agrees with some results for high-$z$ galaxies \citepads{2020MNRAS.498.5541A}. In terms of sSFR, both ratios do a good job of separating high and low sSFR of galaxies in a zigzag pattern across the observational sample region. This supports the idea that the [\ion{C}{II}]/[\ion{O}{III}] ratio can be used for starbursting systems \citepads{2021MNRAS.505.5543V}.

We have shown that both diagnostic diagrams track the behaviour of the physical parameters presented in the simulated galaxies using the luminosities of the main FIR lines. Our model agrees with the observational data in most of the parameter space, and in some cases with other simulations (e.g. \textsc{SIGAME}). Therefore, it is reasonable to expect that the physical model could be used in inferring the physical parameters in FIR diagnostic diagrams to trace different physical parameters.

Comparing our model predictions with new observational data will give an idea of what kind of physical parameters are expected in those galaxies. At the same time, our model can constrain the expected luminosities of FIR lines when no other measurement is available. However, we emphasise that the modelled FIR lines can also be used to study other types of problems. For example, ratios such as [\ion{N}{II}]/[\ion{C}{II}] or other diagnostic diagrams can be used to characterise different types of galaxies (e.g.\ Ultra-luminous infrared galaxies (ULIRGS, where $\mathrm{L}_{\mathrm{IR}} > 10^{12}\,\mathrm{L}_\sun$)) in the local Universe \citepads[e.g.][]{2013ApJ...776...38F} and at high-$z$ \citepads[e.g.][]{2020MNRAS.494.4090C}. Another example is that the ratio between the two [\ion{O}{III}] lines can be used to test the efficiency of the black hole feedback in galaxies, as was done with the IllustrisTNG simulations \citepads{2021arXiv210210752I}. Similarly, ratios like [\ion{O}{III}]/[\ion{N}{III}] or [\ion{O}{III}]/[\ion{N}{II}] can be used as a metallicity indicator \citep[e.g.][]{2011A&A...526A.149N,2018MNRAS.473...20R,2021A&A...652A..23F}, which will be explored in future studies.

\section{Discussion}\label{sec:systematics}

\subsection{Model assumptions}

Although our model emission-line luminosities and ratios are in good agreement with observational results and other simulations, our physically motivated model is based on a number of assumptions. The effect of some of these assumptions is generally not visible in these types of predictions due to uncertainty in the observations and our current understanding of some of the physical processes involved. However, these assumptions may be important for models that try to predict the FIR line emission of galaxies, especially at high-$z$. In this section we highlight the most important assumptions. 

For neutral clouds and \textsc{Hii} regions, we assume static spherical geometries to describe the densities and temperatures within those environments. This assumption is made for simplicity, although we know that these structures may not be spherical. Physical processes such as radiation destroy spherical geometries \citep{2010A&A...523A...6D,2018MNRAS.475.1508P}, which may lead to rough or incorrect line luminosity predictions \citepads{2020MNRAS.497.4718D}. However, approximations using mass distributions (Eq.~15 in \citetalias{2021A&A...645A.133R}) may smooth out these differences, since cloud masses follow scaling relations 
that seem to be valid for observations at different redshifts \citepads{2019NatAs...3.1115D}. 

A problem with our modelled line luminosities from \textsc{Hii} regions is that we assume a fixed density (${\sim} 30\, \mathrm{cm}^{-3}$). We can increase the luminosity by increasing this density and vice-versa. The use of different densities could be important when comparing the contributions of DIG and \textsc{Hii} regions, especially in lines such as [\ion{O}{III}] and [\ion{N}{III}] (see Figs.~\ref{fig:Cont_OIII} and \ref{fig:Cont_NIII}). This DIG--\textsc{Hii} region balance is still unclear in ionised emission lines, and although some estimates exist from optical wavelengths \citepads[e.g.][]{2019MNRAS.487...79P}, a change in this balance can lead to different metallicities, which may affect high-$z$ studies \citepads{2017ApJ...850..136S}. In this work, we calibrate the DIG with observational data at $z=0$, which may also bias the balance between these two ISM phases. Fortunately, the results presented in this work show that these assumptions seem to be in agreement with the observations, which may represent a likely first step in understanding the DIG--\textsc{Hii} region balance. 

Finally, our predictions depend on \textsc{Cloudy} lookup tables, which can give different emissivities depending on the assumed initial abundances or dust configurations \citepads{2020MNRAS.497.4857P}. Different photoionisation models could lead to different interpretations of physical parameters coming from line ratios, such as metallicity or ionisation parameter \citepads[e.g.][]{2022A&A...659A.112J}. Furthermore the intrinsic thermodynamics may not be the same in terms of cooling and heating functions in cosmological simulations and photoionisation models like  \textsc{Cloudy} \citepads{2021arXiv210901674R}. Therefore, some care must be taken when interpreting the line luminosities predicted by our model. 

\subsection{The use of \textsc{EAGLE}}

In this work we use the \textsc{EAGLE} simulation as a proxy of what we would expect to see in the Universe. However, by using a cosmological simulation like \textsc{EAGLE}, our line luminosity predictions of the ISM model are expected to inherit the limitations of the simulation. An example of these limitations is the lack of starburst-like galaxies within \textsc{EAGLE} \citepads{2019A&A...624A..98W}. As we discussed in \citetalias{2021A&A...645A.133R}, this restricts our comparison at $z=0$ mainly to SFR below $10\,\mathrm{M_{\sun}} \, \mathrm{yr}^{-1}$ but also limits the comparisons at other redshifts \citepads{2017MNRAS.472..919K}. To compare our predictions with high SFR observations, we extrapolate the linear relations in the range of $-3.5 < \log(\mathrm{SFR}) <3.5$. However, care must be taken when using this extrapolation.  

Another important physical property within \textsc{EAGLE} that affects our predictions is the gas metallicity, which is usually studied through the gas-phase mass–metallicity relation (MZR). \citetads{2021MNRAS.503.3309B} show that the MZR in \textsc{EAGLE} galaxies, as measured by \citetads{2020MNRAS.496L..33Z} at $z=0$, does not behave similarly to other cosmological simulations or semi-analytical models. As shown in Fig. 4 of \citetads{2021MNRAS.503.3309B}, the metallicities in \textsc{EAGLE} have values around $12+\log(\mathrm{O/H}) {\sim}9$ for stellar masses between $9<\log(\mathrm{M}_{\ast} [\mathrm{M_{\sun}}])<11$, which is high compared to observations with metallicities going from $12+\log(\mathrm{O/H}) {\sim}8.5$ to $12+\log(\mathrm{O/H}) {\sim}9.2$ in the same stellar mass range. This may affect the metallicity that can be recovered from FIR lines if only $z=0$ is used. However, MZR in \textsc{EAGLE} depends on resolution and box-size used due to the assumed AGN and star formation feedback processes \citepads{2017MNRAS.472.3354D}. In Figs.~\ref{fig:Harikane_pars} and \ref{fig:DDFIR_pars}, we present the predicted luminosities in different boxes and redshifts studied in this work and find that some of the FIR line ratios can be useful to infer the metallicity. Therefore, it will be important to compare the FIR line predictions that trace metallicity with observations in the future more consistently.

Although we base our predictions on \textsc{EAGLE}, we expect that similar physical models and/or cosmological simulations can be used to understand the gas properties with FIR emission lines. For example, \citetads{2021ApJ...922...88O} show that these kinds of predictions can be obtained with a different gas fragmentation scheme using \textsc{SIGAME}, which are similar to observations and our predictions. However, most of the galaxies in \textsc{SIGAME} have higher SFR than those studied in this work. This difference in the sample of simulated galaxies comes from choosing a different simulation (\textsc{SIMBA} instead of \textsc{EAGLE}), which allows for the formation of starburst-like systems. Therefore, although some physical assumptions are different for each model and are limited by the simulation used \citep[e.g.][]{2015ApJ...813...36V,2018A&A...609A.130L,2019MNRAS.482.4906P,2019MNRAS.487.1689P,2020ApJ...905..102L}, we expect predictions to behave similarly to those presented in this work. In the future, we will compare results from different simulations, such as \textsc{SIMBA} and \textsc{IllustrisTNG} in an efficient way to reduce the bias that the initial assumptions of the simulations may introduce.  

\subsection{Observational data from samples}

For our comparison between observations and our model line luminosities, we have collected a heterogeneous sample of observed galaxies with FIR emission line information. We have transformed the luminosities of the lines to the same reference cosmology \citepads{2014A&A...571A..16P} and we have estimated the SFR in most of the cases following the $\mathrm{L}_{\mathrm{IR}}$--SFR relation described by \citetads{2012ARA&A..50..531K}. However, there are some other important issues that may affect this heterogeneous sample.

The most important issue may come from the initial mass function (IMF) assumed to estimate the SFRs. Unfortunately, IMF information is not always present in papers. Assuming a star-formation law that takes into account a standard IMF is a possible solution, as we have tried in this work. However, this may add additional uncertainties. For example, \textsc{EAGLE} uses an instantaneous SFR, which is different from the SFR averaged over the last 10 or 100 Myr typically used in observations. Another problem is that some of the SFR (or $\mathrm{L}_{\mathrm{IR}}$) may come from spectral energy distribution (SED) models rather than empirical laws using FIR photometric bands, like IRAS, PACS or SPIRE. In addition, the IR luminosity definition can have different flavours (e.g.\ $\mathrm{L}_{\mathrm{IR}}$, $\mathrm{L}_{\mathrm{FIR}}$ or $\mathrm{L}_{\mathrm{TIR}}$) that use different wavelengths ranges to estimate luminosities, adding to the spread of SFR estimates \citepads{2020ARA&A..58..661F} and leading to differences of factors between 1.1 and 1.7 \citepads{2014A&A...568A..62D,2018A&A...609A.130L}. Another possibility is to use other SFR tracers, such as H$\alpha$-based, UV-based, or radio-based SFRs as shown in many studies \citep[e.g.][]{2016MNRAS.461.1898W,2019A&A...631A.109W}. However, in some cases, such as in high-$z$ galaxies, it can be difficult to obtain such SFRs estimates and other lines like [\ion{C}{II}] come into play as a SFR tracer \citepads{2020A&A...643A...1L}, as we have also shown in this work.

In addition to theoretical considerations, some of the FIR line measurements have systematic uncertainties. For example, calibration pipelines or the use of different instruments may affect the comparison, as shown by \citetads{2018ApJ...867..140L} in the case of \object{H-ATLAS\,J091043.0-000322}, discussed in Sect.~\ref{subsec:comobs}. Furthermore, different-sized (or -shaped) apertures in the observational sample may affect the analysis of the fixed-size aperture we selected for the simulated galaxies (30\,pkpc). For example, the balance between DIG and \textsc{Hii} regions as main contributors to the ionised lines could depend on the selected aperture, as indicated by \citetads{2021MNRAS.508.1582M}. Finally, gravitational lensing can introduce a large uncertainty in the luminosities, which can be reduced by a factor of 30 to 40 in some galaxies when corrected by the magnification factor \citep[e.g.\ in \object{Eyelash} and \object{SPT-S J041839-4751.8,}][respectively]{2018MNRAS.481...59Z,2020Natur.584..201R}. In addition, these magnification factors may change depending on the observed region of the galaxy \citepads{2018ApJ...867..140L}. In this work, we highlight those galaxies to warn about these potential complications, but in general, they show good agreement with our predictions.

\section{Summary and Conclusions} \label{sec:concl}

We have modelled FIR emission lines by post-processing \textsc{EAGLE} cosmological simulations using \textsc{Cloudy}. We have predicted the luminosities of the eight most important lines in the FIR up to $z=6$, using a physically motivated model that traces four different ISM phases: dense molecular gas, neutral atomic gas, diffuse ionised gas and \textsc{Hii} regions. We have also collected a sample of observed galaxies from the literature with which to compare with our predictions. Our main conclusions are as follows. 

\begin{enumerate}
   \item Predictions from our model replicate observed galaxies in the SFR--FIR line luminosity relationship to an average degree of 0.5 \,dex over the range $z=0$--$6$, which is reasonable considering the observational measurements errors. We also compare with different models showing similar level of agreement. We model the SFR--FIR line luminosity relationship for each of the eight lines with a linear relation, each of which shows a slight evolution with redshift. 
   \item We have presented the expected contributions of each ISM phase to each FIR line. These contributions change as a function of SFR. For the [\ion{C}{II}] 158\,$\mu$m line, the main contributor is the neutral atomic gas, with considerable contributions from \textsc{Hii} regions at $z=1$--$4$ and the DIG at $z<2$, which may be related to metallicity. For the [\ion{N}{II}] lines at 122 and 205\,$\mu$m, the DIG contributes more than \textsc{Hii} regions in the local Universe, but the opposite is true at high-$z$, where \textsc{Hii} regions seem to dominate over the DIG. For the [\ion{O}{I}] lines at 63 and 145\,$\mu$m, the contribution of dense molecular gas is important in the local Universe. However, the atomic gas is dominant at high-$z$. Finally, for the [\ion{O}{III}] lines at 52 and 88\,$\mu$m and [\ion{N}{III}] at 57\,$\mu$m, we show that \textsc{Hii} regions dominate, with important contribution from the DIG at low SFR in the local Universe.
   \item Our predictions indicate that [\ion{C}{II}] may not be a good SFR tracer for starburst galaxies, since the [\ion{C}{II}]/SFR ratio seems to decrease as a function of the offset from the star-forming main-sequence. However, compared to the other FIR lines, [\ion{C}{II}] seems to be the best SFR tracer due to its weak redshift evolution. [\ion{O}{III}] and [\ion{O}{I}] may also be good SFR tracers. Nonetheless, our predictions of [\ion{O}{III}] at $z=0$ may be underestimated, and more observations of [\ion{O}{I}] are necessary at $z<4$ to confirm our predictions.
   \item We compare our predictions in two diagnostic diagrams, and we find reasonable agreement with observations. We compare $L_{\mathrm[\ion{O}{III}]}$/SFR and $L_{\mathrm[\ion{C}{II}]}$/SFR ratios and find that mock galaxies at high-$z$ tend to have higher $L_{\mathrm[\ion{O}{III}]}$/SFR ratios and slightly lower $L_{\mathrm[\ion{C}{II}]}$/SFR ratios than galaxies in the local Universe. We also compare the [\ion{C}{II}]/[\ion{O}{III}] and [\ion{N}{II}]/[\ion{O}{I}] ratios and find that mock galaxies at high-$z$ tend to have lower [\ion{C}{II}]/[\ion{O}{III}] and [\ion{N}{II}]/[\ion{O}{I}] ratios than galaxies in the local Universe.
   \item Finally, we have examined the impact of physical parameters on these diagnostic diagrams. When we compare physical parameters to line luminosities, we find that $L_{\mathrm[\ion{C}{II}]}$/SFR and $L_{\mathrm[\ion{O}{III}]}$/SFR ratios trace hydrogen density and ISRF well in the mock galaxies. However, these ratios are not good metallicity tracers, because $L_{\mathrm[\ion{O}{III}]}$/SFR does not evolve linearly with metallicity and $L_{\mathrm[\ion{C}{II}]}$/SFR does not change with metallicity. Furthermore, we find that [\ion{C}{II}]/[\ion{O}{III}] and [\ion{N}{II}]/[\ion{O}{I}] ratios can be good metallicity and sSFR tracers. For example, [\ion{C}{II}]/[\ion{O}{III}] can trace metallicities close to solar and [\ion{N}{II}]/[\ion{O}{I}] below solar. On the other hand, we can identify systems with different sSFRs by means of both [\ion{C}{II}]/[\ion{O}{III}] and [\ion{N}{II}]/[\ion{O}{I}] ratios, which can be very useful for improving calibrations of [\ion{C}{II}] as a SFR tracer.
\end{enumerate}
    
In the future we expect to use this model and its predictions to understand the effect that AGN can have on the ISM, as well as the physical parameters traced by these lines and their ratios. We make our model predictions and collected observational sample publicly available to allow potential users to compare with their work and/or interpret new observations. We envisage that our predictions will also be useful in planning for future FIR missions.

\begin{acknowledgements}
The authors thank Karen Pardos Olsen for providing the data from the \textsc{SIGAME} framework. This research has made use of the 3Mdb database\footnote{\url{https://sites.google.com/site/mexicanmillionmodels}} \citepads{2015RMxAA..51..103M} from which initial experiments to model \textsc{Hii} regions were performed. This research made use of Astropy,\footnote{\url{http://www.astropy.org}} a community-developed core Python package for Astronomy \citep{2013A&A...558A..33A,2018AJ....156..123A}. This research made use of several Python packages, among them: numpy \citep{harris2020array}, pandas \citep{mckinney-proc-scipy-2010} and matplotlib \citep{Hunter:2007}. This research has made use of the SIMBAD database, operated at CDS, Strasbourg, France. This research has made use of the NASA/IPAC Extragalactic Database (NED), which is funded by the National Aeronautics and Space Administration and operated by the California Institute of Technology. This research has made use of NASA’s Astrophysics Data System Bibliographic Services. The EAGLE simulations were performed using the DiRAC-2 facility at Durham, managed by the ICC, and the PRACE facility Curie based in France at TGCC, CEA, Bruy\`{e}res-le-Ch\^{a}tel. We would like to thank the Center for Information Technology of the University of Groningen for their support and for providing access to the Peregrine high performance computing cluster.
\end{acknowledgements}

\bibliographystyle{aa}
\bibliography{ReferenceADS,Manual}


\begin{appendix}

\section{Observational sample}\label{app:obs_sample}

We have collected measurements from the literature of the FIR emission lines predicted in this work. The observational sample is a heterogeneous selection of galaxies that covers the redshift range between 0 and 6.5. We present references of the works used in this sample of galaxies together with the number of measurements available per line in Table~\ref{tab:tab_red}. When possible we recalculated the luminosities and SFRs onto the $\Lambda$CDM cosmology used in this work \citepads{2014A&A...571A..16P}. When the cosmology is not explicitly mentioned, we assume it is the same as used in this work, so no corrections are applied. We flag those measurements where a magnification factor is involved in estimating a luminosity due to gravitational lensing. We estimate SFR in most of the literature samples from the infrared luminosity ($\mathrm{L}_{\mathrm{IR}}$) using the relation of \citetads{2012ARA&A..50..531K}, which assumes a similar IMF (Kroupa) to the one used in \textsc{EAGLE} (Chabrier). Unless stated otherwise, we use the same infrared luminosities as published in the respective works. In cases where $\mathrm{L}_{\mathrm{IR}}$ is unavailable or is unreliable due to measurement error, we use SFR estimates from other works. Additionally, we remove strong AGN galaxies (i.e.\ QSO and Blazar). This observational sample is available as a supplementary material in a Zenodo repository, at \url{https://doi.org/10.5281/zenodo.6576202}.

\longtab[1]{
    \begin{longtable}{c cccccccc cc}
\caption{References and number of measurements per line with a redshift range for the observational sample, after AGN dominated galaxies have been removed.\label{tab:tab_red}}\\
\hline
\hline
 & \multicolumn{8}{c}{Line measurements} & & \\
\cline{2-9} 
Reference & {[\ion{O}{III}]} & {[\ion{N}{III}]} & {[\ion{O}{I}]} &
{[\ion{O}{III}]} & {[\ion{N}{II}]} & {[\ion{O}{I}]} & {[\ion{C}{II}]} & {[\ion{N}{II}]} & \multicolumn{2}{c}{Redshift}  \\
\cline{10-11}
& 52 $\mu$m & 57 $\mu$m & 63 $\mu$m & 88 $\mu$m & 122 $\mu$m & 145 $\mu$m & 158 $\mu$m & 205 $\mu$m & Min. & Max. \\
\hline
\endfirsthead
\caption{continued.}\\
\hline\hline
 & \multicolumn{8}{c}{Line measurements} & & \\
\cline{2-9} 
Reference & {[\ion{O}{III}]} & {[\ion{N}{III}]} & {[\ion{O}{I}]} &
{[\ion{O}{III}]} & {[\ion{N}{II}]} & {[\ion{O}{I}]} & {[\ion{C}{II}]} & {[\ion{N}{II}]} & \multicolumn{2}{c}{Redshift}  \\
\cline{10-11}
& 52 $\mu$m & 57 $\mu$m & 63 $\mu$m & 88 $\mu$m & 122 $\mu$m & 145 $\mu$m & 158 $\mu$m & 205 $\mu$m & Min. & Max. \\
\hline
\endhead
\hline
\endfoot
{\citetads{2008ApJS..178..280B}} & 45 & 41 & 158 & 87 & 100 & 47 & 223 & $\cdots$ & 0.0 & 1.13 \\
{\citetads{2010ApJ...724..957S}} & $\cdots$ & $\cdots$ & $\cdots$ & $\cdots$ & $\cdots$ & $\cdots$ & 9 & $\cdots$ & 1.12 & 1.96 \\
{\citetads{2011MNRAS.415.3473V}} & 1 & 1 & 1 & 1 & 1 & 1 & 1 & $\cdots$ & 3.04 & 3.04 \\
{\citetads{2012ApJ...752....2D}} & $\cdots$ & $\cdots$ & $\cdots$ & $\cdots$ & $\cdots$ & $\cdots$ & $\cdots$ & 1 & 3.94 & 3.94 \\
{\citetads{2012MNRAS.427..520C}} & $\cdots$ & $\cdots$ & 6 & $\cdots$ & $\cdots$ & $\cdots$ & $\cdots$ & $\cdots$ & 1.1 & 1.62 \\
{\citetads{2013ApJ...776...38F}} & 22 & 23 & 23 & $\cdots$ & 23 & 23 & 23 & $\cdots$ & 0.04 & 0.26 \\
{\citetads{2014A&A...568A..62D}} & $\cdots$ & $\cdots$ & 10 & 4 & $\cdots$ & $\cdots$ & 19 & $\cdots$ & 1.1 & 6.6 \\
{\citetads{2014ApJ...796...63M}} & $\cdots$ & $\cdots$ & $\cdots$ & $\cdots$ & $\cdots$ & 8 & 15 & 12 & 0.22 & 2.96 \\
{\citetads{2015ApJ...799...13B}} & $\cdots$ & $\cdots$ & 7 & $\cdots$ & $\cdots$ & $\cdots$ & 8 & $\cdots$ & 1.41 & 2.0 \\
{\citetads{2015ApJ...799...21S}} & 8 & 14 & 26 & 26 & 23 & 23 & 26 & 12 & 0.0 & 0.3 \\
{\citetads{2015ApJ...801...72R}} & $\cdots$ & $\cdots$ & 14 & $\cdots$ & $\cdots$ & 25 & 29 & $\cdots$ & 0.0 & 0.04 \\
{\citetads{2015ApJ...806..260F}} & $\cdots$ & $\cdots$ & $\cdots$ & 1 & 1 & $\cdots$ & $\cdots$ & $\cdots$ & 2.81 & 2.81 \\
{\citetads{2015MNRAS.449.2883G}} & $\cdots$ & $\cdots$ & $\cdots$ & $\cdots$ & $\cdots$ & $\cdots$ & 20 & $\cdots$ & 2.12 & 5.7 \\
{\citetads{2015Natur.522..455C}} & $\cdots$ & $\cdots$ & $\cdots$ & $\cdots$ & $\cdots$ & $\cdots$ & 11 & $\cdots$ & 5.15 & 5.69 \\
{\citetads{2016A&A...586L...7B}} & $\cdots$ & $\cdots$ & $\cdots$ & $\cdots$ & $\cdots$ & $\cdots$ & 1 & 1 & 4.77 & 4.77 \\
{\citetads{2016AJ....151...14C}} & $\cdots$ & $\cdots$ & 5 & 4 & 5 & $\cdots$ & 5 & $\cdots$ & 0.0 & 0.0 \\
{\citetads{2016ApJ...819...69Z}} & $\cdots$ & $\cdots$ & $\cdots$ & $\cdots$ & $\cdots$ & $\cdots$ & $\cdots$ & 196 & 0.0 & 0.06 \\
{\citetads{2016ApJ...824..146Z}} & $\cdots$ & $\cdots$ & 7 & $\cdots$ & $\cdots$ & $\cdots$ & 8 & $\cdots$ & 0.11 & 0.2 \\
{\citetads{2016ApJ...827...34O}} & $\cdots$ & $\cdots$ & $\cdots$ & $\cdots$ & $\cdots$ & $\cdots$ & 2 & $\cdots$ & 4.42 & 4.43 \\
{\citetads{2016ApJ...829...93K}} & $\cdots$ & $\cdots$ & $\cdots$ & $\cdots$ & $\cdots$ & $\cdots$ & $\cdots$ & 214 & 0.0 & 0.25 \\
{\citetads{2016ApJ...832..151P}} & $\cdots$ & $\cdots$ & $\cdots$ & $\cdots$ & $\cdots$ & $\cdots$ & 3 & 3 & 5.29 & 5.65 \\
{\citetads{2016ApJ...832..209U}} & $\cdots$ & $\cdots$ & $\cdots$ & $\cdots$ & $\cdots$ & $\cdots$ & 1 & $\cdots$ & 2.29 & 2.29 \\
{\citetads{2016ApJS..226...19F}} & 28 & 50 & 141 & 115 & 98 & 81 & 178 & 186 & 0.0 & 0.44 \\
{\citetads{2016MNRAS.462L...6K}} & $\cdots$ & $\cdots$ & $\cdots$ & $\cdots$ & $\cdots$ & $\cdots$ & 1 & $\cdots$ & 6.03 & 6.03 \\
{\citetads{2017ApJ...837...12W}} & 8 & 4 & 5 & $\cdots$ & $\cdots$ & $\cdots$ & $\cdots$ & $\cdots$ & 1.03 & 3.27 \\
{\citetads{2017ApJ...846...32D}} & $\cdots$ & $\cdots$ & 226 & 148 & 71 & $\cdots$ & 229 & $\cdots$ & 0.0 & 0.09 \\
{\citetads{2017ApJ...846..105O}} & $\cdots$ & $\cdots$ & $\cdots$ & $\cdots$ & $\cdots$ & $\cdots$ & 34 & $\cdots$ & 5.15 & 7.6 \\
{\citetads{2017Natur.545..457D}} & $\cdots$ & $\cdots$ & $\cdots$ & $\cdots$ & $\cdots$ & $\cdots$ & 4 & $\cdots$ & 6.07 & 6.59 \\
{\citetads{2018A&A...609A.130L}} & $\cdots$ & $\cdots$ & $\cdots$ & $\cdots$ & $\cdots$ & $\cdots$ & 55 & $\cdots$ & 4.22 & 7.91 \\
{\citetads{2018ApJ...856..174V}} & $\cdots$ & $\cdots$ & $\cdots$ & 1 & $\cdots$ & $\cdots$ & $\cdots$ & $\cdots$ & 3.13 & 3.13 \\
{\citetads{2018ApJ...859...12G}} & $\cdots$ & $\cdots$ & $\cdots$ & $\cdots$ & $\cdots$ & $\cdots$ & 3 & $\cdots$ & 4.42 & 4.44 \\
{\citetads{2018ApJ...861...94H}} & $\cdots$ & 47 & 55 & 46 & 56 & 50 & 56 & $\cdots$ & 0.0 & 0.13 \\
{\citetads{2018ApJ...867..140L}} & 1 & 1 & 1 & 1 & 1 & $\cdots$ & 2 & $\cdots$ & 1.78 & 1.78 \\
{\citetads{2018ApJ...869L..22W}} & $\cdots$ & $\cdots$ & $\cdots$ & 1 & $\cdots$ & $\cdots$ & 1 & $\cdots$ & 6.08 & 6.08 \\
{\citetads{2018MNRAS.473...20R}} & 1 & 1 & 1 & 1 & 1 & 1 & 1 & $\cdots$ & 2.96 & 2.96 \\
{\citetads{2018MNRAS.481...59Z}} & 9 & $\cdots$ & 25 & 35 & 38 & 35 & 33 & 13 & 1.03 & 3.63 \\
{\citetads{2018MNRAS.481.1976Z}} & $\cdots$ & $\cdots$ & $\cdots$ & $\cdots$ & $\cdots$ & $\cdots$ & 9 & $\cdots$ & 1.73 & 1.94 \\
{\citetads{2019A&A...626A..23C}} & $\cdots$ & 3 & 30 & 36 & 15 & 12 & 41 & 3 & 0.0 & 0.04 \\
{\citetads{2019A&A...631A.167D}} & $\cdots$ & $\cdots$ & $\cdots$ & 1 & 1 & 1 & 1 & 1 & 4.22 & 4.22 \\
{\citetads{2019ApJ...876....1T}} & $\cdots$ & $\cdots$ & $\cdots$ & 1 & $\cdots$ & $\cdots$ & 1 & 1 & 4.34 & 4.34 \\
{\citetads{2019ApJ...882...10N}} & $\cdots$ & $\cdots$ & $\cdots$ & $\cdots$ & $\cdots$ & $\cdots$ & 7 & $\cdots$ & 6.03 & 6.59 \\
{\citetads{2019ApJ...883L..29L}} & $\cdots$ & $\cdots$ & $\cdots$ & $\cdots$ & 1 & $\cdots$ & 1 & 1 & 4.69 & 4.69 \\
{\citetads{2019MNRAS.488.1489H}} & $\cdots$ & $\cdots$ & $\cdots$ & $\cdots$ & $\cdots$ & $\cdots$ & $\cdots$ & 1 & 2.55 & 2.55 \\
{\citetads{2020A&A...643A...2B}} & $\cdots$ & $\cdots$ & $\cdots$ & $\cdots$ & $\cdots$ & $\cdots$ & 117 & $\cdots$ & 4.41 & 5.87 \\
{\citetads{2020ApJ...889L..11R}} & $\cdots$ & $\cdots$ & 1 & $\cdots$ & $\cdots$ & $\cdots$ & $\cdots$ & $\cdots$ & 6.03 & 6.03 \\
{\citetads{2020ApJ...896...93H}} & $\cdots$ & $\cdots$ & $\cdots$ & 1 & 1 & $\cdots$ & 1 & $\cdots$ & 6.03 & 6.03 \\
{\citetads{2020ApJ...898...33C}} & $\cdots$ & $\cdots$ & $\cdots$ & $\cdots$ & $\cdots$ & $\cdots$ & 1 & 1 & 4.24 & 4.24 \\
{\citetads{2020MNRAS.494.4090C}} & $\cdots$ & $\cdots$ & $\cdots$ & $\cdots$ & $\cdots$ & $\cdots$ & 27 & 37 & 3.07 & 5.81 \\
{\citetads{2020Natur.581..269N}} & $\cdots$ & $\cdots$ & $\cdots$ & $\cdots$ & $\cdots$ & $\cdots$ & 1 & $\cdots$ & 4.26 & 4.26 \\
{\citetads{2020Natur.584..201R}} & $\cdots$ & $\cdots$ & $\cdots$ & $\cdots$ & $\cdots$ & $\cdots$ & 1 & $\cdots$ & 4.22 & 4.22 \\
{\citetads{2021A&A...647A.194F}} & $\cdots$ & $\cdots$ & $\cdots$ & $\cdots$ & $\cdots$ & $\cdots$ & 2 & $\cdots$ & 4.54 & 4.57 \\
{\citetads{2021ApJ...907..122M}} & $\cdots$ & $\cdots$ & $\cdots$ & $\cdots$ & $\cdots$ & $\cdots$ & 5 & $\cdots$ & 4.62 & 4.64 \\
{\citetads{2021ApJ...909..130R}} & $\cdots$ & $\cdots$ & $\cdots$ & $\cdots$ & $\cdots$ & $\cdots$ & 1 & $\cdots$ & 1.84 & 1.84 \\
{\citetads{2021ApJ...911...99F}} & $\cdots$ & $\cdots$ & $\cdots$ & $\cdots$ & $\cdots$ & $\cdots$ & 2 & $\cdots$ & 6.07 & 6.07 \\
{\citetads{2021ApJ...913...41L}} & $\cdots$ & $\cdots$ & $\cdots$ & $\cdots$ & $\cdots$ & 1 & $\cdots$ & $\cdots$ & 4.7 & 4.7 \\
{\citetads{2021MNRAS.507.3952R}} & $\cdots$ & $\cdots$ & $\cdots$ & $\cdots$ & $\cdots$ & $\cdots$ & 5 & $\cdots$ & 4.23 & 4.77 \\
\end{longtable}

}

We now comment on some references for which special attention is needed.


\paragraph[]{\citetads{2008ApJS..178..280B}: We use the median line flux in galaxies with more than one measurement. $\mathrm{L}_{\mathrm{IR}}$ is calculated using the IRAS 60 and 100$\,\mu$m fluxes as described in \citetads{2008ApJS..178..280B}. For two galaxies (\object{3C\,368} and \object{Z\,25-7}), we adopt the preferred Local Group velocity in NED to estimate the distance to these galaxies \citepads{1996AJ....111..794K}. For 19 galaxies (\object{DDO\,50}, \object{IC\,10}, \object{IC\,1613}, \object{IC\,342}, \object{M\,33}, \object{M\,81}, \object{M\,82}, \object{Maffei\,2}, \object{NGC\,0185}, \object{NGC\,0247}, \object{NGC\,0300}, \object{NGC\,1569}, \object{NGC\,4236}, \object{NGC\,4569}, \object{NGC\,6503}, \object{NGC\,6822}, \object{NGC\, 6946}, \object{VCC\,1043} and \object{VCC\,92}), we use the redshift-independent distances\footnote{These distances come from methods that use standard candles or rules, such as Cepheids and globular clusters.} from NED \citep{1999MNRAS.304..595G,2010ApJ...715..833O,2013AJ....146...86T,2014ApJ...782....4K}. Finally for \object{3C\,368}, we use the SFR from \citetads{2015A&A...575A..80P}, and for \object{NGC\,4038} and \object{NGC\,4039}, we use the SFRs from \citetads{2018ApJ...861...94H}.}

\paragraph{\citetads{2010ApJ...724..957S}: We adopt magnification factors of two and eight for \object{SMM\, J22471-0206} and \object{Hbootes03}, respectively. Due to the uncertain magnification in all other galaxies, we flag their luminosities as a precaution. For five galaxies (\object{2XMM\,J094144.6+385440}, \object{3C\,065}, \object{3C\,368}, \object{IRAS\, F10026+4949} and \object{IRAS\, F22231-0512}), we adopt the SFRs from other works \citep{2014A&A...568A..62D,2015A&A...575A..80P,2015MNRAS.448...75K,2021ApJ...910...44V}.}

\paragraph{\citetads{2012ApJ...752....2D}: We use updated values for the magnification factors: for \object{QSO\,J0831+5245} we adopt the upper limit of eight \citepads{2016A&A...587A..43S}, and for \object{[LWB2009]\,MM184222+593828} we adopt a value of 12 \citepads{2011ApJ...739L..30L}.}

\paragraph{\citetads{2013ApJ...776...38F}: We adopt the SFR value from \citetads{2016ApJ...829...93K} for \object{IRAS\,00397-1312}.}

\paragraph{\citetads{2014A&A...568A..62D}: We use the magnification factors of other works in cases where the lensing models are better, or for consistency with other references used in this work, for the following galaxies: \object{SDP\,81} \citepads{2011MNRAS.415.3473V}, \object{QSO\,J0831+5245} \citepads{2016A&A...587A..43S} and \object{[CRR2012]\, HLS\, J091828.6+514223} \citepads{2018A&A...609A.130L}. In addition, we take the magnification factor and SFR for \object{G15.v2.779}, \object{HFLS3} and \object{SMM\, J22471-0206} from \citetads{2020ApJ...898...33C}, \citetads{2014ApJ...790...40C} and \citetads{2015ApJ...806..260F}, respectively. For the following galaxies, we take SFR values from the literature: \object{2XMM\,J094144.6+385440} \citepads{2015MNRAS.448...75K}, \object{3C\,065} \citepads{2015A&A...575A..80P}, \object{3C\,368} \citepads{2015A&A...575A..80P}, \object{3C\,446} \citepads{2021ApJ...910...44V}, \object{IRAS\, F22231-0512} \citepads{2021ApJ...910...44V}, \object{SMM\, J02399-0136} \citepads{2015ApJ...806..260F}, \object{IRAS\, F10026+4949} \citepads{2013A&A...549A.125R} and \object{SWIRE\,J104704.97+592332.3} \citepads{2010ApJ...724..957S}.}

\paragraph{\citetads{2014ApJ...796...63M}: For \object{HXMM01}, we adopt the magnification factor from \citetads{2017ApJ...837...12W}. For \object{HLock01}, we adopt the magnification factor from \citetads{2011ApJ...738..125G}.}

\paragraph{\citetads{2015ApJ...799...13B}: We assume that the cosmology used is the one from \citetads{2003ApJS..148..175S}, as they mention that their work is a continuation of \citetads{2010ApJ...724..957S}. We assume that all galaxies in this sample are magnified \citepads{2018MNRAS.481.1976Z}.}

\paragraph{\citetads{2015ApJ...799...21S}: $\mathrm{L}_{\mathrm{IR}}$ is calculated with the IRAS 60 and 100$\,\mu$m fluxes as described in \citetads{2008ApJS..178..280B}.}

\paragraph{\citetads{2015ApJ...801...72R}: We use the redshift from NED and the infrared flux reported in their tables to estimate $\mathrm{L}_{\mathrm{IR}}$.}

\paragraph{\citetads{2015ApJ...806..260F}: Data for \object{QSO\,J0831+5245} comes from \citetads{2010ApJ...714L.147F}, where we have adopted the magnification factor from \citetads{2016A&A...587A..43S}. Although this galaxy is part of the sample, we do not use it for the comparison in this work because the SFR estimated from the $\mathrm{L}_{\mathrm{IR}}$ is unreliable due to the contribution of the AGN. Reported luminosities for [\ion{N}{II}] come from ZEUS for \object{QSO\,J0831+5245}  and \object{SMM\, J02399-0136}, and ALMA for \object{Cloverleaf} (\object{QSO\,J1415+1129}).}

\paragraph{\citetads{2015MNRAS.449.2883G}: We take the magnification values from \citetads{2016ApJ...826..112S}; we flag those as limits when there is no magnification factor available (e.g.\ for \object{SPT-S\,J051258-5935.6}). For \object{SPT-S\,J055138-5057.9} we adopt the magnification factor from \citetads{2020MNRAS.494.4090C}, while for \object{SPT-S\,J053816-5030.8} we adopt the magnification factor (18.8) reported by \citet{2016ApJ...826..112S}.}

\paragraph{\citetads{2016AJ....151...14C}: $\mathrm{L}_{\mathrm{IR}}$ is calculated using the IRAS 60 and 100$\,\mu$m fluxes as described in \citetads{2008ApJS..178..280B}. For \object{DDO\,155} we use the estimated SFR from H$\alpha$ by \citetads{2020AJ....160...66P}.}

\paragraph{\citetads{2016ApJ...819...69Z}: We obtain redshifts from NED when available. For five galaxies (\object{M\,82}, \object{NGC\,1569}, \object{NGC\,2976}, \object{NGC\,3077} and \object{NGC\,4569}) we use the redshift-independent distances from NED \citep{2013AJ....146...86T,2014ApJ...782....4K}. For \object{NGC\,3557} we use the preferred redshift distance. Most of the galaxies in this sample do not have an estimated SFR, so we use the SFRs from \citetads{2008ApJS..178..280B} and \citetads{2017ApJ...846...32D} for most of the missing estimates. We use SFRs from \citetads{2016ApJ...829...93K} for \object{UGC\,2369} and \object{NGC\,5010}, and \citetads{2015ApJS..217....1T} for \object{IC\,4518A}.}

\paragraph{\citetads{2016ApJ...824..146Z}: Some of the galaxies are identified to have second velocity components in their spectra; therefore we use the median luminosity value in those cases.}

\paragraph{\citetads{2016ApJ...827...34O}: The pair of galaxies in this reference is treated as one galaxy (\object{H-ATLAS\,J000307.2-330250}), as the second component does not exist in SIMBAD. We use the IR luminosities derived from SED fitting from \citetads{2021MNRAS.500.3667S}.}

\paragraph{\citetads{2016ApJ...829...93K}: We assume that the additional regions in \object{NGC\,6946} and \object{NGC\,4038} (\object{Antennae}) are also part of the galaxy. We use SFRs from other references for the following galaxies: \object{3C\,405} \citepads{2008ApJS..178..280B}; \object{Mrk\,1298} \citepads{2014ApJS..214...23S}; \object{IC\,4518A} \citepads{2015ApJS..217....1T}; \object{3C\,315} and \object{3C\,433} \citepads{2016AJ....151..120W}; \object{ESO\,255-IG007}, \object{NGC\,2976} and \object{VV\,705} \citepads{2016ApJ...819...69Z}; \object{3C\,305} \citepads{2016ApJS..226...19F}; \object{3C\,31} \citepads{2016ApJ...818..182V}; \object{IRAS\,08355-4944}, \object{IRAS\, F01417+1651}, \object{MCG-03-34-064}, \object{NGC\,877} and \object{VV340a} \citepads{2017ApJ...846...32D}; and, \object{Antennae} and \object{NGC\,4151} \citepads{2018ApJ...861...94H}.}

\paragraph{\citetads{2016ApJ...832..209U}: We assume a magnification factor of 30 for \object{IRAS\, F10214+4724}. When correcting the $\mathrm{L}_{\mathrm{IR}}$ for this magnification factor, the SFR derived from the \citetads{2012ARA&A..50..531K} relation is similar to the SFR estimated by \citetads{2006AJ....132.2398E}. We use the  magnification factor of eight as an upper limit for \object{QSO\,J0831+5245} \citepads{2016A&A...587A..43S}. We assume this work uses the cosmology of \citetads{2009ApJS..180..330K}.}

\paragraph{\citetads{2016ApJS..226...19F}: $\mathrm{L}_{\mathrm{IR}}$ is calculated with the IRAS 60 and 100$\,\mu$m fluxes as described in \citetads{2008ApJS..178..280B}. For \object{LEDA\,101275} and \object{Mrk\,463E} we use the IR luminosity reported by \citetads{2011PASP..123.1011A}. For eight galaxies (\object{M\,82}, \object{NGC\,1569}, \object{NGC\,2976}, \object{NGC\,3077} and \object{NGC\,4569}) we use redshift-independent distances from NED \citepads{2013AJ....146...86T}. For \object{NGC\,4569} and \object{NGC\,6946} we adopt distances reported by \citetads{2008ApJ...683...78C} and \citetads{2014AJ....148..107R}, respectively. We use SFRs from other references for the following galaxies: \object{3C\,405} \citepads{2008ApJS..178..280B}; \object{IC\,4518A} \citepads{2015ApJS..217....1T}; \object{2E\,4728}, \object{3C\,317} and \object{MCG+05-33-005} \citepads{2014MNRAS.444L..63F}; \object{3C\,33}, \object{3C\,234}, \object{3C\,315} and \object{3C\,433} \citepads{2016AJ....151..120W}; \object{LEDA\,3098117}, \object{LEDA\,4666674}, \object{NGC\,7592W} and \object{Z\,468-2} \citepads{2017ApJ...846...32D}; \object{Centaurus\, A} and \object{Circinus galaxy} \citepads{2018ApJ...861...94H}; \object{Mrk 266B} \citetads{2019AJ....158..169S}; and, \object{ESO\,141-55}, \object{2MASX\,J00535615-7038045}, \object{IRAS\,03450+0055} and \object{UGC\,12138}	\citetads{2022MNRAS.510..687R}.}

\paragraph{\citetads{2016MNRAS.462L...6K}: We use only the information of \object{[ZFM2015]\, A383\, 5.1}, as observations of the other galaxy (\object{MS0451-H}) are only upper limits and are not identified as a galaxy in SIMBAD.}

\paragraph[]{ \citetads{2017ApJ...846...32D}: Redshifts are taken from NED to convert from the WMAP cosmology \citepads{2009ApJS..180..330K} used for the sample in \citetads{2009PASP..121..559A} to the Planck cosmology used in this work. We obtain  IR luminosities from the sample webpage\footnote{GOALS sample at \url{https://goals.ipac.caltech.edu}}.}

\paragraph{\citetads{2017ApJ...846..105O}: We remove two galaxies that are not identified in SIMBAD (\object{MS0451-H} and \object{
[CFP2010]\,BDF\,3299} sub-region).}

\paragraph{\citetads{2018A&A...609A.130L}: For some galaxies we use a different magnification factor to those reported in \citetads{2018A&A...609A.130L}. For \object{H-ATLAS\,J142413.9+022304} and \object{SPT-S\,J034510-4725.6} we use the magnification factor used by \citetads{2020ApJ...898...33C}, and for \object{[CLM2003]\,J022802.97-041618.3} we use the values by \citetads{2017ApJ...846..105O}. We use the IR luminosity value for \object{AzTEC 3} from \citetads{2021MNRAS.500.3667S}. For galaxies coming from \citetads{2016ApJ...833...71A} we use the IR luminosities of \citetads{2017ApJ...847...21F}. We use the reported SFR instead of converting from the IR luminosities for \object{[CLM2003]\,J022802.97-041618.3} and \object{[WMH2013]\,5}. Finally, we use the SFR from \citetads{2020Natur.581..269N} for \object{ALMA\,J081740.86+135138.2}.}

\paragraph{\citetads{2018ApJ...856..174V}: We use a magnification factor of seven so the dust disk of \object{H-ATLAS\,J113526.3-014605} can be consistent with other high-$z$ SMG.}

\paragraph{\citetads{2018ApJ...859...12G}: The data presented with this reference is an improvement of the results presented in \citetads{2012MNRAS.427.1066S}.}

\paragraph{\citetads{2018ApJ...867..140L}: We adopt the IR luminosity from \citetads{2013ApJ...779...25B}. Two velocity components are measured in the [\ion{C}{II}] line of \object{SPD\,11}. We adopt the magnification factor estimated for the dominant (red) component.}

\paragraph{\citetads{2018MNRAS.473...20R}: We adopt a magnification factor of 10.9 according to \citetads{2011ApJ...738..125G}.}

\paragraph{\citetads{2018MNRAS.481...59Z}: We use the magnification factor reported in their Table 4. For \object{H-ATLAS\,J084933.4+021443} and \object{SDP\,81}, we adopt 2.8 and 25, respectively, as magnification factors \citep{2017ApJ...837...12W,2011MNRAS.415.3473V}.}

\paragraph{\citetads{2019A&A...626A..23C}: We use distances reported by \citetads{2013A&A...557A..95R} in most of the galaxies. For ten galaxies (\object{ESO\,495-21}, \object{Mrk\,209}, \object{NGC\,1140}, \object{NGC\,1569}, \object{NGC\,1705}, \object{NGC\,4214}, \object{NGC\,5253}, \object{NGC\,625}, \object{UGC\,4483} and \object{UGC\,6456}) we use redshift-independent distances from NED \citepads{2013AJ....146...86T}. For \object{ESO\,350-38} and \object{HS\,0822+3542} we use the preferred redshift values from NED. For \object{UGCA\,116} we use the distance reported in \citetads{2009ApJ...696.1668B}. We use the reported distance used in \citetads{2019A&A...626A..23C} for \object{Mrk\,33} and discard the information from the \object{LMC} and \object{SMC}. For \object{2MASX\,J12390403+3920437} we use the SFR reported by \citetads{2017A&A...599A..71D}.}

\paragraph{\citetads{2019A&A...631A.167D}: We use the magnification factor of 32.7 by \citetads{2016ApJ...826..112S}.}

\paragraph{\citetads{2019ApJ...876....1T}: We assume the cosmology used is a $\Lambda$CDM cosmology with $\Omega= 0.27$ and $H_0 =70$ km s$^{-1}$ Mpc$^{-1}$.}

\paragraph{\citetads{2019ApJ...882...10N}: For \object{[NBW2019]\,J0842+1218C2} we assume that the upper limit for SFR is 100 M$_{\sun}$ yr$^{-1}$.}

\paragraph{\citetads{2019ApJ...883L..29L}: We use the IR luminosity values and [\ion{C}{II}] measurements from \citetads{2006ApJ...645L..97I}.}

\paragraph{\citetads{2019PASJ...71..109H}: The reported values for the SFR are assumed to be upper limits, as those come from QSOs.}

\paragraph{\citetads{2020A&A...643A...2B}: These objects have not been ingested in SIMBAD, but we use the coordinates of the closest object. However, this can lead to incorrect identifications. For example, the closest object to the coordinates of \object{VUDS-COSMOS-5100541407} is a star (\object{COSMOS\,877137}). Caution is therefore required. We use SFRs from SED fitting except for cases where it seems to be overestimated. For \object{COSMOS-DEIMOS-873756} and \object{VUDS-COSMOS-510596653} we assume the SFRs are upper limits, where for \object{VUDS-COSMOS-510596653} we use the estimated IR luminosity and convert it to SFR.}

\paragraph{\citetads{2020ApJ...904..130V}: We do not include the companions of the galaxies, as those are not yet identified in SIMBAD.}

\paragraph{\citetads{2020MNRAS.494.4090C}: We use the SFRs from \citetads{2015MNRAS.449.2883G} for most of the galaxies. We obtain IR luminosities for \object{SPT-S\,J234942-5638.2}, \object{SPT-S\,J235339-5010.1} and \object{SPT-S\,J235718-5153.7} from \citetads{2012ApJ...756..101G}. For \object{SPT-S\,J020258-6121.2}, \object{SPT-S\,J045859-5805.1}, \object{SPT-S\,J045912-5942.4} and \object{SPT-S\,J204823-5520.7} we obtain IR luminosities and magnification factors from \citetads{2020ApJ...905...85S}. For \object{SPT-S\,J231124-5450.5} we use the magnification factor from \citetads{2016ApJ...826..112S}. For \object{SPT-S\,J235149-5722.2} we use the IR luminosity from \citetads{2017ApJ...847...21F} as an upper limit. Finally, we discard the galaxies without coordinates information in SIMBAD.}

\paragraph{\citetads{2020Natur.584..201R}: We adopt the values that take into account the magnification of 32.3.}

\paragraph{\citetads{2021ApJ...907..122M}: We adopt the values that take into account the magnification. These galaxies are not yet included in SIMBAD.}

\paragraph{\citetads{2021ApJ...911...99F}: We adopt the values that  take into account the magnification (their Tables 3 and 4). These galaxies are not yet included in SIMBAD.}

\paragraph{\citetads{2021MNRAS.507.3952R}: We adopt the values that take into account the magnification. For five galaxies (\object{SPT-S\,J011308-4617.7}, \object{SPT-S\,J034510-4725.7}, \object{SPT-S\,J044143-4605.5}, \object{SPT-S\,J213244-5803.1} and \object{SPT-S\,J214654-5507.9}), we adopt the magnification factors from \citetads{2020MNRAS.494.4090C}.}

\section{Dataset of estimated emission line luminosities}\label{app:estimations_dataset}

Predictions of the eight emission lines discussed in this work (Table~\ref{tab:lines}) are available at the CDS and in a Zenodo repository at \url{https://doi.org/10.5281/zenodo.6576202}. The dataset contains the total line luminosities as well as the contributions of the different ISM phases. We show ten rows from the dataset containing these luminosities in Table~\ref{tab:lums_est} as an example of the format and content. We also present a table of the physical parameters involved in the predicted line luminosities.  We show ten rows from the dataset containing the physical parameters in Table~\ref{tab:phys_est} as an example of the format and content. The first columns of these tables matches the Group Number in the \textsc{EAGLE} database \citepads{2016A&C....15...72M}, which is the unique identifier of the \textsc{FoF} (Friends-of-Friends) halo of a given galaxy (unique per snapshot). Although we have applied our physical model on the sample of 8\,227 galaxies simulated with \textsc{EAGLE}, the dataset contains 8\,224 galaxies since one of the galaxies does not have enough gas for the estimates, and the other two galaxies fail to predict luminosities in a reasonable amount of computational time.

\begin{sidewaystable}
\caption{Example of dataset with the line luminosity estimates derived from this work.}\label{tab:lums_est}
\centering
    \begin{tabular}{ccccccccccccc}
\hline
\hline
Group & Sim. & z & L$_{\mathrm{NIII_{57}}}$ & L$_{\mathrm{NIII_{57},DIG}}$ & L$_{\mathrm{NIII_{57},HII}}$ & L$_{\mathrm{NIII_{57},ato}}$ & L$_{\mathrm{NIII_{57},mol}}$ & L$_{\mathrm{OIII_{88}}}$ & L$_{\mathrm{OIII_{88},DIG}}$ & L$_{\mathrm{OIII_{88},HII}}$ & L$_{\mathrm{OIII_{88},ato}}$ & L$_{\mathrm{OIII_{88},mol}}$ \\
Number &  &  & dex(L$_{\odot}$) & dex(L$_{\odot}$) & dex(L$_{\odot}$) & dex(L$_{\odot}$) & dex(L$_{\odot}$) & dex(L$_{\odot}$) & dex(L$_{\odot}$) & dex(L$_{\odot}$) & dex(L$_{\odot}$) & dex(L$_{\odot}$) \\
\hline
688 & Recal & 0 & 2.93 & 2.93 & $\cdots$ & $\cdots$ & $-$4.93 & 3.51 & 3.51 & $\cdots$ & $\cdots$ & $\cdots$ \\
11 & Recal & 1 & 6.33 & 5.52 & 6.26 & $-$0.42 & $-$3.19 & 6.89 & 6.17 & 6.80 & 0.06 & $-$3.56 \\
1 & Recal & 5 & 6.48 & 4.55 & 6.48 & 0.60 & $\cdots$ & 7.20 & 5.23 & 7.20 & 1.06 & $\cdots$ \\
2442 & Ref100 & 0 & 6.44 & 4.21 & 6.44 & $-$0.67 & $-$3.13 & 6.90 & 4.83 & 6.89 & $-$0.22 & $-$3.35 \\
4634 & Ref100 & 1 & 5.30 & 4.97 & 5.03 & $-$0.34 & $-$2.56 & 5.90 & 5.60 & 5.59 & 0.18 & $-$2.86 \\
6207 & Ref100 & 2 & 6.40 & 4.82 & 6.38 & 0.70 & $-$4.07 & 7.10 & 5.46 & 7.09 & 1.26 & $-$4.07 \\
5662 & Ref100 & 3 & 6.69 & 4.49 & 6.69 & 0.39 & $-$4.41 & 7.39 & 5.17 & 7.39 & 0.96 & $-$4.70 \\
4194 & Ref100 & 4 & 6.41 & 4.18 & 6.40 & 0.33 & $\cdots$ & 7.08 & 4.91 & 7.08 & 0.85 & $\cdots$ \\
229 & Ref100 & 4 & 5.93 & 5.83 & 5.26 & 1.18 & $-$2.95 & 6.55 & 6.50 & 5.60 & 1.75 & $-$3.31 \\
332 & Ref100 & 5 & 6.61 & 4.14 & 6.60 & 0.85 & $\cdots$ & 7.28 & 4.86 & 7.28 & 1.37 & $-$4.70 \\
\hline
\end{tabular}
\end{sidewaystable}

\begin{sidewaystable}
\caption{Example of physical parameters dataset derived from this work.}\label{tab:phys_est}
\centering
    \begin{tabular}{ccccccccccccc}
\hline
\hline
Group & Sim. & z & M$_{\mathrm{\ast}}$ & M$_{\mathrm{gas}}$ & M$_{\mathrm{neutral}}$ & Age$_{\mathrm{\ast}}$ & log$(Q)$ & ISRF & SFR & $\rm{Z/Z}_{\odot}$ & Pressure & n$(\mathrm{H})_{\mathrm{cloud}}$ \\
Number &  &  & dex(M$_{\odot}$) & dex(M$_{\odot}$) & dex(M$_{\odot}$) & dex(Gyr) & dex(1 / s) & Habing & dex(M$_{\odot}$ / yr) &  & dex(K / cm$^3$) & dex(1 / cm$^3$) \\
\hline
688 & Recal & 0 & 8.154 & 7.208 & 6.930 & 0.852 & 34.769 & 0.530 & $-$2.511 & 0.356 & 4.151 & 0.594 \\
11 & Recal & 1 & 10.186 & 9.665 & 9.273 & 0.260 & 35.761 & 0.804 & $-$0.409 & $-$0.344 & 7.816 & 0.854 \\
1 & Recal & 5 & 8.732 & 10.103 & 9.947 & $-$0.656 & 44.669 & 1.496 & 0.269 & $-$1.586 & 8.777 & 1.473 \\
2442 & Ref100 & 0 & 9.847 & 9.735 & 9.625 & 0.784 & 38.941 & 0.687 & $-$0.319 & $-$0.073 & 5.975 & 0.406 \\
4634 & Ref100 & 1 & 9.458 & 9.625 & 9.271 & 0.296 & 40.203 & 2.253 & 0.123 & $-$0.127 & 7.756 & 1.979 \\
6207 & Ref100 & 2 & 9.151 & 9.846 & 9.410 & $-$0.098 & 41.948 & 1.828 & 0.087 & $-$0.562 & 7.196 & 1.361 \\
5662 & Ref100 & 3 & 9.057 & 9.854 & 9.566 & $-$0.161 & 42.137 & 1.925 & 0.271 & $-$0.757 & 7.534 & 1.560 \\
4194 & Ref100 & 4 & 8.682 & 9.814 & 9.449 & $-$0.430 & 44.019 & 1.898 & 0.064 & $-$1.385 & 7.391 & 1.472 \\
229 & Ref100 & 4 & 10.371 & 10.088 & 9.754 & $-$0.346 & 40.174 & 1.466 & $-$0.067 & $-$1.629 & 7.803 & 1.348 \\
332 & Ref100 & 5 & 8.735 & 10.115 & 9.926 & $-$0.690 & 45.148 & 1.757 & 0.420 & $-$1.413 & 7.981 & 1.563 \\
\hline
\end{tabular}

\end{sidewaystable}

\section{Linear regressions for lines}\label{app:linregcoef}

Assuming that FIR luminosities change with SFR and redshift ($z$), we fit the estimated data to obtain linear relations in terms of these parameters for each FIR emission line. At any $z$ there is a linear relation of the form
\begin{eqnarray} 
    \displaystyle \log(\mathrm{L}_{\mathrm{line}}) &=& c_{0} + c_{1} \log(\mathrm{SFR}) + c_{2} \log(1+z) + \nonumber\\ 
    && \displaystyle \quad c_{3} \log(\mathrm{SFR}) \log(1+z),
    \label{eq:linreg}
\end{eqnarray}
with SFR in units of M$_{\sun}$ yr$^{-1}$ and line luminosities in L$_{\sun}$. The values for the coefficients presented in Table~\ref{tab:linreg} are obtained from the combination of \textsc{Recal-L0025N0752} and \textsc{Ref-L100N1504} mock data. We also estimate the coefficients of the [\ion{O}{III}] and [\ion{N}{III}] line relations using only galaxies where \textsc{Hii} regions are dominant (the contribution is higher than 50\%). 

\begin{table*}
    \centering
    \caption{Linear relations derived from this work for each of the FIR emission lines using Eq.~\ref{eq:linreg}}
    \begin{tabular}{cccccc}
\hline
\hline
& \multicolumn{4}{c}{Coefficients} & \\
\cline{2-5}
log(L$_{\rm{line}}$) & $c_{0}$ & $c_{1}$ & $c_{2}$ & $c_{3}$ & 1$\sigma$ \\
\hline
{[\ion{O}{III}]} 52 $\mu$m & 6.72$\pm$0.02 & 1.54$\pm$0.02 & $-$0.15$\pm$0.03 & $-$0.41$\pm$0.03 & 0.63 \\
{[\ion{N}{III}]} 57 $\mu$m & 6.38$\pm$0.02 & 1.54$\pm$0.02 & $-$0.39$\pm$0.03 & $-$0.35$\pm$0.03 & 0.61 \\
{[\ion{O}{I}]} 63 $\mu$m & 6.13$\pm$0.01 & 1.03$\pm$0.01 & 0.70$\pm$0.02 & 0.09$\pm$0.02 & 0.38 \\
{[\ion{O}{III}]} 88 $\mu$m & 6.84$\pm$0.02 & 1.46$\pm$0.02 & $-$0.07$\pm$0.03 & $-$0.35$\pm$0.03 & 0.60 \\
{[\ion{N}{II}]} 122 $\mu$m & 5.59$\pm$0.01 & 1.13$\pm$0.01 & $-$0.68$\pm$0.02 & 0.06$\pm$0.02 & 0.46 \\
{[\ion{O}{I}]} 145 $\mu$m & 4.96$\pm$0.01 & 1.10$\pm$0.01 & 0.69$\pm$0.02 & $-$0.15$\pm$0.02 & 0.41 \\
{[\ion{C}{II}]} 158 $\mu$m & 6.55$\pm$0.01 & 0.72$\pm$0.01 & 0.14$\pm$0.01 & 0.24$\pm$0.01 & 0.24 \\
{[\ion{N}{II}]} 205 $\mu$m & 5.48$\pm$0.01 & 0.93$\pm$0.01 & $-$0.82$\pm$0.02 & 0.24$\pm$0.02 & 0.42 \\
\hline
& \multicolumn{4}{c}{Coefficients for \textsc{Hii} regions} & \\
\hline
{[\ion{O}{III}]} 52 $\mu$m & 6.99$\pm$0.02 & 1.11$\pm$0.02 & $-$0.57$\pm$0.03 & 0.25$\pm$0.04 & 0.53 \\
{[\ion{N}{III}]} 57 $\mu$m & 6.65$\pm$0.02 & 1.15$\pm$0.02 & $-$0.82$\pm$0.03 & 0.26$\pm$0.04 & 0.51 \\
{[\ion{O}{III}]} 88 $\mu$m & 7.11$\pm$0.02 & 1.06$\pm$0.02 & $-$0.48$\pm$0.03 & 0.26$\pm$0.04 & 0.50\\
\hline
\end{tabular}
    \label{tab:linreg}
\end{table*}

\end{appendix}

\end{document}